\documentclass[prd,preprint,showpacs,preprintnumbers,amsmath,amssymb]{revtex4}

\usepackage{mathrsfs}
\DeclareMathAlphabet{\mathpzc}{OT1}{pzc}{m}{it}

\usepackage{ulem}
\usepackage[english]{babel}

\usepackage{amsmath}
\usepackage{amssymb}

\usepackage{mathrsfs}
\DeclareMathAlphabet{\mathpzc}{OT1}{pzc}{m}{it}

\usepackage{graphicx}

\usepackage{cancel} 
\usepackage{paralist}
\usepackage{booktabs}
\usepackage{hyperref}
\usepackage{xspace}

\usepackage{graphicx,color,verbatim,acronym,epsfig,subfigure}
\usepackage{slashed,skak}
\usepackage{latexsym,amsfonts,amssymb,amsmath,makeidx,mathrsfs,multirow,lineno,bm,bbm}

\begin{document}
\title{FIMP dark matter mediated by massive gauge boson around the phase transition period and the gravitational waves production}
\author{Ligong Bian$^{2,3}$}
\author{Yi-Lei Tang$^{1}$}
\email{tangylei@mail.sysu.edu.cn}
\author{Ruiyu Zhou$^{4}$}
\email{zhoury@cqupt.edu.cn}

\affiliation{
    $^1$~School of Physics, Sun Yat-Sen University, Guangzhou 510275, China\\
	$^2$~Department of Physics, Chongqing University, Chongqing 401331, China,\\
	$^3$~Chongqing Key Laboratory for Strongly Coupled Physics, Chongqing 401331, China\\
    $^4$~School of Science, Chongqing University of Posts and Telecommunications, Chongqing 400065, P. R. China
}
\date{\today}
\begin{abstract}
We study the feebly interacting massive particle dark matter whose production processes are significantly affected by the phase evolution and the complicated thermal corrections to the vector boson. We calculate the freeze-in processes to obtain the correct dark matter relic density by enumerating all the possible $1 \leftrightarrow 2$ and $2 \leftrightarrow 2$ processes. The predicted gravitational waves emitted by the first-order phase transitions and the cosmic strings are evaluated.
\end{abstract}

\graphicspath{{figure/}}

\maketitle
\baselineskip=16pt

\pagenumbering{arabic}

\vspace{1.0cm}
\tableofcontents

\newpage

\section{Introduction}

Dark matter is considered to account for about 84\% of the matter in our universe\cite{Planck:2018vyg}. A natural way to explain the relic abundance is to attribute the creation of the dark matter particles to the thermal plasma in the early epoch of our universe. If the dark matter particles are initially assigned to be in thermal equilibrium with the standard model (SM) sectors, and then ``freeze-out'' from the plasma as the universe expands and cools down, such kind of dark matter is usually called the weak interaction massive particle (WIMP, for a review, see \cite{Bertone:2004pz}).  On the other hand, if the dark matter interacts feebly with the early plasma and is created (or ``freeze-in'') gradually from a void initial condition,  such kind of dark matter can be called a feebly-interacting massive particle (FIMP)\cite{McDonald:2001vt, Choi:2005vq, Kusenko:2006rh, Petraki:2007gq, Hall:2009bx, Bernal:2017kxu}.

In the WIMP case, the dark matter freezes out roughly around the temperature $T_f \sim \frac{m_{\chi}}{26}$, where $m_{\chi}$ is the mass of the dark matter particle.  Practical evaluation shows that usually the thermal corrections become negligible so that the usual methodology based upon the zero-temperature theories becomes a very good approximation to calculate the relic abundance. On the other hand, FIMPs are created much earlier when $\frac{T}{m_{\chi}} \sim 0.3$-$3$ in a relatively higher temperature, so the evolution of the phases at the early universe, including the phase transition processes may affect the dark matter production processes. In particular, the external leg's effective masses may vary significantly as both the vacuum expectation values (VEVs) and the thermal corrections evolve. The internal mediator's propagators for the scattering process can also change during this period.  As a result, the production processes of the dark matter particles can be switched on and/or off due to the threshold effects, finally altering the interaction rates evidently during the freeze-in processes.

If vector boson participates in the freeze-in interactions~\cite{Essig:2011nj, Chu:2011be, Davidson:2000hf, Vogel:2013raa, An:2018nvz, Dvorkin:2019zdi, Heeba:2019jho, Hambye:2019dwd, Bhattacharyya:2018evo, Das:2021nqj}, more complexities arise due to the separation of the longitudinal and the transverse degrees of freedom. For the massive vector boson, part of the goldstone boson also decouple from the longitudinal polarization degree of freedom.  One has to evaluate the thermally averaged interaction rates carefully by considering all these degrees of freedom separately\cite{Tang:2019kwx}. Furthermore, the thermally corrected dispersion relations of the vector bosons are no longer Lorentz-invariant, thus accumulates the complexity of the phase space integration.

In this paper, we rely on a model including the fermionic dark matter $\chi$ and a $U(1)^{\prime}$ gauge boson $A^{\prime}$. Two Higgs singlets $\Phi_{s}$ and $\Phi_{w}$ are introduced to break the gauge symmetry spontaneously. As the temperature drops, $\Phi_{w}$ and $\Phi_s$ acquire VEVs successively, and phase transitions then occur to endow a mass term to the $A^{\prime}$ through the VEVs.  The dark matter is feebly charged under the $U(1)^{\prime}$ group, so that the gauge boson participates the freeze-in processes. $\Phi_{w}$ is also assigned with a proper charge to formulate a Yukawa coupling with the dark matter to account for the interaction between the longitudinal polarization of the $A^{\prime}$ and the dark matter particle.  In this paper, we consider both the thermal masses and the VEV-induced masses of the $A^{\prime}$, and their effects on the dark matter production rate.

The discovery of the gravitational waves by LIGO\cite{LIGOScientific:2016aoc} initiates a new method to verify some new physics models through observing their predicted production of stochastic gravitational waves in the early universe.  First-order phase transitions associated with dark matter production~\cite{Jiang:2015cwa, Chala:2016ykx, Chao:2017vrq, Beniwal:2017eik, Huang:2017rzf, Huang:2017kzu, Hektor:2018esx, Baldes:2018emh, Madge:2018gfl, Beniwal:2018hyi, Bian:2018mkl, Bian:2018bxr, Shajiee:2018jdq, Mohamadnejad:2019vzg, Kannike:2019mzk, Paul:2019pgt, Chen:2019ebq, Barman:2019oda, Chiang:2019oms, Borah:2020wut, Kang:2020jeg, Pandey:2020hoq, Han:2020ekm, Alanne:2020jwx, Wang:2020wrk, Ghosh:2020ipy, Huang:2020crf, Chao:2020adk, Mazumdar:2018dfl} can emit gravitational waves to be probed by the future space-based gravitational wave interferometers e.g., as LISA\cite{LISA:2017pwj}, TianQin\cite{TianQin:2015yph, Hu:2018yqb, TianQin:2020hid}, Taiji\cite{Hu:2017mde, Ruan:2018tsw}, DECIGO\cite{Seto:2001qf, Kudoh:2005as}, and BBO{\cite{Ungarelli:2005qb, Cutler:2005qq}.
In our paper, the spontaneously symmetry breaking of the local $U(1)^\prime$ symmetry can yield the formation of cosmic strings for the scenario of  $\langle \Phi_w \rangle \gg \langle \Phi_s \rangle$, which is essential for a considerable or dominate contribution of the dark matter production rate from the longitudinal vector boson\footnote{For a study on heavy dark matter together with gravitational waves from topological defects such as cosmic strings and domain walls can be found in Ref.~\cite{Bian:2021vmi,Deng:2020dnf}.}. These cosmic strings then collide and self-interact to form loops, and the loops finally disappears, with the legacy of significant gravitational waves formulated via cusp, kink and kink-kink collisions\cite{Gouttenoire:2019rtn, Gouttenoire:2019kij, Auclair:2019wcv}.   
We estimate the possibility to probe these gravitational waves emitted from the cosmic strings and first-order phase transitions.

This paper is organized as follows: our model is described in Sec.~\ref{sec:modte};  the methodology for the evaluation of phase transition and gravitational waves are given in Sec.~\ref{sec:ptgw}; we go into details for our calculation of the freeze in production process of the fermionic dark matter in Sec.~\ref{FreezeInRates}; we comment on the phenomenological constraints on the model in Sec.~\ref{sec:phecon}; numerical results of the dark matter and gravitational wave productions for some benchmarks are given in Sec.\ref{sec:numres}; the Sec.~\ref{sec:sum} is devoted to the summary and future prospect; and some details for the readers are given in Appendix \ref{sec:thermass},\ref{ThermalCorrectionGaugeFermion},\ref{intrate},\ref{DetailsFreezeIn}.

\section{Model description}
\label{sec:modte}
In this paper, besides the SM particles, we introduce two dark Higgs singlets of $\Phi_{s}$, $\Phi_{w}$, and one Dirac fermionic $\chi$  field that are all charged under the $U(1)_{\text{dark}}$ gauge group with the corresponding gauge field $A_{\mu}^{\prime}$. $\chi$ contains two Weyl components, which should always appear in pair to elude the anomaly.  We impose a $Z_2$ symmetry under which $\chi$ is $Z_2$-odd, while all the other particles are $Z_2$-even. The $U(1)_{\text{dark}}$ charge carried by $\chi$  is denoted by $t_{\chi}$, and the $U(1)_{\text{dark}}$ charge carried by $\Phi_s$, $\Phi_w$ is denoted by $t_s=1$ and $t_w$. For the purpose of the freeze-in scenario,  $t_{\chi} \ll 1$ so $\chi$ could not directly interact with the $\Phi_s$. Assigning $t_w = 2 t_{\chi}$ gives rise to the possible tenuous Yukawa coupling between $\chi$ and $\Phi_w$. After $\Phi_{w}$ acquires the VEV, the two Weyl components of $\chi$ split, and the lighter one becomes the dark matter candidate, with its stability guaranteed by the $Z_2$ symmetry. With the above setups, the total Lagrangian corresponding to the dark sector is written below,
\begin{eqnarray}
\mathcal{L} \supset \mathcal{L}_{\text{kin}} + \mathcal{L}_{\chi m} + \mathcal{L}_Y - V(H, \Phi_s, \Phi_w),
\end{eqnarray}
where
\begin{eqnarray}
\mathcal{L}_{\text{kin}} &=& -\frac{1}{4} F_{\mu \nu}^{\prime} F^{\prime \mu \nu} + D_{\mu} \Phi_s (D^{\mu} \Phi_s)^\dagger + D_{\mu} \Phi_w (D^{\mu} \Phi_w)^\dagger + i \bar{\chi} D_{\mu} \gamma^{\mu} \chi, \nonumber \\
\mathcal{L}_{\chi m} &=& m_{\chi} \overline{\chi} \chi, \nonumber \\
\mathcal{L}_Y &=& \frac{\sqrt{2} y_\chi}{2} \Phi_w \bar \chi \chi^C + h.c..
\end{eqnarray}
Here, $F_{\mu \nu}^{\prime} = \partial_{\mu} A_{\nu}^{\prime} - \partial_{\nu} A_{\mu}^{\prime}$, $D_{\mu}= \partial_{\mu} + i t g_D A'_{\mu}~$,$~y_\chi \ll 1$,  and $g_D$ is the dark gauge coupling constant.  For simplicity, we define $g_{\chi} = t_{\chi} g_D$, and $g_{w} = t_w g_D = 2 t_{\chi} g_D$, so $D_{\mu} \chi= \partial_{\mu} \chi + i g_{\chi} A'_{\mu} \chi$ and $D_{\mu} \Phi_w= \partial_{\mu} \Phi_w + i g_w A'_{\mu} \Phi_w$. The potential term is given by
\begin{eqnarray}
V(H,\Phi_s,\Phi_w)&= & \mu_0^2 (H^\dagger H) + \lambda(H^\dagger H)^2+\lambda_{sh} ( \Phi_s^* \Phi_s H^\dagger H)+ \lambda_{wh}( \Phi_w^* \Phi_w H^\dagger H) \nonumber \\ 
&+& \mu_s^2 \Phi_s^* \Phi_s + \mu_w^2 \Phi_w^* \Phi_w + \lambda_s (\Phi_s^* \Phi_s)^2+ \lambda_w (\Phi_w^* \Phi_w)^2 + \lambda_{sw} (\Phi_s^* \Phi_s)(\Phi_w^* \Phi_w),
\end{eqnarray}
where $H$ is the SM Higgs doublet. 

We expand the scalar fields around their classical backgrounds as,
\begin{eqnarray}
  H &=&
  \begin{pmatrix}
    G^+ + \tilde{G}^+
    \\
    \frac{ h + i G^0 + \tilde{h}+\tilde{G}^0 }{\sqrt{2}}
  \end{pmatrix}, \nonumber \\
  \Phi_s &=& \phi_s + i \phi_{s\eta}\; + \frac{\tilde{\phi}_s+i \tilde{\phi}_{s \eta}}{\sqrt{2}}, \nonumber \\
  \Phi_w &=& \phi_w + i \phi_{w\eta} +\frac{\tilde{\phi}_w + i \tilde{\phi}_{w\eta}}{\sqrt{2}},
\end{eqnarray}
where $h$, $G^0$, $G^+$, $\phi_s$, $\phi_{s \eta}$, $\phi_w$, $\phi_{w \eta}$ are background fields, and the corresponding $\tilde{h}$, $\tilde{G}^0$, $\tilde{G}^+$, $\tilde{\phi}_s$, $\tilde{\phi}_{s \eta}$, $\tilde{\phi}_w$, $\tilde{\phi}_{w \eta}$ are ``particles''.
The $Z_2$-odd $\chi$ can be decomposed into two Weyl spinors
\begin{eqnarray}
\chi = \left[ \begin{array}{c}
\chi_L \\
i \sigma^2 \chi_R^*
\end{array} \right],
\end{eqnarray}
and the mass term can be written by
\begin{eqnarray}
\mathcal{L} \supset \frac{1}{2} \left[ \chi_L^T ~~ \chi_R^T \right] \left[ \begin{array}{cc}
\delta m & m_{\chi} \\
m_{\chi} & \delta m
\end{array} \right] \left[ \begin{array}{c}
\chi_L \\
\chi_R
\end{array} \right] + \text{h.c.}, \label{ChiMassMatrix}
\end{eqnarray}
where $\delta m = \sqrt{2} y_{\chi} \phi_w$. Diagonalizing (\ref{ChiMassMatrix}) with
\begin{eqnarray}
\tilde{\chi}_1 &=& \frac{i}{\sqrt{2}}(\chi_L-\chi_R), \nonumber \\
\tilde{\chi}_2 &=& \frac{1}{\sqrt{2}}(\chi_L+\chi_R),
\end{eqnarray}
gives rise to
\begin{eqnarray}
\mathcal{L} \supset \frac{1}{2} \left[ \tilde{\chi}_1^T ~~ \tilde{\chi}_2^T \right] \left[ \begin{array}{cc}
m_{\chi} - \delta m & 0\\
0 & m_{\chi} +\delta m
\end{array} \right] \left[ \begin{array}{c}
\chi_1 \\
\chi_2
\end{array} \right] + \text{h.c.}. \label{ChiMassDiagonalized}
\end{eqnarray}
It would be more convenient to define two 4-dimensional Majorana spinors
\begin{eqnarray}
\chi_1 = \left[ \begin{array}{c}
\tilde{\chi}_1 \\
i \sigma^2 \tilde{\chi}_1^*
\end{array} \right],  \chi_2 = \left[ \begin{array}{c}
\tilde{\chi}_2 \\
i \sigma^2 \tilde{\chi}_2^*
\end{array} \right],
\end{eqnarray}
and then we have
\begin{eqnarray}
\chi &=& \frac{\chi_1 - i \chi_2}{\sqrt{2}}, \nonumber \\
\chi^C &=& \frac{\chi_1 + i \chi_2}{\sqrt{2}}.
\end{eqnarray}
Therefore, the Yukawa and gauge interactions can be reduced to
\begin{eqnarray}
\mathcal{L} \supset &\frac{y_{\chi}}{2}& \left( -\overline{\chi}_1 \tilde{\phi}_w \chi_1 + \overline{\chi}_2 \tilde{\phi}_w \chi_2  + \overline{\chi}_1 \tilde{\phi}_{w \eta} \chi_2 +  \overline{\chi}_2 \tilde{\phi}_{w \eta} \chi_1 \right) + i t_{\chi} g_D  \overline{\chi}_1 A\!\!\!/^{\prime} \chi_2\;.
\end{eqnarray}




We are interested in the phase transition and dark matter freeze-in production process mainly around TeV-scale, and we discuss the phase transition evaluations in two scenarios. In the scenario I, $v_w \approx v_s \sim $ TeV scale and both $\Phi_s$ and $\Phi_w$ appear in the phase evaluation processes.  In this scenario, $\delta m \ll m_{\chi}$ so that $\chi_{1,2}$ can be treated as a pair of pseudo-Dirac particles. The scenario II to be discussed is $v_w \gg v_s$, as well as $\mu_w^2 \gg \mu_s^2$, the cosmic string produced after the $\phi_w$ acquires VEV and the spontaneously breakdown of the $U(1)_{\rm dark}$, in which $\Phi_w$ decouple from our TeV-scale phase transition evaluations, except its Goldstone remains $\phi_{w \eta}$ which contributes to the longitudinal polarization of $A^{\prime}$.  
 In this scenario, $\phi_w$ changes little in the TeV-scale temperature, then it can be regarded as a constant, and is assigned with its zero-temperature value $v_w$. 
It is then convenient to write $\Phi_w$ into the nonlinear form $\Phi_w = v_w  e^{i \phi_{w \eta}/v_w}$. The Yukawa term then becomes
\begin{eqnarray}
\mathcal{L}_Y &\approx& y_{\chi} v_w  e^{i \phi_{w \eta}/v_w} \bar{\chi} \chi^C+h.c. \nonumber \\
&\approx& y_{\chi} v_w \bar{\chi} \chi^C + i y_{\chi} \bar{\chi} \chi^C \phi_{w \eta} - \frac{y_{\chi}}{2 v_w} \phi_{w \eta}^2 \bar{\chi} \chi^C \phi_{w \eta}^2 + h.c..
\end{eqnarray}
We see clearly the $\chi$-splitting mass term above, as well as the higher order 3- and 4-point effective vertices. The $v_w$ in this scenario can become extraordinary large, thus amplify the $\delta m$ to split $\chi_{1,2}$ into completely two majorana fermions. Sometimes $\delta m > m_{\chi}$ to induce an additional minus sign in the first eigenvalue of (\ref{ChiMassDiagonalized}). We are going to illustrate our manipulation of it in our later discussions.
The last thing we want to emphasize is that there is also a tiny coupling between $\chi$ and $\phi_s$ induced by the faint mixing (denoted by $V_{sw}$) between the $\phi_w$ and $\phi_s$ sectors. We just parametrize such an interaction with the effective coupling
\begin{eqnarray}
\mathcal{L}_{\chi \chi \phi_s} = y_{\chi} V_{sw} \phi_{s} \bar{\chi} \chi^C + h.c.\;. \label{SWMixing}
\end{eqnarray}

\section{First order phase transition and the production of the gravitational waves}
\label{sec:ptgw}
In this section, we write down the methodology for calculations of phase transition and gravitational waves produced during first-order phase transition process and from the cosmic strings decay.

\subsection{Finite temperature effective potential}

For the study of the phase transition in scenario I, with the standard methodology, we adopt the thermal one-loop effective potential~\cite{Quiros:1999jp},
\begin{eqnarray}
V_{eff}(h,\phi_s,\phi_w,T) &=& V_{\rm 0} (h,\phi_s,\phi_w)+ V_{\rm CW}(h,\phi_s,\phi_w) + V^{\rm c.t}_{1}(h,\phi_s,\phi_w) \nonumber \\
&+& V_{1}^{T}(h,\phi_s,\phi_w,T) + V_{1}^{\rm daisy}(h,\phi_s,\phi_w,T) \;. \label{Vtotal}
\end{eqnarray}
The $V_{\rm 0} (h,\phi_s,\phi_w)$ and $V_{\rm CW}(h,\phi_s,\phi_w)$ are the tree-level potential and the 1-loop Coleman-Weinberg potential, with $V^{\rm c.t}_{1}(h,\phi_s,\phi_w)$ to keep the zero temperature vacuum from shifting. The finite temperature correction is described by the term of $V_{1}^{T}(h,\phi_s,\phi_w,T)$, and the Daisy-correction term $V_{1}^{\rm daisy}(h,\phi_s,\phi_w,T)$.

Rotating the fields to expand along the $\phi_{s \eta} = \phi_{w \eta} = 0$ hyper plane, we obtain the tree-level potential,
\begin{eqnarray}
V_{\rm 0} (h,\phi_s,\phi_w)=\frac{\lambda h^4  }{4}+\frac{1}{2}  (\lambda_{sh} \phi_s^2+\lambda_{wh} \phi _{w}^2+\mu_0^{2}) h^2 +\lambda_s \phi_s^4+ (\lambda_{sw} \phi_{s}^2+\mu_w^{2})\phi_{w}^2+\lambda_w \phi_{w}^4+\mu_s^{2} \phi _{s}^2.	\label{V0}
\end{eqnarray}
At the zero temperature, considering the stationary point conditions,
\begin{align}
\label{eq:mini}
\left.\frac{d V_{0}(h, \phi_s, \phi_w)}{d h}\right|_{h=v_h}=0\;,\left.\frac{d V_{0}(h, \phi_s, \phi_w)}{d \phi_s}\right|_{\phi_s=v_s}=0\;,\left.\frac{d V_{0}(h, \phi_s, \phi_w)}{d \phi_w}\right|_{\phi_w=v_w}=0\;,
\end{align}
we get
\begin{eqnarray}
\mu_0^2 &=& -\lambda  v_h^2-\lambda_{sh} v_s^2-\lambda_{wh} v_w^2 \nonumber \\
\mu_s^2 &=& -\lambda_{sh}v_h^2/2 -2 \lambda_{s} v_s^2 - \lambda_{sw} v_w^2 \nonumber \\
\mu_w^{2} &=& -\lambda_{wh} v_{h}^2/2-\lambda_{sw} v_{s}^2-2 \lambda_{w} v_{w}^2\;. \label{mu2vev}
\end{eqnarray}
In this paper, we assign $v_h$,  $v_s$ and $v_w$ as well as all the other coupling constants as our input parameters, and utilize Eq.~(\ref{mu2vev}) to evaluate $\mu_{0,s,w}^2$.

The Coleman-Weinberg contribution is given by~\cite{Coleman:1973jx}	
\begin{align}
	V_{\rm CW}(h,\phi_s,\phi_w)= \sum_{i} \frac{g_{i}(-1)^{F}}{64\pi^2}  m_{i}^{4}(h,\phi_s,\phi_w)\left(\mathrm{Ln}\left[ \frac{m_{i}^{2}(h,\phi_s,\phi_w)}{\Lambda^2} \right] - C_i\right)\,,
	\label{eq:oneloop}
\end{align}
where $F=0 \; (1)$ for bosons (fermions), $\Lambda$ is the $\overline{\text{MS}}$ renormalization scale, $g_{i}= \{1,1,1,1,1,1,2,6,3,-12 \}$ for the $\{h, \eta ,\phi_s,\phi_{s\eta},\phi_w,\phi_{w\eta},G^{\pm},W,Z,T\}$ in this model, and $C_i=5/6$ for gauge bosons and $C_i=3/2$ for scalar fields and fermions. $\Lambda$ is a renormalization scale to be fixed to $\Lambda=3$ TeV in this paper. 
 The field-dependent Higgs mass matrix is given by,
\begin{eqnarray}
  && M^{2} = \label{eq:mass:matrix} \\
  && \begin{pmatrix}
    3 \lambda h^2  +\lambda_{sh} \phi_s^2 +\lambda_{wh} \phi_w^2 +\mu_0^{2} & 2  \lambda_{sh} h \phi_s & 2\lambda_{wh} h \phi_w
    \\
    2  \lambda_{sh} h \phi_s & \lambda_{sh} h^2 + 2 (6 \lambda_{s} \phi_{s}^2+\mu_s^{2}+\lambda_{sw} \phi_w^2) & 4 \lambda_{sw} \phi_s \phi_w
    \\
   2 \lambda_{wh} h \phi_{w} & 4 \lambda_{sw} \phi_s \phi_w & 2 \mu_w^2+\lambda_{wh} h^2+2 \lambda_{sw} \phi_{s}^2+12 \lambda_{w} \phi_{w}^2
  \end{pmatrix}.\; \nonumber
\end{eqnarray}
The field dependent dark photon mass is given by
\begin{eqnarray}
m_{A^{\prime}} = \sqrt{2 ( g_D^2 \phi_s^2 + g_w^2 \phi_w^2)}. \label{mA}
\end{eqnarray}

The slight shift of the tree-level VEVs induced by $V_{\rm CW}$ is canceled by the ``counter-terms" (CT)\cite{Cline:2011mm}
\begin{align}
V^{\rm c.t}_{1}=\delta \lambda_s \phi_s^4-\delta \mu_s^2 \phi_s^2-\delta \mu_w^2 \phi_w^2 + \delta \lambda_w \phi_w^4 +\delta \lambda h^4-\delta \mu_0^2 h^2+\delta \lambda_{sh} h^2 \phi_s^2 + \delta \lambda_{wh} h^2 \phi_w^2 + \delta \lambda_{sw} \phi_w^2 \phi_s^2\;,
\end{align}
with the relevant coefficients determined by,
\begin{align}
&\frac{\partial V^{\rm c.t}_{1}}{\partial h} = -\frac{\partial V_{\rm CW}}{\partial h}\;,
\frac{\partial^{2} V^{\rm c.t}_{1}}{\partial h \partial h} = -\frac{\partial^{2} V_{\rm CW}}{\partial h \partial h}\;,\nonumber\\
&\frac{\partial V^{\rm c.t}_{1}}{\partial \phi_{s(w)}} = -\frac{\partial V_{\rm CW}}{\partial \phi_{s(w)}}\;,
\frac{\partial^{2} V^{\rm c.t}_{1}}{\partial \phi_{s(w)} \partial \phi_{s(w)}} = -\frac{\partial^{2} V_{\rm CW}}{\partial \phi_{s(w)} \partial \phi_{s(w)}}\;,\nonumber\\
&\frac{\partial^{2} V^{\rm c.t}_{1}}{\partial h \partial \phi_{s(w)}} = -\frac{\partial^{2} V_{\rm CW}}{\partial h \partial \phi_{s(w)}}\;,\frac{\partial^{2} V^{\rm c.t}_{1}}{\partial \phi_s \partial \phi_w} = -\frac{\partial^{2} V_{\rm CW}}{\partial \phi_s \partial \phi_w}\; \label{CounterTerms}
\end{align}
evaluated at the EW minimum of $\{ h=v_{h}, \phi_{s}=v_{s}, \phi_w=v_w \}$. The logarithmic IR divergences encountered in (\ref{CounterTerms}) take the form\cite{Cline:1996mga, Casas:1994us,  Elias-Miro:2014pca, Martin:2014bca}
\begin{eqnarray}
\frac{\partial m_G^2}{\partial \phi_i} \frac{\partial m_G^2}{\partial \phi_j} \ln\frac{m_G^2}{\Lambda^2}\;,
\end{eqnarray}
where $\phi_i$ can be any scalar field, and $G$ is one Goldstone mass term. We follow Ref.~\cite{Cline:2011mm} to replace the Nambu-Goldstone boson masses with $\Lambda_{\rm IR}$ in (\ref{CounterTerms}). In this paper, we adopt $\Lambda_{\rm IR}=200$ GeV.

The one-loop finite temperature corrections are given by~\cite{Dolan:1973qd}
\begin{align}
\label{potVth}
 V_{1}^{T}(h,\phi_s,\phi_w, T) = \frac{T^4}{2\pi^2}\, \sum_i n_i J_{B,F}\left( \frac{ m_i^2(h,\phi_{s},\phi_w)}{T^2}\right)\;,
\end{align}
where the functions $J_{B,F}$ are
\begin{align}
\label{eq:jfunc}
J_{B,F}(y) = \pm \int_0^\infty\, dx\, x^2\, \ln\left[1\mp {\rm exp}\left(-\sqrt{x^2+y}\right)\right]\; ,
\end{align}
with $y\equiv m_{i}^2(h,\phi_{s})/T^2$, and the upper (lower) sign corresponds to bosonic (fermionic) contributions, respectively.
The thermal integrals $J_{B,F}$ given by Eq.~(\ref{eq:jfunc}) can be expressed as an infinite sum of modified Bessel functions of the second kind $K_{n} (x)$ with $n=2$~\cite{Anderson:1991zb},
\begin{align}
\label{Bessel_JFJB}
J_{B,F}(y) = \lim_{N \to +\infty} \mp \sum_{l=1}^{N} {(\pm1)^{l}  y \over l^{2}} K_{2} (\sqrt{y} l)\;.
\end{align}

The daisy term  $V_{1}^{\rm daisy}(h,\phi_s,\phi_w,T)$ is given by\cite{Carrington:1991hz, Arnold:1992rz}
\begin{eqnarray}
V_{1}^{\rm daisy}(h,\phi_s,\phi_w,T)=-\frac{T}{12 \pi} \sum_{i={\rm bosons}} n_i \left[ ( m_i^2(h,\phi_{s},\phi_w)+c_i(T))^{\frac{3}{2}} - ( m_i^2(h,\phi_{s},\phi_w))^{\frac{3}{2}} \right]\;, \label{DaisyTerms}
\end{eqnarray}
where the finite temperature corrections are calculated as
\begin{align}
c_{h}(T)&=\frac{1}{48} T^2 (9 g^2+3 g'^2+4 (6 \lambda +\lambda _{sh}+\lambda_{wh})+12 y_{t}^2)\;,\\
c_{s}(T)&=\frac{1}{12} T^2 (3 g_D^2+ (4\lambda_{s}+2 \lambda_{sh}+\lambda_{sw}))\;,\\
c_{w}(T)&=\frac{1}{12} T^2 (\lambda_{sw}+4 \lambda_{w}+2 \lambda_{wh})\;,\\
c_{A'}(T)&=\frac{1	}{3} g_D^2 T^2, \\
c_{B}(T)&=\frac{11}{6} g_1^2 T^2, \\
c_{W}(T)&=\frac{11}{6} g_2^2 T^2, \label{ThermalMass}
\end{align}
where $g_1$ and $g_2$ are the SM $U(1)_Y \times SU(2)_L$ gauge couplings. The definitions of the $m_i^2(h, \phi_s, \phi_w)+c_i(T)$ in the mixing situation are the eigenvalues of (\ref{ThermalMass}), with the diagonal elements added with the $c_i(T)$ defined in (\ref{ThermalMass}). The details of the mixture of the vector bosons are illustrated in Appendix.~\ref{sec:thermass}.

For the studies of phase transition in the scenario II, the evaluation of temperature dependent effective potential is actually similar with Eqs.(\ref{Vtotal})-(\ref{ThermalMass}) in scenario I, with all of the $\phi_w$, $\lambda_{w, sw, wh}$ terms removed. More explicitly,
after integrating out the $\phi_w$, the $\phi_w$ mediated processes also converges into point-like interactions. This eliminates the $\phi_w$ involved terms, while shifting the $\lambda_{h, s, hs}$ and $\mu^2_{0, s}$ in (\ref{V0}) into $\tilde{\lambda}_{h, s, hs}$ and $\tilde{\mu}^2_{0, s}$.  For simplicity, we neglect the ``tilde'' without confusion to write down the potential from the aspect of effective theory,
\begin{eqnarray}
V_{\rm 0, \phi_w \!\!\!\!\!\!\!\!--} (h,\phi_s,\phi_w)=\frac{\lambda h^4  }{4}+\frac{1}{2}  (\lambda_{sh} \phi_s^2+ \mu_0^{2}) h^2 +\lambda_s \phi_s^4 + \mu_s^{2} \phi _{s}^2. \label{V0_2fields}
\end{eqnarray}
Therefore, the third row and column in Eq.~(\ref{eq:mass:matrix}) also disappears,
\begin{eqnarray}
 M^{2}_{\rm 0, \phi_w \!\!\!\!\!\!\!\!--}  = 
 \begin{pmatrix}
    3 \lambda h^2  +\lambda_{sh} \phi_s^2 +\lambda_{wh} \phi_w^2 +\mu_0^{2} & 2  \lambda_{sh} \phi_h \phi_s  \\
    2  \lambda_{sh} h \phi_s & \lambda_{sh} h^2 + 2 (6 \lambda_{s} \phi_{s}^2+\mu_s^{2}+\lambda_{sw} \phi_w^2)
  \end{pmatrix}\;.\label{eq:mass:matrix} 
\end{eqnarray}
For the field dependent mass $m_{A^{\prime}}$, (\ref{mA}) becomes
\begin{eqnarray}
m_{A^{\prime}} = \sqrt{2 (g_D^2 \phi_s^2 + g_w^2 v_w^2)}\;. \label{mAPrime}
\end{eqnarray}
Notice that although $g_w \ll g_D$, the extremely large $v_w \gg \phi_s$ might still contribute significantly to the gauge boson's mass.

Since the minimum of the effective potential  $(h, \phi_s, \phi_w)$ evolve as the temperature drops, it is necessary to study the phase evolution and transition structures of the system. We utilize both the \texttt{CosmoTransitions}\cite{Wainwright:2011kj} and \texttt{PhaseTracer}\cite{Athron:2020sbe} by making independent programs to find out the phases as well as the transition processes among them for the cross-validation, and will only adopt the data when the results from both programs are consistent.

\subsection{Bubble nucleation temperature $T_n$ and the percolation temperature $T_p$}

For a study on first-order phase transition, one has to compute the bubble nucleation temperature $T_n$,  and the percolation temperature $T_p$, that are usually somewhat lower than the critical temperature $T_c$ when two vacua are degenerate. The bubble nucleation temperature $T_n$ can be estimated by\cite{Moreno:1998bq}
\begin{eqnarray}
\int_{t_c}^{t_n} dt \frac{\Gamma}{H^3} = \int_{T_n}^{T_c} dT \frac{\Gamma}{H^4 T} = 1,
\end{eqnarray}
which means that at temperatures lower than the critical temperature, at least one bubble should be created inside per-unit Hubble volume at the bubble nucleation temperature $T_n$. The bubble nucleation rate $\Gamma$ is defined by\cite{Wang:2020jrd}
\begin{eqnarray}
\Gamma \sim T^4 \left( \frac{S}{2 \pi}\right)^4 e^{-S},
\end{eqnarray}
where $A$ is an $\mathcal{O}(1)$ constant, and $S=\text{min}\lbrace S_4, S_3/T \rbrace$ is the action of the bubble solution. Usually in our model around $T_{c,n,p}$, $S_3/T < S_4$ so that we only display the $S_3$ definition
\begin{eqnarray}
S_3 = 4 \pi \int_0^{\infty} dr r^2 \left[ \frac{1}{2} \frac{d \phi_i}{d r} \frac{d \phi_i}{d r} + V_{\text{eff}}(\phi_i T) \right],
\end{eqnarray}
where $\phi_i(r) = (h, \phi_s, \phi_i)$ is the ``bounce solution'' acquired from the equations of motion
\begin{eqnarray}
\frac{d^2 \phi_i}{d r^2} + \frac{2}{r} \frac{d \phi_i}{d r} = \frac{\partial V_{\text{eff}}}{\partial \phi_i},
\end{eqnarray}
with the boundary conditions
\begin{eqnarray}
\left. \frac{d \phi_i}{d r} \right|_{r=0}= \phi_{i \text{phase 1}}, ~~\left. \frac{d \phi_i}{d r} \right|_{r=\infty}= \phi_{i \text{phase 2}}, 
\end{eqnarray}
between the two phases $\phi_{i \text{phase 1}}$ and $\phi_{i \text{phase 2}}$ during the transition. 

The definition of the percolation time $t_p$ is given by\cite{Leitao:2012tx, Guth:1979bh, Guth:1981uk}
\begin{eqnarray}
P(t) \simeq 0.71,
\end{eqnarray}
where
\begin{eqnarray}
P(t) = \exp\left[ -\frac{4 \pi}{3} \int_{t_c}^t dt^{\prime} \Gamma(t^{\prime}) a^3(t^{\prime}) r^3(t, t^{\prime})\right].
\end{eqnarray}
Here $a(t^{\prime})$ is the scale factor of the Friedmann-Robertson-Walker (FRW) metric, $r(t, t^{\prime})$ is the comoving radius of a bubble given by
\begin{eqnarray}
r(t, t^{\prime}) = \int_{t^{\prime}}^{t} d\tau \frac{v_b}{a(\tau)},
\end{eqnarray}
where $v_b$ is the velocity of the bubble wall. After $t_p$ is evaluated, one can solve the $T_p$ through the equation
\begin{eqnarray}
H^2 = \frac{8 \pi}{3 M_{\text{pl}}^2} g^{\star} T^4
\end{eqnarray}
by replacing $H$ with $1/(2 t_p)$ during the radiation dominant epoch. Here $g^{\star}$ is the effective degrees of freedom of the plasma, which is approximated by $g^{\star} \simeq 106$.

\subsection{Gravitational waves from the first-order phase transition}
To evaluate the gravitational wave spectrum emitted during the first-order phase transition, one has to acquire the phase transition strength parameter of $\alpha$, and the phase transition duration parameter of $\beta$, which are defined to be
\begin{eqnarray}
\alpha &=& \frac{\rho_{\text{vac}}}{\rho_{\text{rad}}}, \nonumber \\
\beta &= & \frac{dS}{dt} = H T \frac{dS}{dT}, \label{alphabetaForGW}
\end{eqnarray}
where $\rho_{\text{rad}}=\pi^2 g_* T^4/30$ is the plasma energy density,  and\cite{Enqvist:1991xw}
\begin{eqnarray}
\rho_{\text{vac}} = V_{\text{eff}}(\phi_{\text{phase }1})-V_{\text{eff}}(\phi_{\text{phase }2}) - T \frac{\partial}{\partial T} [V_{\text{eff}}(\phi_{\text{phase }1})-V_{\text{eff}}(\phi_{\text{phase }2})]
\end{eqnarray}
is the released vacuum energy during the phase transition. Both $\alpha$ and $\beta$ can be calculated at the phase transition temperature of either $T^*=T_n$ or $T^*=T_p$ for slightly different results. In this paper, we adopt $T^*=T_p$, however we still use the symbol $T^*$, as well as the $H^* = H(T^*)$  in our following displayed equations for the purpose of generality.

We then follow Ref.~\cite{Wang:2020jrd} to evaluate the gravitational wave from the first-order phase transition by summing up the contributions from bubble collision, sound wave and turbulence,
\begin{eqnarray}
\Omega_{\text{GW}}=\Omega_{\text{co}}+\Omega_{\text{sw}}+\Omega_{\text{turb}}.
\end{eqnarray}

\subsubsection{The bubble walls collision contributions}
The bubble walls collision term $\Omega_{\text{co}}$ from the ``envelope approximation'' results is given by\cite{Huber:2008hg, Caprini:2015zlo, Kamionkowski:1993fg}\footnote{Here, we note that recent simulations suggest that the scalar oscillation stage continuely contribute GW production, See Ref.~\cite{Di:2020ivg,Cutting:2020nla,Cutting:2018tjt}. }
\begin{eqnarray}
h^2\Omega_{\rm co}(f) \simeq 1.67\times10^{-5}\left(\frac{\beta}{H_*}\right)^{-2}\left(\frac{\kappa_{\phi}\alpha}{1 + \alpha}\right)^2\left(\frac{100}{g_{\star}}\right)^{1/3}\frac{0.11v_b}{0.42 + v_b^2}\frac{3.8(f/f_{\rm co})^{2.8}}{1 + 2.8(f/f_{\rm co})^{3.8}} \,\,,
\end{eqnarray}
For Jouguet detonations, we adopt the Chapman-Jouguet condition of the wall velocity $v_b$ as below~\cite{Steinhardt:1981ct},
\begin{eqnarray}\label{eq:bubblespeed}
v_b=\frac{1/ \sqrt{3}+\sqrt{\alpha ^2+2 \alpha /3}}{1+\alpha }\;.
\end{eqnarray}
with the peak frequency $f_{\rm co}$ locating at
\begin{eqnarray}
f_{\rm co} \simeq 1.65\times10^{-5}\text{Hz}\frac{\beta}{H_*}\left(\frac{0.62}{1.8 - 0.1v_b + v_b^2}\right)\left(\frac{T_*}{100\rm GeV}\right)\left(\frac{g_*}{100}\right)^{1/6} \,\,.
\end{eqnarray}

\subsubsection{The sound wave contributions}
The sound wave contribution $\Omega_{\text{sw}}$  is given by
\begin{eqnarray}
\Omega h^2_{\rm sw}(f)=2.65 \times 10^{-6}(H_*\tau_{sw})\left(\frac{\beta}{H_*}\right)^{-1} v_b
\left(\frac{\kappa_\nu \alpha }{1+\alpha }\right)^2
\left(\frac{g_*}{100}\right)^{-\frac{1}{3}}
\left(\frac{f}{f_{\rm sw}}\right)^3 \left(\frac{7}{4+3 \left(f/f_{\rm sw}\right)^2}\right)^{7/2}, \label{OmegaSW}
\end{eqnarray}
with the peak frequency being~\cite{Hindmarsh:2013xza,Hindmarsh:2015qta,Hindmarsh:2017gnf}:
\begin{eqnarray}
f_{\rm sw}=1.9 \times 10^{-5} \frac{\beta}{H_*} \frac{1}{v_b} \frac{T_*}{100}\left({\frac{g_*}{100}}\right)^{\frac{1}{6}} {\rm Hz }\;.
\end{eqnarray}
In Eq.~(\ref{OmegaSW}),  the $\tau_{sw}$ shows the duration of the sound wave from the phase transition~\cite{Ellis:2020awk}, which is calculated as
\begin{eqnarray}
\tau_{sw}=min\left[\frac{1}{H_*},\frac{R_*}{\bar{U}_f}\right],
\end{eqnarray}
where $H_*R_*=v_b(8\pi)^{1/3}(\beta/H)^{-1}$, and the root-mean-square (RMS) fluid velocity $\bar{U}_f$ can be approximated as \cite{Hindmarsh:2017gnf, Caprini:2019egz, Ellis:2019oqb}
\begin{eqnarray}
\bar{U}_f^2\approx\frac{3}{4}\frac{\kappa_\nu\alpha}{1+\alpha}\;. \label{Uf}
\end{eqnarray}
The $\kappa_\nu$ factor in (\ref{Uf}) indicates the latent heat transferred into the kinetic energy of plasma,  which is given by~\cite{Espinosa:2010hh}
\begin{eqnarray}
\kappa_{\nu}=\frac{\sqrt{\alpha}}{0.135+\sqrt{0.98+\alpha}}.
\end{eqnarray}

\subsubsection{The turbulence contributions}
The magnetic hydrodynamic turbulence term $\Omega_{\text{turb}}$ is given by
\begin{eqnarray}
\Omega h^2_{\rm turb}(f)=3.35 \times 10^{-4}\left(\frac{\beta}{H_*}\right)^{-1}
\left(\frac{\varepsilon \kappa_\nu \alpha }{1+\alpha }\right)^{\frac{3}{2}}
\left(\frac{g_*}{100}\right)^{-\frac{1}{3}}
v_b
\frac{\left(f/f_{\rm turb}\right)^3\left(1+f/f_{\rm turb}\right)^{-\frac{11}{3}}}{\left[1+8\pi f a_0/(a_* H_*)\right]}\;,
\end{eqnarray}
with the peak frequency\cite{Caprini:2009yp}
\begin{eqnarray}
f_{\rm tur}=2.7  \times 10^{-5} \frac{\beta}{h_*} \frac{1}{v_b} \frac{T_*}{100}\left({\frac{g_*}{100}}\right)^{\frac{1}{6}} {\rm Hz }\;.
\end{eqnarray}
The efficiency factor $\varepsilon \approx 0.1$,  redshift of the frequency is obtained as
\begin{eqnarray}
	h_{\ast} = \bigl( 1.65 \times 10^{-5} Hz \bigr) \left( \frac{T_{*}}{100 \rm{GeV}} \right) \left( \frac{g_{\ast}}{100} \right)^{1/6}\;.
\end{eqnarray}

\subsection{Cosmic strings and gravitational waves}
In the scenario II, cosmic strings start to form after $\phi_w$ begins to acquire its VEV. These strings collide and self-interact into loops, and then shrinks to leave us the gravitational waves formulated via cusp, kink and kink-kink collisions.  The spectrum can be expressed as
\begin{eqnarray}\label{CS_GWs}
\Omega_{\rm GW}(f)h^2=\frac{8\pi h^2}{3M_{\rm Pl}^2H_0^2}\int_0^{t_0}dt\left(\frac{a(t)}{a(t_0)}\right)^3\int_0^\infty d\ell\,n_{\rm CS}(\ell,t)P_{\rm GW}\left(\frac{a(t_0)}{a(t)}f,\ell\right).
\end{eqnarray}
Particularlly, Ref.~\cite{Auclair:2019wcv} transforms Eq.~(\ref{CS_GWs}) from the Nambu-Goto string~Ref.\cite{Vachaspati:1984gt} into~Ref.\cite{Auclair:2019wcv,Blanco-Pillado:2017oxo}
\begin{eqnarray}
\Omega_{\rm GW}(f)h^2=\frac{8\pi h^2}{3M_{\rm Pl}^2H_0^2}G\mu^2 f\sum_{n=1}^\infty C_n(f)P_n,
\end{eqnarray}
where $n=1$, 2, ..., labels the radiation frequencies $\omega_n=2\pi n/(\ell/2)$.  The dimensionless parameter $G\mu$ is
\begin{eqnarray}\label{Gmu}
G\mu\sim \frac{2 v_w^2}{M_{\rm Pl}^2},
\end{eqnarray}
where $v_w$ is the VEV of the $U(1)_{\rm dark}$ scalar field which spontaneously breaks down.  $P_n$ is the corresponding average loop power spectrum with its numerical results adopted from Ref.~\cite{Blanco-Pillado:2017oxo},  and $C_n$ is given by
\begin{eqnarray}\label{Cn}
C_n=\frac{2n}{f^2}\int_0^{\infty}\frac{dz}{H(z)(1+z)^6}n_{\rm CS}\left(\frac{2n}{(1+z)f},t(z)\right),
\end{eqnarray}
where the integration of the time parameter $t$ has been transformed to the redshift parameter $z$.  
To evaluate the integration in Eq.~(\ref{Cn}),  we need the cosmic time $t(z)$ and Hubble constant $H(z)$ to be expressed from the redshift parameter $z$ to become
\begin{eqnarray}
t(z)=\int_z^\infty\frac{dz'}{H(z')(1+z')};\quad H(z)=H_0\sqrt{\Omega_r\mathcal{G}(z)(1+z)^4+\Omega_m(1+z)^3+\Omega_\Lambda},
\end{eqnarray}
with the current abundances of the radiation, matter and dark energy given by~\cite{Planck:2018vyg}
\begin{eqnarray}
\Omega_r=9.1476\times10^{-5},\quad \Omega_m=0.308,\quad\Omega_\Lambda=1-\Omega_r-\Omega_m.
\end{eqnarray}
The function $\mathcal{G}(z)$ is given by
\begin{equation}
\mathcal{G}(z)=\frac{g_*(t)}{g_*(t_0)}\left(\frac{g_S(t_0)}{g_S(t)}\right)^{4/3}\approx\begin{cases}~1,&z<10^9;\\~0.83,&10^9<z<2\times10^{12};\\~0.39,&z>2\times10^{12},\end{cases}
\end{equation}
while the cosmic string number density for the loops produced in radiation dominated era but survive until matter domination is given by~\cite{Blanco-Pillado:2013qja}:
\begin{eqnarray}
n_{\rm CS}(\ell,t)=\begin{cases}~n_{\rm CS}^r(\ell,t)=\frac{0.18}{t^{3/2}(\ell+\Gamma G\mu t)^{5/2}},&(\ell\leqslant0.1\,t);\\~n_{\rm CS}^{r,m}(\ell,t)=\frac{0.18\, t_{\rm eq}^{1/2}}{t^{2}(\ell+\Gamma G\mu t)^{5/2}} &(\ell\leqslant0.09\, t_{\rm eq}-\Gamma G\mu t);\\~n_{\rm CS}^m(\ell,t)=\frac{0.27-0.45(\ell/t)^{0.31}}{t^2(\ell+\Gamma G\mu t)^2},&(\ell\leqslant0.18\, t),\end{cases}
\end{eqnarray}
where $\Gamma=50$~\cite{Blanco-Pillado:2017oxo}, and $t_{\rm eq}=2.25\times10^{36}~{\rm GeV}^{-1}$ which is the matter-radiation equality time.

\section{Freeze-in processes} \label{FreezeInRates}

The dark matter particles can be produced through both decay and annihilation processes.   To calculate the relic density, we rely on the Boltzmann equation. Assuming that all the other particles except the $\chi_{1,2}$ are in equilibrium with the plasma, and ignoring the feedback of the dark matter particles annihilating into the plasma due to the extreme smallness of the $\chi_{1,2}$ abundances compared with their equilibrium values, the Boltzmann equation is given by
\begin{eqnarray}
s H z \frac{d Y_{\chi}}{d  x} = 2 \gamma_{\rm tot}, \label{Boltzmann}
\end{eqnarray}
where $Y_{\chi}=Y_{\chi_1}+Y_{\chi_2}=\frac{n_{\chi_1}+n_{\chi_2}}{s}$ is the total dark matter particle number density normalize by the entropy density, $\gamma_{\rm tot}$ is the summation over all the ``rates'', and $x=\frac{|m_{\chi_1}|}{T}$ is the dimensionless parameter measuring the evolution of time.

To calculate the $\gamma_{\rm tot}$,  we need to sum over all of the $1 \leftrightarrow 2$ and $2 \leftrightarrow 2$ processes taking into account the thermal corrections on the external legs. The thermal corrections to the dark sector particles $\chi_{1,2}$ are neglected. Besides the VEV-dependent mass terms, all the other particles receive thermal corrections on their dispersion relations.  These cause the complicated threshold effects, and the production rates change significantly during the freeze-in processes.  

After the electroweak phase transition when $H$ acquires a nonzero VEV,  there will be intricate mixings of both the gauge bosons and the Higgs bosons between the SM and dark sectors. This is extremely hard to manipulate.  Fortunately, in our interested parameter space when $m_{\chi} \gg 100 \text{ GeV}$, the freeze-in processes basically cease when $x=\frac{m_{\chi}}{T} \gtrsim 3$, which is still well above the electroweak phase transition temperature $T_{\text{ew}} \sim 100 \text{ GeV}$. Therefore, we neglect the dark matter production below the electroweak phase transition. 

For the gauge boson and the SM fermions, we adopt the hard thermal loop (HTL) results to evaluate the phase space of the final states.  Goldstone equivalence gauge is also utilized for the convenience to decompose the degrees of freedom of the vector boson $A^{\prime}$.  The details of the HTL corrections to the vector boson and the SM fermions are illustrated in Appendix.~\ref{ThermalCorrectionGaugeFermion}. 

The Higgs boson masses are extracted from Eq.~(\ref{Vtotal}).  Since we only consider the processes above the electroweak phase transition so that $h=0$, and there the mixings between the SM Higgs doublet and the $\phi_{s,w}$ vanish. Therefore,
\begin{eqnarray}
m_{H}^2 &=& \frac{\partial^2 V_{eff}}{\partial h^2},  \label{MH} \\
M_{s,w}^2 &=& \frac{1}{2} \left( \begin{array}{cc}
\frac{\partial^2 V_{eff}}{\partial \phi_s^2} & \frac{\partial^2 V_{eff}}{\partial \phi_s \partial \phi_w} \\
\frac{\partial^2 V_{eff}}{\partial \phi_w \partial \phi_s} & \frac{\partial^2 V_{eff}}{\partial \phi_w^2}
\end{array} \right). \label{Msw}
\end{eqnarray}
Since before the electroweak phase transition, all the elements of the SM Higgs doublet are degenerate, so $m_{H}$ is the mass for all of the SM Higgs bosons. Diagonalizing (\ref{Msw}) gives two mass eigenstates mixed from the $\tilde{\phi}_{s,w}$. We use $\phi_{1,2}$ to represent the two eigenstates, and $m_{1,2}$ to denote the corresponding masses. The mixing matrix elements are assigned with
\begin{eqnarray}
V = \left( \begin{array}{cc}
V_{1s} & V_{1w} \\
V_{2s} & V_{2w}
\end{array} \right),
\end{eqnarray}
so that
\begin{eqnarray}
\text{diag}[m_1^2, ~m_2^2]=V M_{s,w}^2 V^{\dagger}. \label{MassEigenScalar}
\end{eqnarray}

When $\phi_s \neq 0$ and $\phi_w \neq 0$, the $ \frac{\partial^2 V_{eff}}{\partial \phi_s \partial \phi_w}\neq 0$, so we need to diagonalize Eq.(~\ref{Msw}) and calculate the mass eigenvalues and mixing matrix.  In this case, the masses of $\phi_{s\eta}$ and $\phi_{w\eta}$ also vanish. This is because besides the gauged $U(1)_{\text{dark}}$ group, there is an additional global $U(1)$ symmetry which is also broken to generate another Goldstone boson. The two massless states recombine into
\begin{eqnarray}
\phi_{A^{\prime}\eta} &=& \frac{g_D \phi_s}{\sqrt{g_D^2 \phi_s^2+g_w^2 \phi_w^2}} \tilde{\phi}_{s \eta} + \frac{g_w \phi_w}{\sqrt{g_D^2 \phi_s^2 +g_w^2 \phi_w^2}} \tilde{\phi}_{w \eta}, \nonumber \\
\phi_{G\eta} &=& -\frac{g_w \phi_w}{\sqrt{g_D^2 \phi_s^2+g_w^2 \phi_w^2}} \tilde{\phi}_{s \eta} +  \frac{g_D \phi_s}{\sqrt{g_D^2 \phi_s^2+g_w^2 \phi_w^2}} \tilde{\phi}_{w \eta},
\end{eqnarray}
or one can warp the coefficients by parametrizing them into $U_{(A^{\prime}, G)(s, w)}$,
\begin{eqnarray}
\phi_{A^{\prime}\eta} &=& U_{A^{\prime}s} \tilde{\phi}_{s \eta} + U_{A^{\prime}w} \tilde{\phi}_{w \eta}, \nonumber \\
\phi_{G\eta} &=& U_{Gs} \tilde{\phi}_{s \eta} + U_{Gw} \tilde{\phi}_{w \eta}.
\end{eqnarray}
Here $\phi_{A^{\prime} \eta}$ connects with the $A^{\prime}$, and will be partly eaten by the longitudinal polarization of $A^{\prime}$, as will be illustrated in Appendix \ref{ThermalCorrectionGaugeFermion}. $\phi_{G\eta}$ is the Goldstone boson corresponding to the global $U(1)$ symmetry. In the zero temperature, it might cause some phenomenology problems such as the Higgs invisible decay. We will discuss these problems later in this paper.

The assignment of the $\phi_{A^{\prime}\eta}$, $\phi_{G \eta}$ masses depends on the VEV structures of $\phi_{s,w}$ . When $\phi_{s,w}$ are both nonzero, both $\phi_{A^{\prime}\eta, G\eta}$ are massless, and if one or both of the $\phi_{s,w}$ become zero,  The counterpart of the non-zero VEV of $\phi_{s,w}$ is massless, while the counterpart of the zero VEV shares the same mass with the corresponding $\phi_s$ or $\phi_w$.

For further usage, we list all the couplings in Tab.~\ref{CouplingsTable}.

\begin{table}
\begin{tabular}{|c|c|c|}
\hline
interactions & Symbol & Coupling constants \\
\hline
$\overline{\chi}_i \chi_i \phi_j$ & $y_{ij}$ & $y_{\chi} V_{j w} (\delta_{2i}-\delta_{1i})$ \\
\hline
$\overline{\chi}_1 \chi_2 \phi_{(A^{\prime}, G) \eta}$ & $y_{(A^{\prime}, G)\eta}$ & $y_{\chi} U_{(A^{\prime}, G) w}$ \\
\hline
$ [(\partial_{\mu} \phi_i) \phi_{(A^{\prime},G)\eta} - \phi_i \partial_{\mu} \phi_{(A^{\prime},G)\eta}] A^{\mu}$ & $g_{i(A^{\prime}, G)}$ & $2 g_w V_{i w} U_{(A^{\prime}, G) w} + 2 g_D V_{i s} U_{(A^{\prime}, G) s}$ \\
\hline
$ A_{\mu}^{\prime} A^{\prime \mu} \phi_i$ & $G_{A^{\prime} A^{\prime} i}$ & $4 g_D^2 V_{1s} \phi_s + 4 g_w^2 V_{1w} \phi_w$ \\
\hline
$\phi_1 \phi_1 \phi_1$ & $A_{111}$ & $\begin{array}{l} 24 V_{1s}^3 \phi_s \lambda_s + 12 V_{1s} V_{1w}^2 \phi_s \lambda_{sw} \\ + 12 V_{1s}^2 V_{1w} \phi_w \lambda_{sw} + 24 V_{1w}^3 \lambda_w\end{array}$ \\
\hline
$\phi_1 \phi_1 \phi_2$ & $A_{112}$ & $\begin{array}{l} 24 V_{1s}^2 V_{2s} \phi_s \lambda_s + 4 V_{2s} V_{1w}^2 \phi_s \lambda_{sw} + 8 V_{1s} V_{1w} V_{2w} \phi_s \lambda_{sw} \\ + 8 V_{1s} V_{2s} V_{1w} \phi_w \lambda_{sw} + 4 V_{1s}^2 V_{2w} \phi_w \lambda_{sw} + 24*V_{1w}^2 V_{2w} \phi_w \lambda_w\end{array}$ \\
\hline
$\phi_1 \phi_2 \phi_2$ & $A_{122}$ & $\begin{array}{l} 24 V_{1s} V_{2s}^2 \phi_s \lambda_s + 8 V_{2s} V_{1w} V_{2w}\phi_s \lambda_{sw} + 4 V_{1s} V_{2w}^2 \phi_s \lambda_{sw} \\ + 4 V_{2s}^2 V_{1w} \phi_w \lambda_{sw} + 8 V_{1s} V_{2s} V_{2w} \phi_w \lambda_{sw} + 24 V_{1w} V_{2w}^2 \phi_w \lambda_w \end{array}$ \\
\hline
$\phi_2 \phi_2 \phi_2$ & $A_{222}$ & $\begin{array}{l} 24 V_{2s}^3 \phi_s \lambda_s + 12 V_{2s} V_{2w}^2 \phi_s \lambda_{sw}\\  + 12 V_{2s}^2 V_{2w} \phi_w \lambda_{sw} + 24 V_{sw}^3 \phi_w \lambda_w \end{array}$ \\
\hline
$\phi_i \phi_{G \eta} \phi_{G \eta}$ & $A_{iGG}$ & $\begin{array}{l} 8 V_{is} \phi_s U_{Gs}^2 \lambda_s + 4 V_{iw} U_{Gs}^2 \phi_w \lambda_{sw} \\+ 4 V_{is} \phi_s U_{Gw}^2 \lambda_{sw} + 8 V_{iw} \phi_w U_{Gw}^2 \lambda_w \end{array}$ \\
\hline
$\phi_i \phi_{G \eta} \phi_{A^{\prime} \eta}$ & $A_{iGA^{\prime}}$ & $\begin{array}{l} 8 V_{is} \phi_s U_{A^{\prime}s} U_{Gs} \lambda_s + 4 V_{iw} V_{A^{\prime} s} V_{Gs} \phi_w \lambda_{sw} \\ + 4  V_{is} \phi_s U_{A^{\prime}w} U_{Gw} \lambda_{sw} + 8 V_{iw} \phi_w U_{A^{\prime}w} U_{Gw} \lambda_w \end{array}$ \\
\hline
$\phi_i \phi_{A^{\prime} \eta} \phi_{A^{\prime} \eta}$ & $A_{i A^{\prime} A^{\prime}}$ & $ \begin{array}{c} 8 V_{is}\phi_s V_{A^{\prime} s}^2 \lambda_s + 4 V_{iw} V_{A^{\prime}s}^2 \phi_w \lambda_{sw} \\ + 4 V_{is} \phi_s V_{A^{\prime}w}^2 \lambda_{sw} + 8 V_{iw} \phi_w V_{A^{\prime}w}^2 \lambda_w \end{array}$ \\
\hline
\end{tabular}
\caption{Couplings formalism and constants to be used.} \label{CouplingsTable}
\end{table}

Now we are ready to calculate the interaction rates $\gamma_{\rm tot}$. This involves evaluating the diagrams in Fig.\ref{OneTwoDiagrams}, \ref{AADiagrams}-\ref{HHDiagrams} and \ref{ChiChi2SMSM} for all the $1 \leftrightarrow 2$, non-SM product $2 \leftrightarrow 2$ and SM product $2 \leftrightarrow 2$ processes respectively. For the brevity of this section, we leave the detailed formulas in the Appendix \ref{DetailsFreezeIn}.

\begin{figure}
\includegraphics[width=0.3\textwidth]{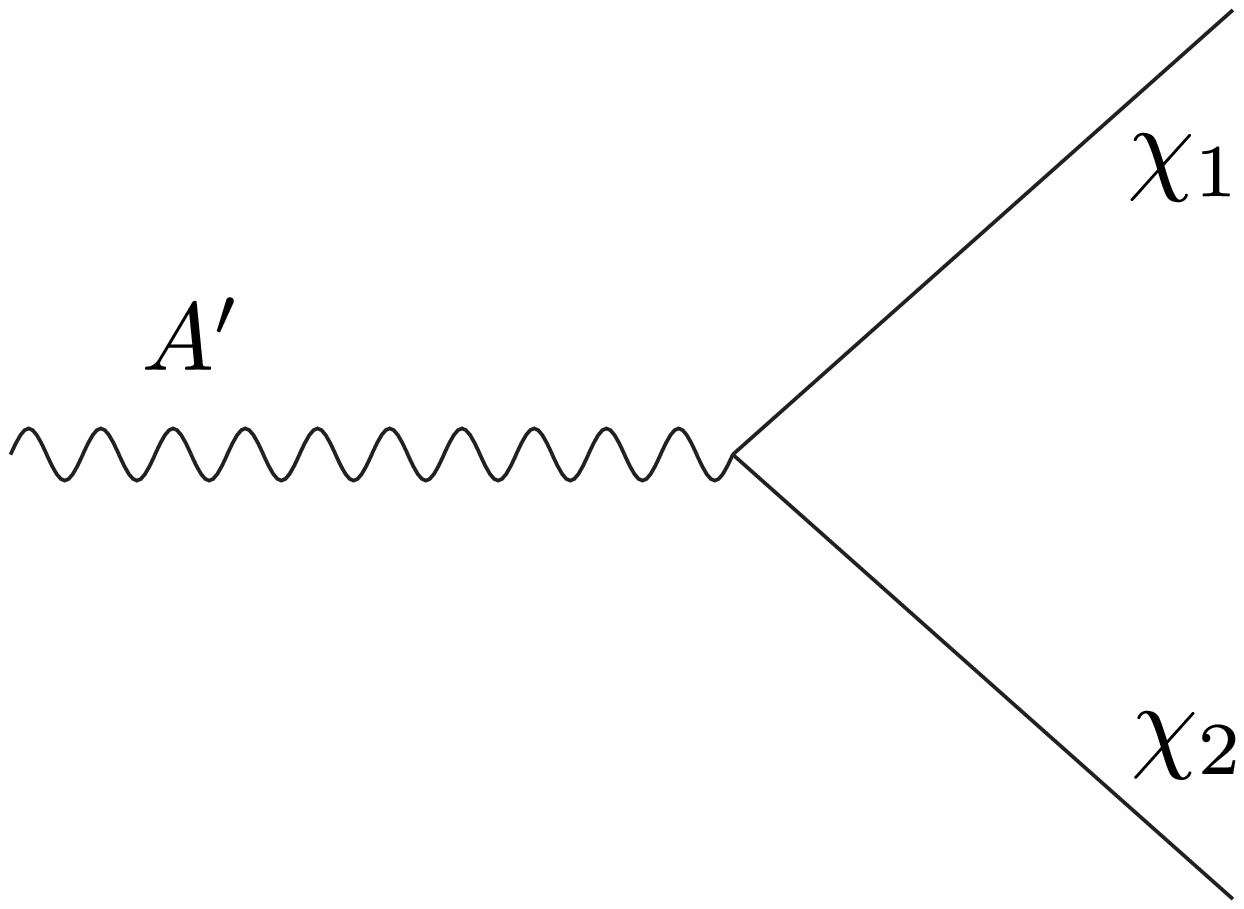}~
\includegraphics[width=0.3\textwidth]{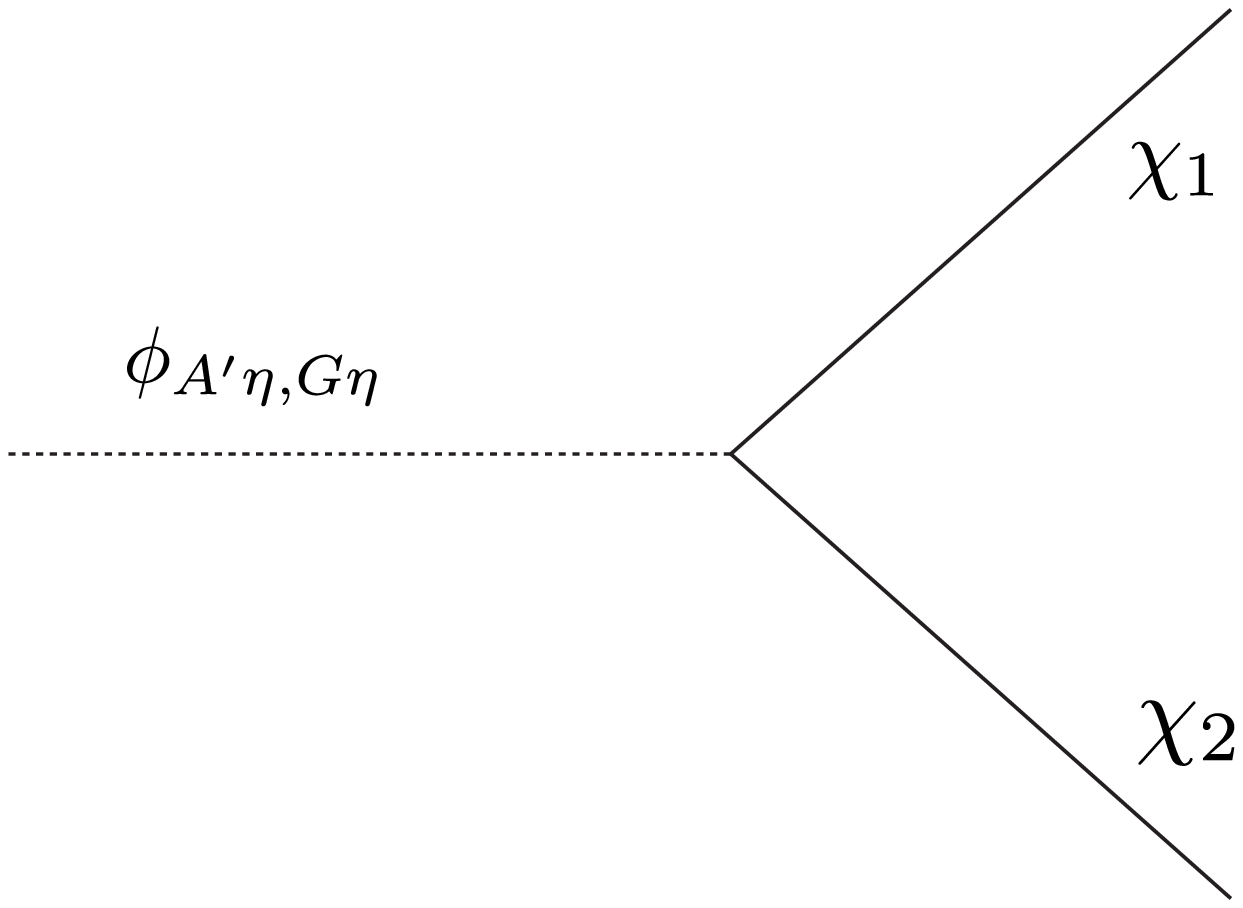}~
\includegraphics[width=0.3\textwidth]{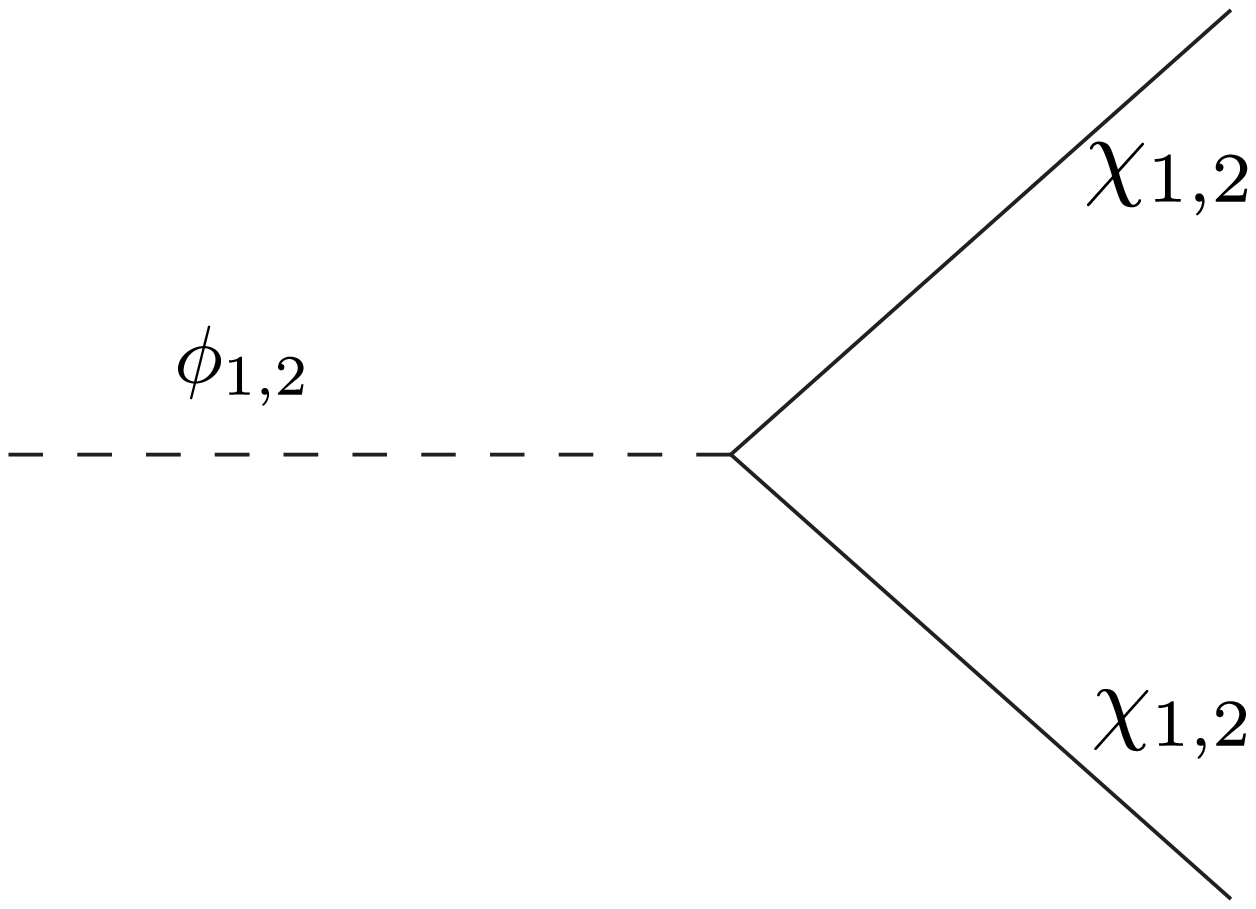}
\caption{$1 \leftrightarrow 2$ diagrams for the production of dark matter. The decayed particle can be $A^{\prime}$ with the accessory $\phi_{A^{\prime} \eta}$, $\phi_{1,2}$, and $\phi_{s \eta, w \eta}$.}
\label{OneTwoDiagrams}
\end{figure}

\begin{figure}
\includegraphics[width=0.2\textwidth]{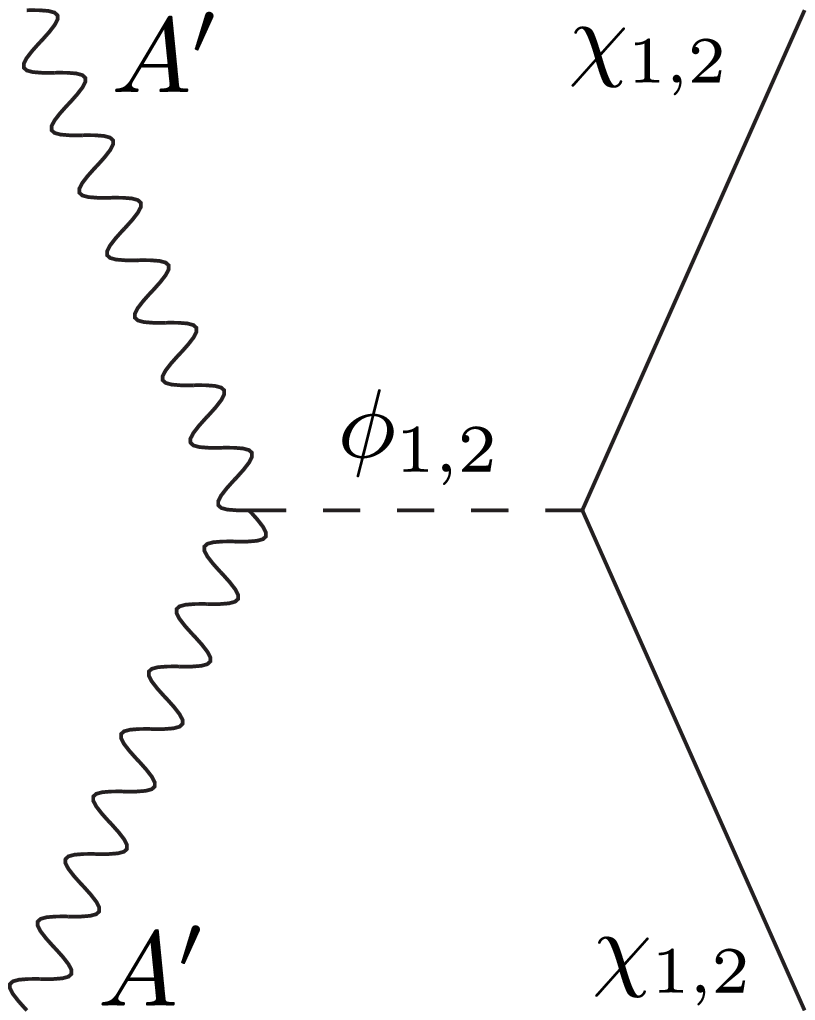}~~
\includegraphics[width=0.2\textwidth]{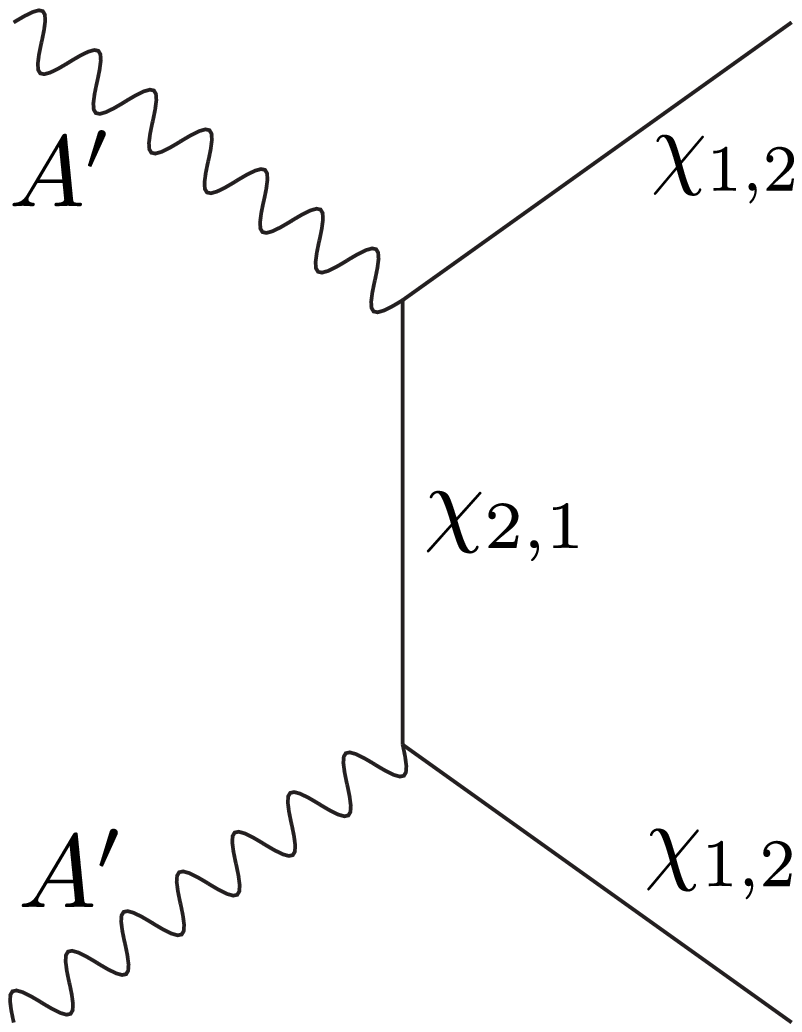}~~
\includegraphics[width=0.2\textwidth]{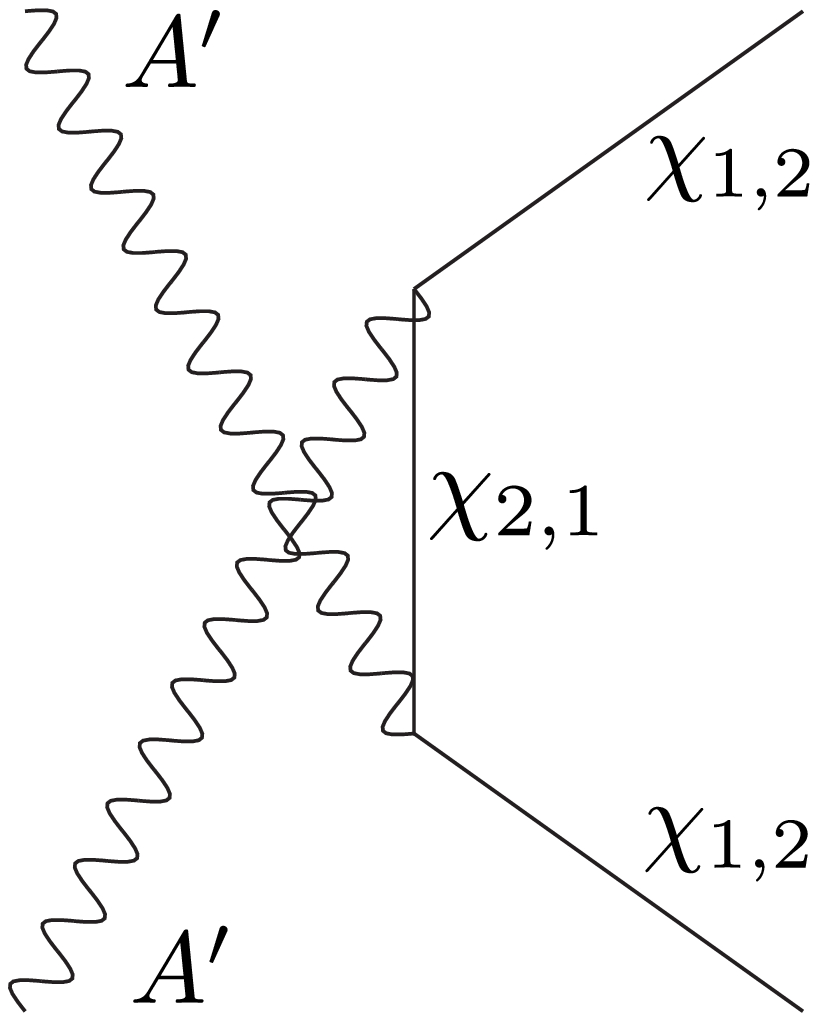}
\caption{Pure vector elements of the $A^{\prime} A^{\prime} \leftrightarrow \chi_i \chi_i$ diagrams.} \label{AADiagrams}
\end{figure}
\begin{figure}
\includegraphics[width=0.2\textwidth]{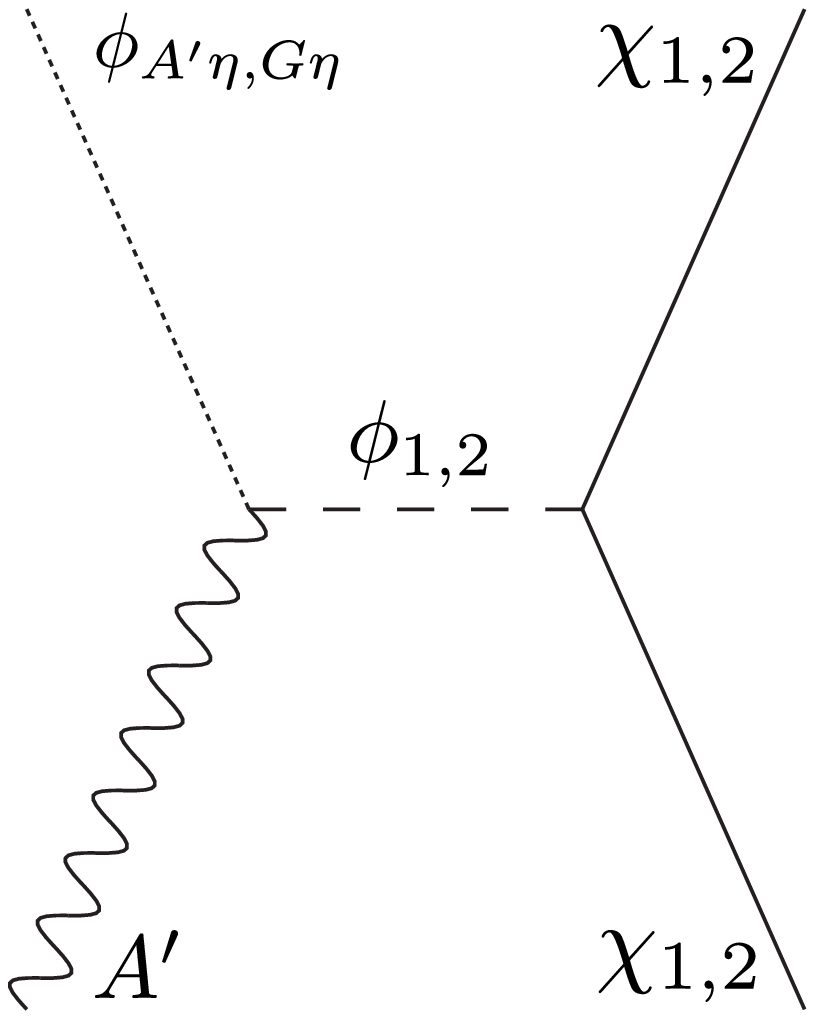}~~
\includegraphics[width=0.2\textwidth]{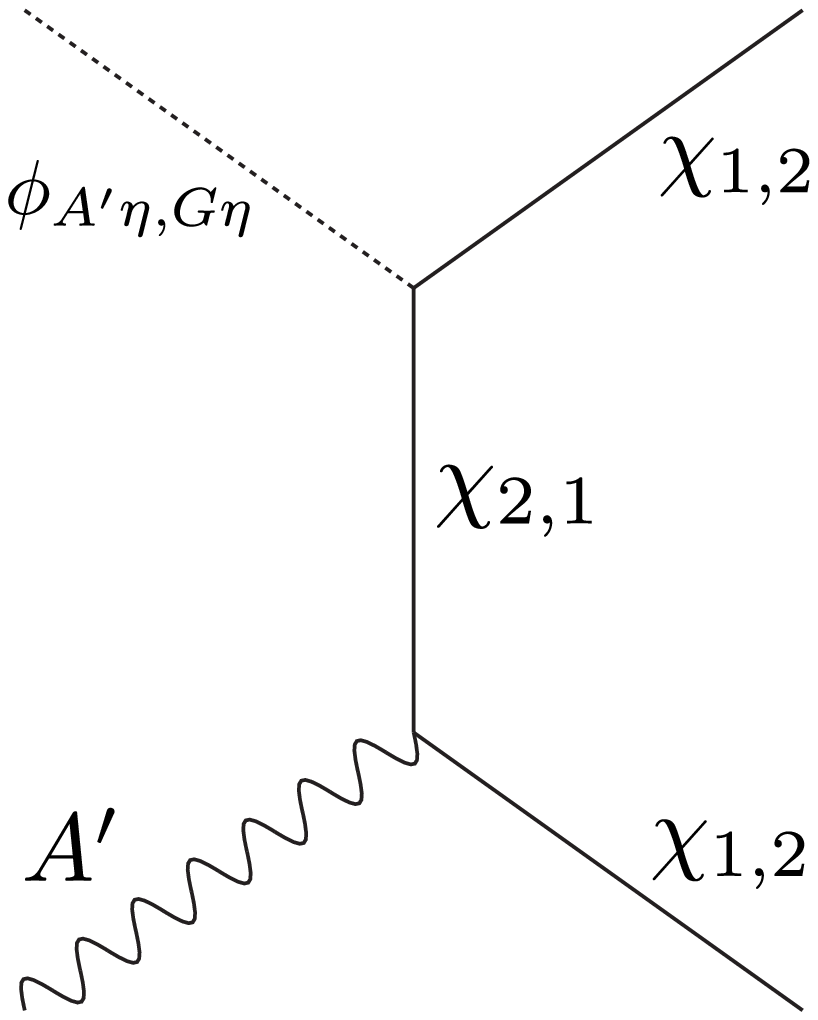}~~
\includegraphics[width=0.2\textwidth]{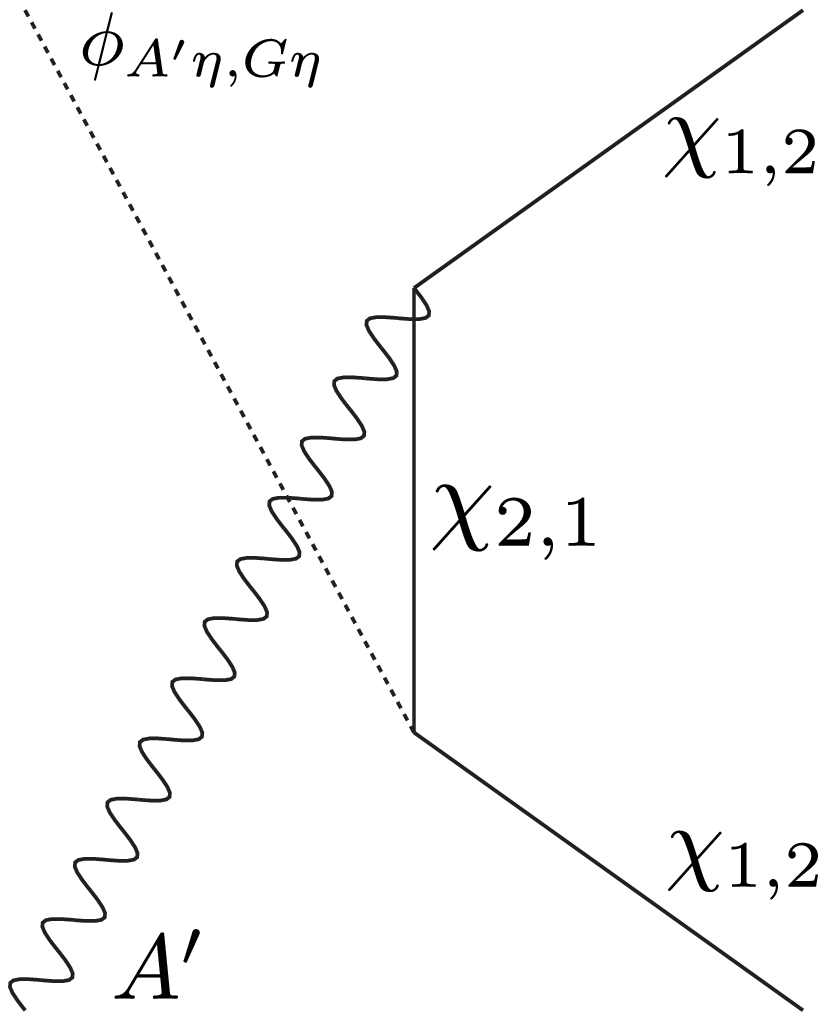}
\caption{$\phi_{A^{\prime}\eta, G\eta} A^{\prime} \leftrightarrow \chi_i \chi_i$ diagrams. The Goldstone degree of freedom of  $A^{\prime} A^{\prime} \leftrightarrow \chi_i \chi_i$ can also be indicated.}
\label{GADiagrams}
\end{figure}
\begin{figure}
\includegraphics[width=0.2\textwidth]{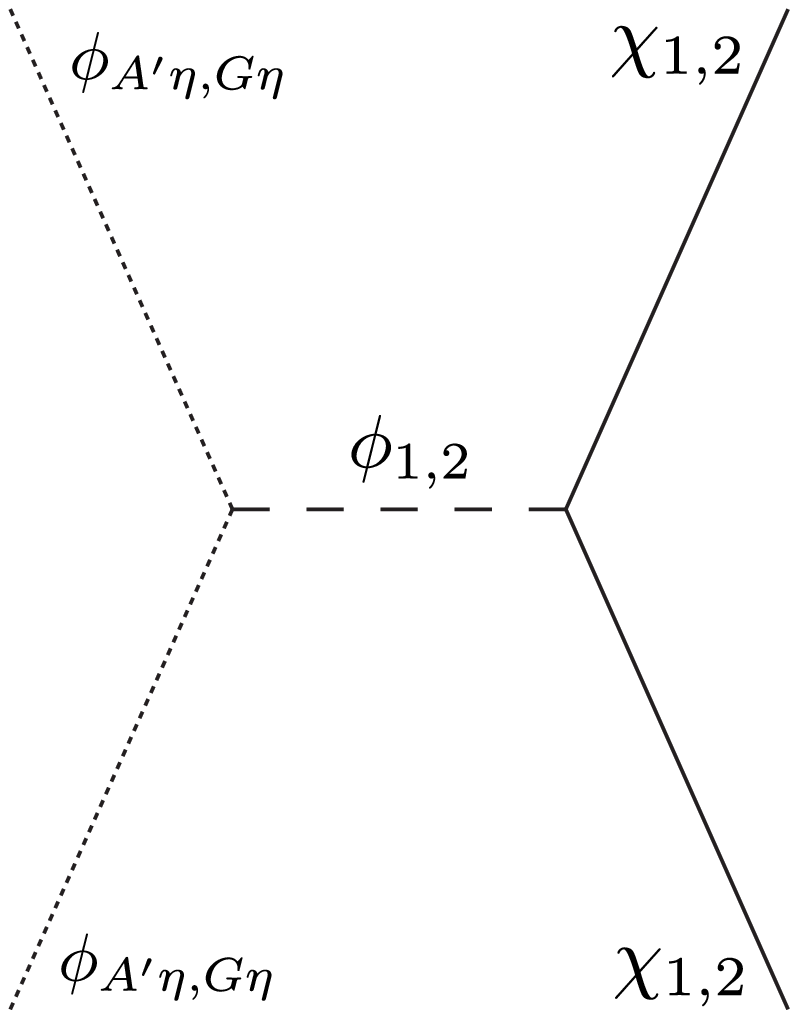}~~
\includegraphics[width=0.2\textwidth]{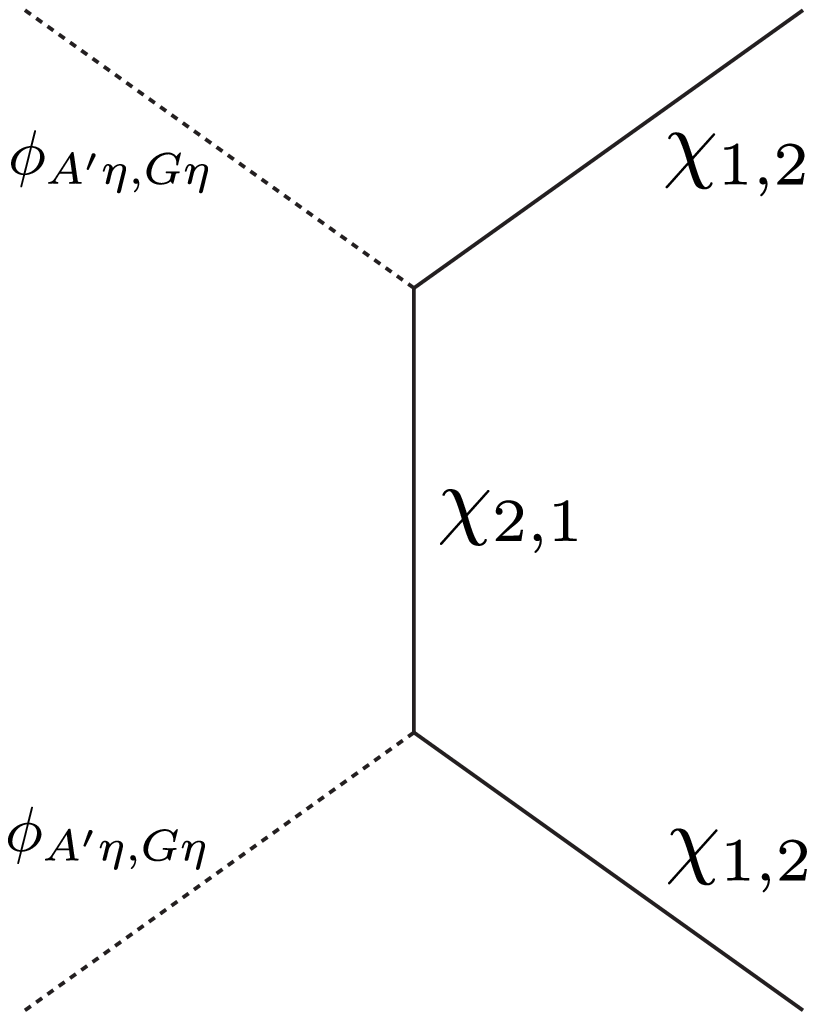}~~
\includegraphics[width=0.2\textwidth]{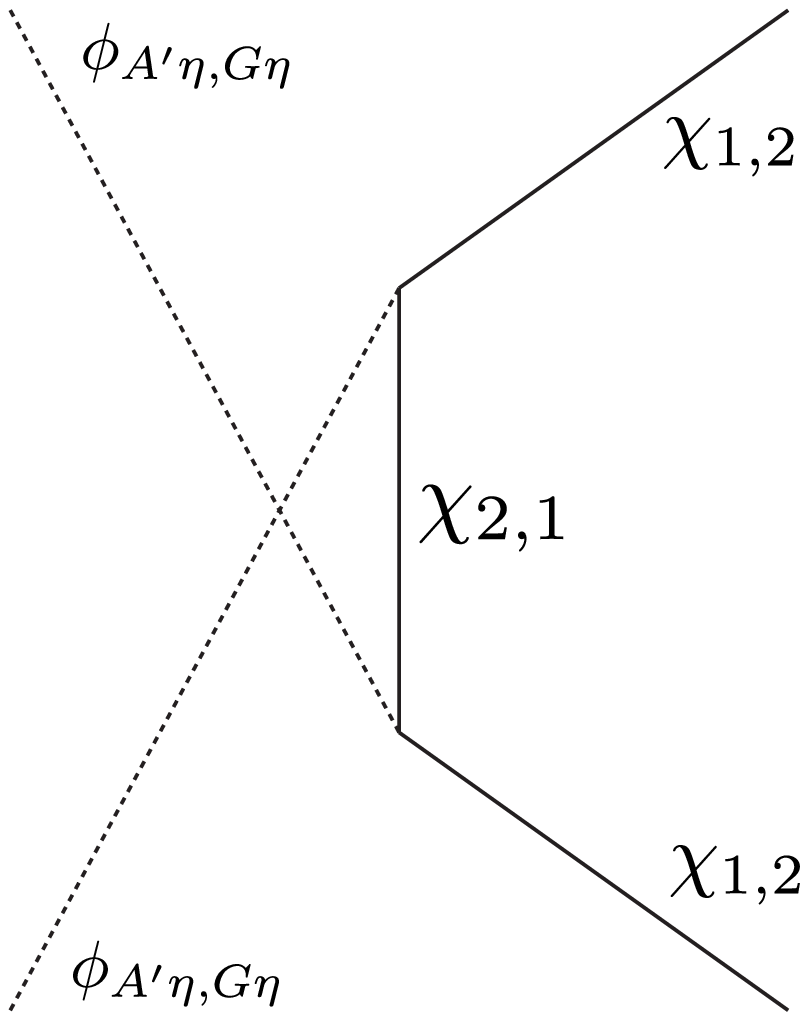}
\caption{$\phi_{A^{\prime}\eta, G\eta} \phi_{A^{\prime}\eta, G\eta}\leftrightarrow \chi_i \chi_i$ diagrams. The Goldstone degree of freedom of  $A^{\prime} A^{\prime} \leftrightarrow \chi_i \chi_i$ can also be indicated.}
\label{GGDiagrams}
\end{figure}

\begin{figure}
\includegraphics[width=0.2\textwidth]{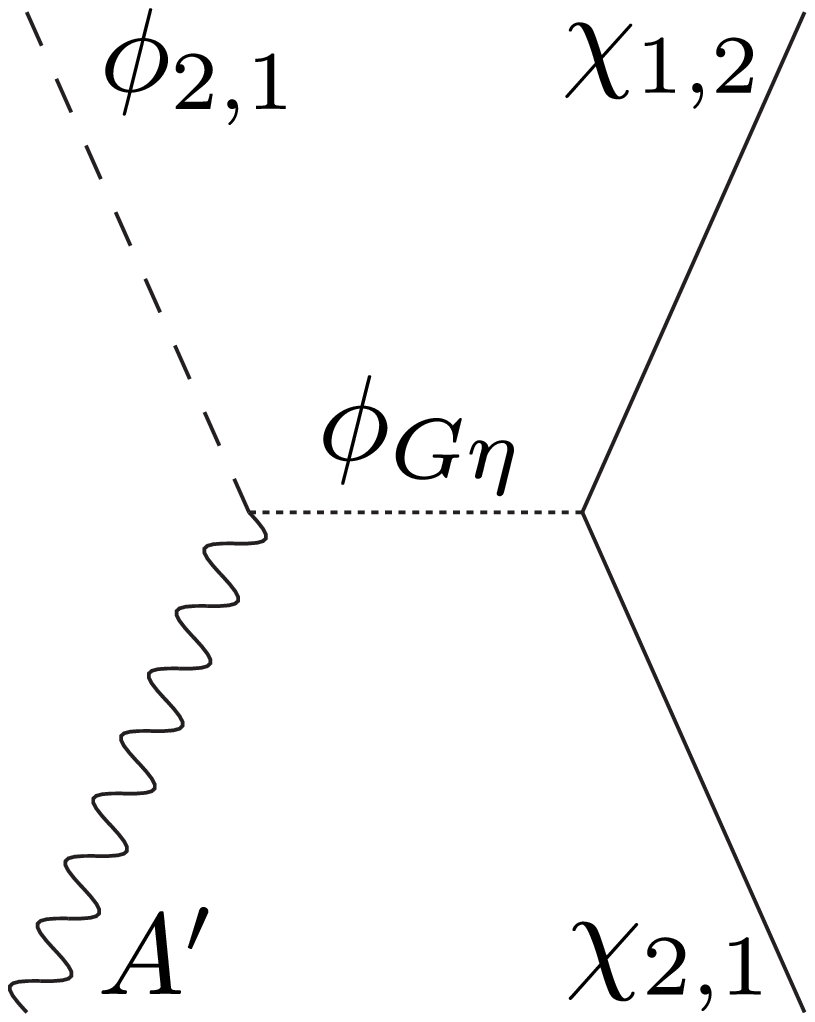}~
\includegraphics[width=0.2\textwidth]{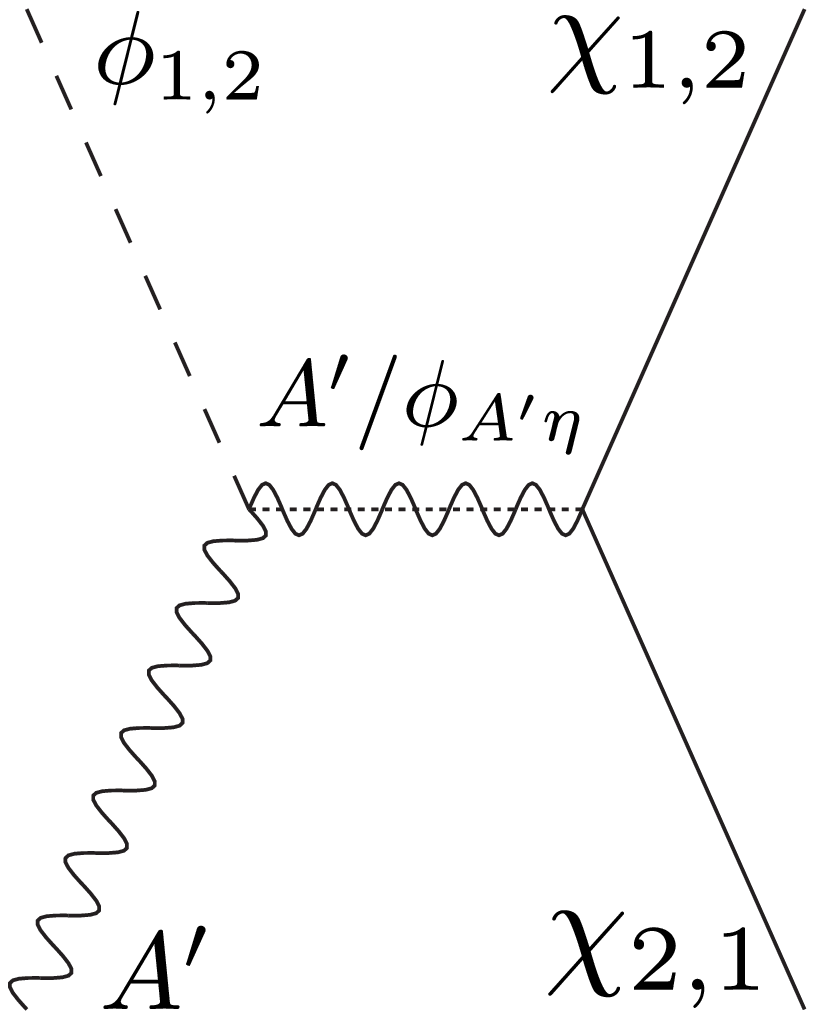}~
\includegraphics[width=0.2\textwidth]{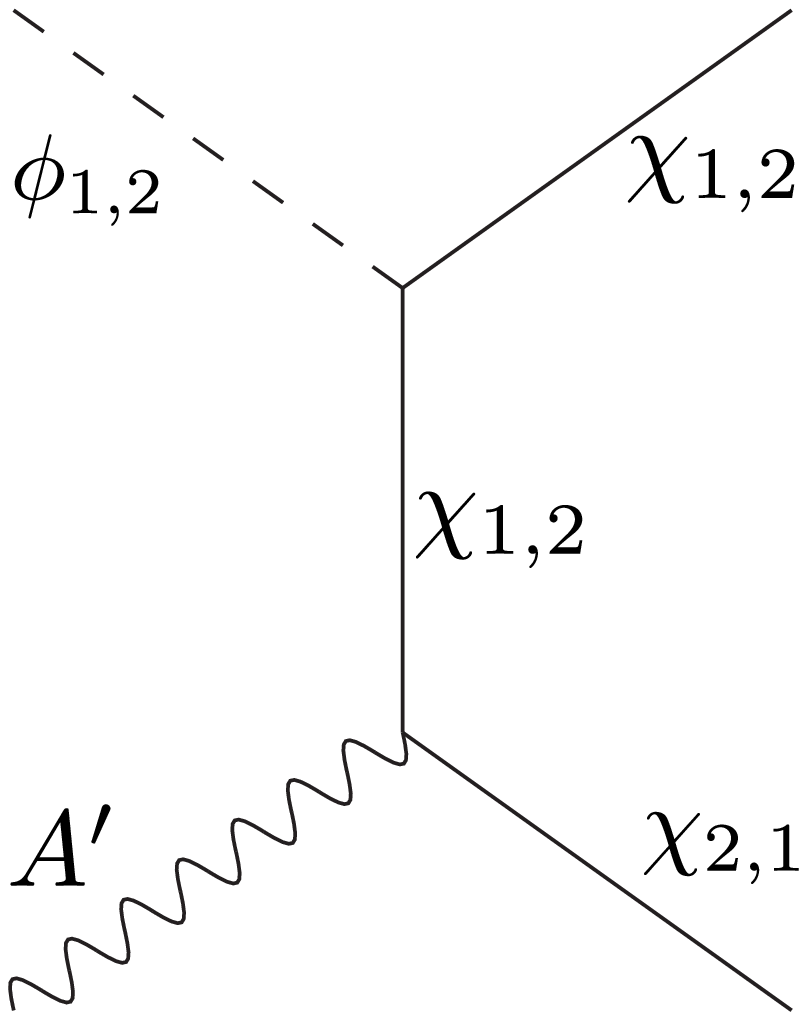}~
\includegraphics[width=0.2\textwidth]{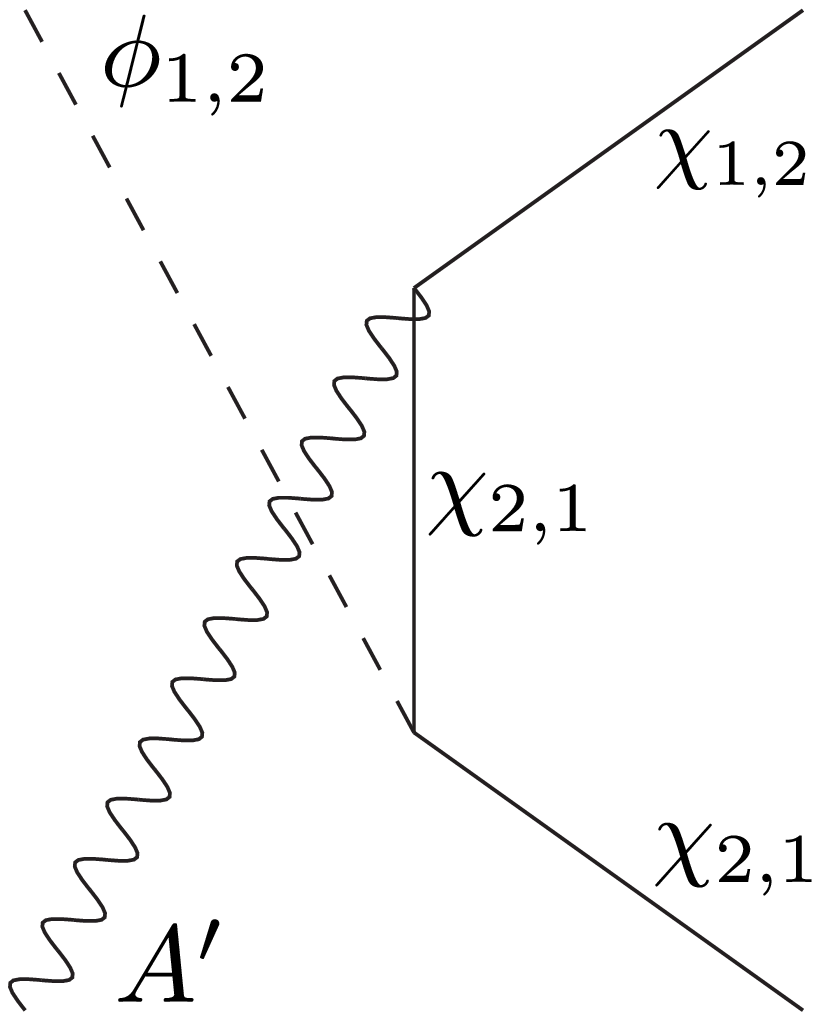}
\caption{$\phi_{1,2} A^{\prime} \leftrightarrow \chi_1 \chi_2$ diagrams. }
\label{HADiagrams}
\end{figure}
\begin{figure}
\includegraphics[width=0.2\textwidth]{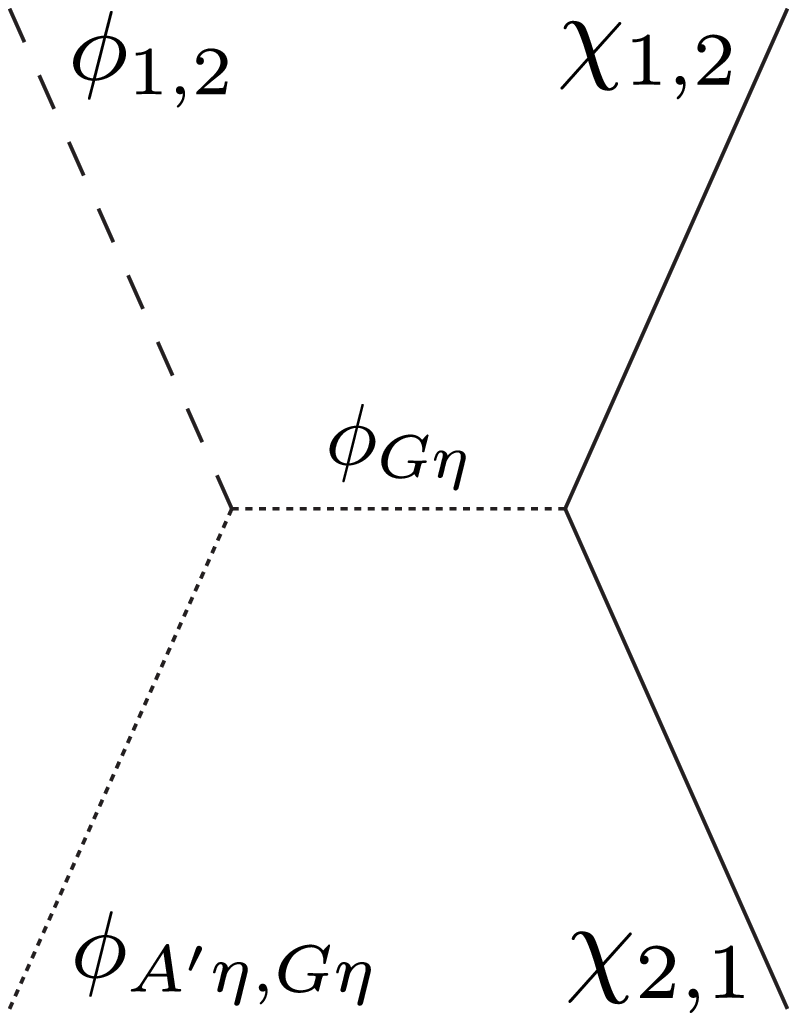}~
\includegraphics[width=0.2\textwidth]{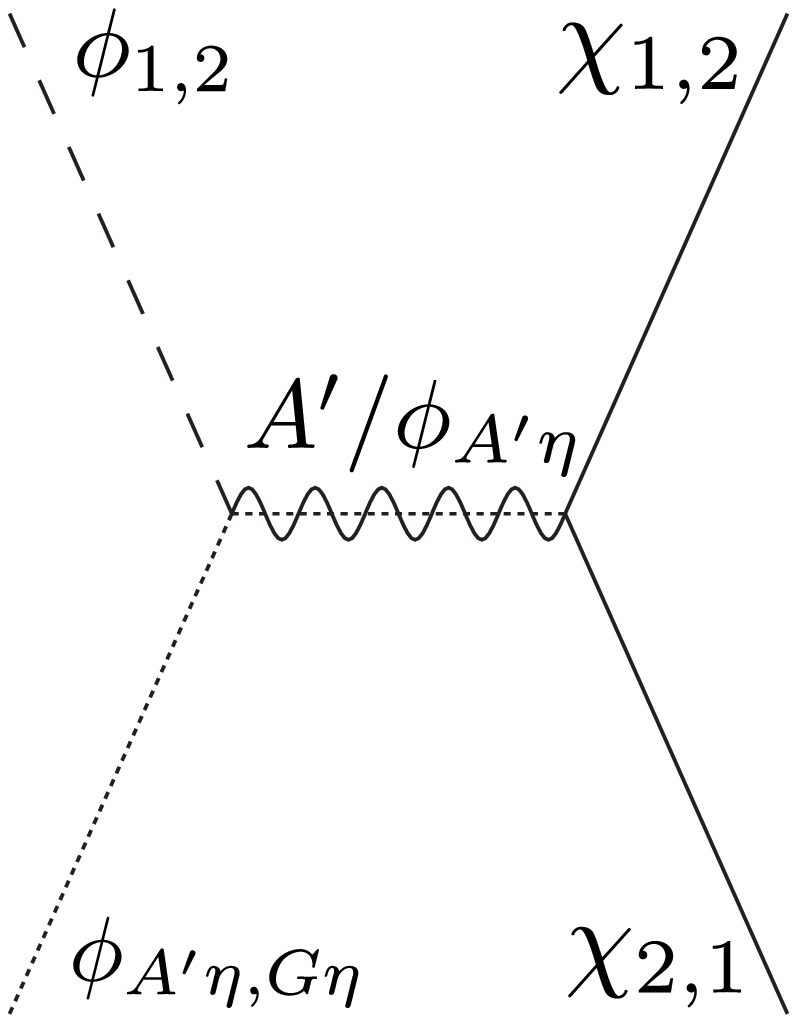}~
\includegraphics[width=0.2\textwidth]{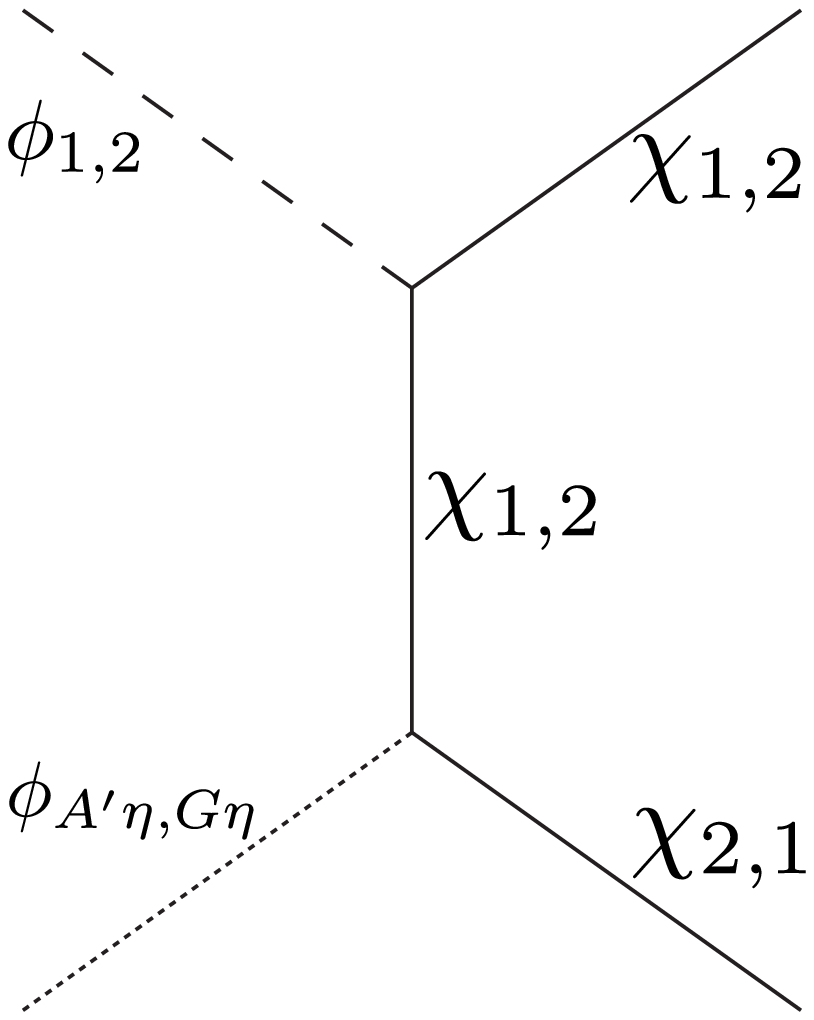}~
\includegraphics[width=0.2\textwidth]{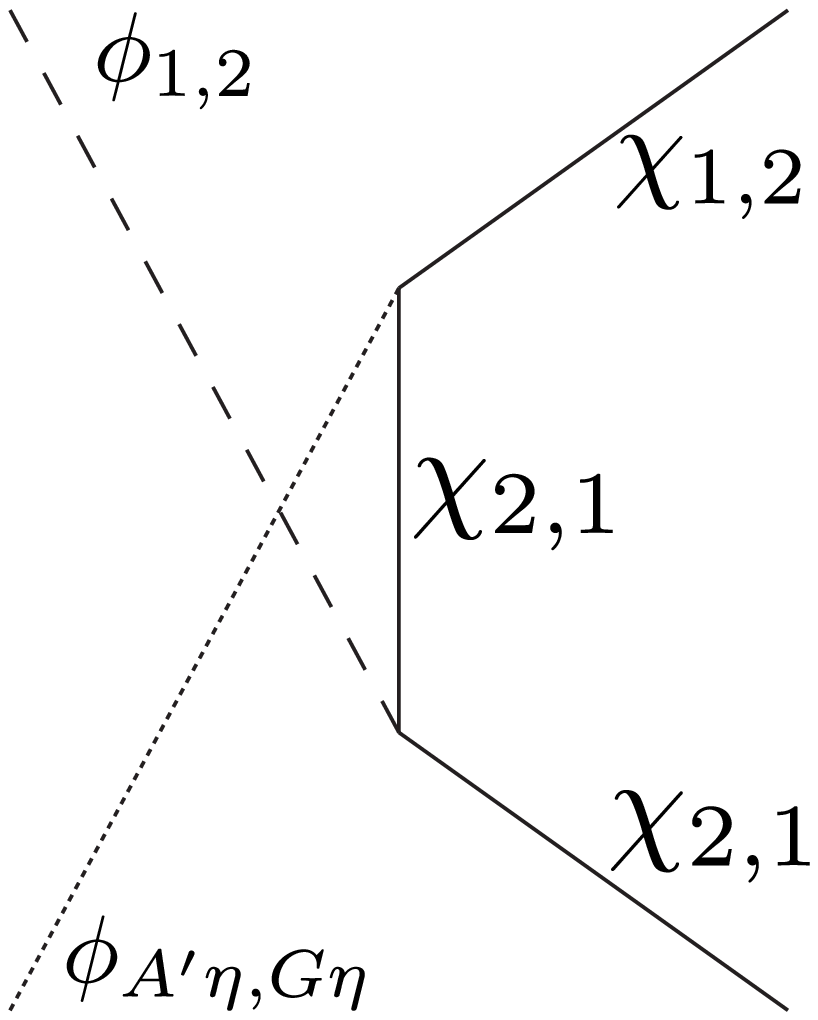}
\caption{$\phi_{A^{\prime}\eta, G\eta} \phi_{1,2} \leftrightarrow \chi_1 \chi_2$ diagrams.  The Goldstone degree of freedom of $A^{\prime} \phi_{1,2} \leftrightarrow \chi_1 \chi_2$ can also be indicated.} \label{GHDiagrams}
\end{figure}
\begin{figure}
\includegraphics[width=0.2\textwidth]{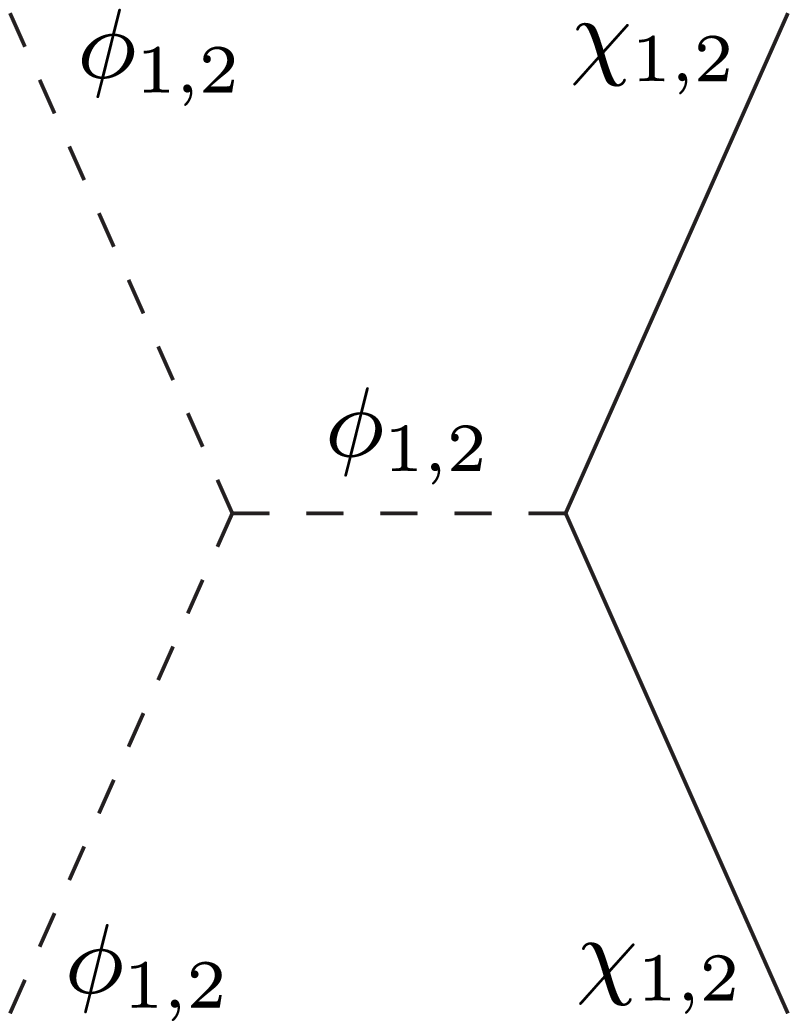}~
\includegraphics[width=0.2\textwidth]{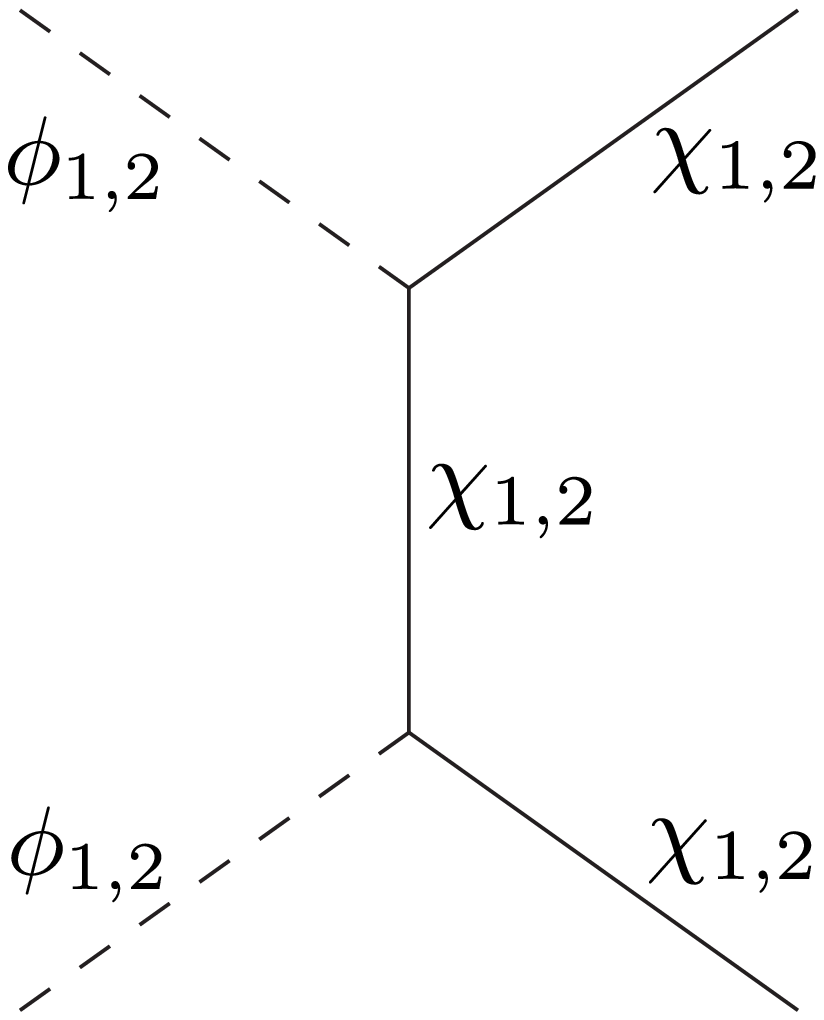}~
\includegraphics[width=0.2\textwidth]{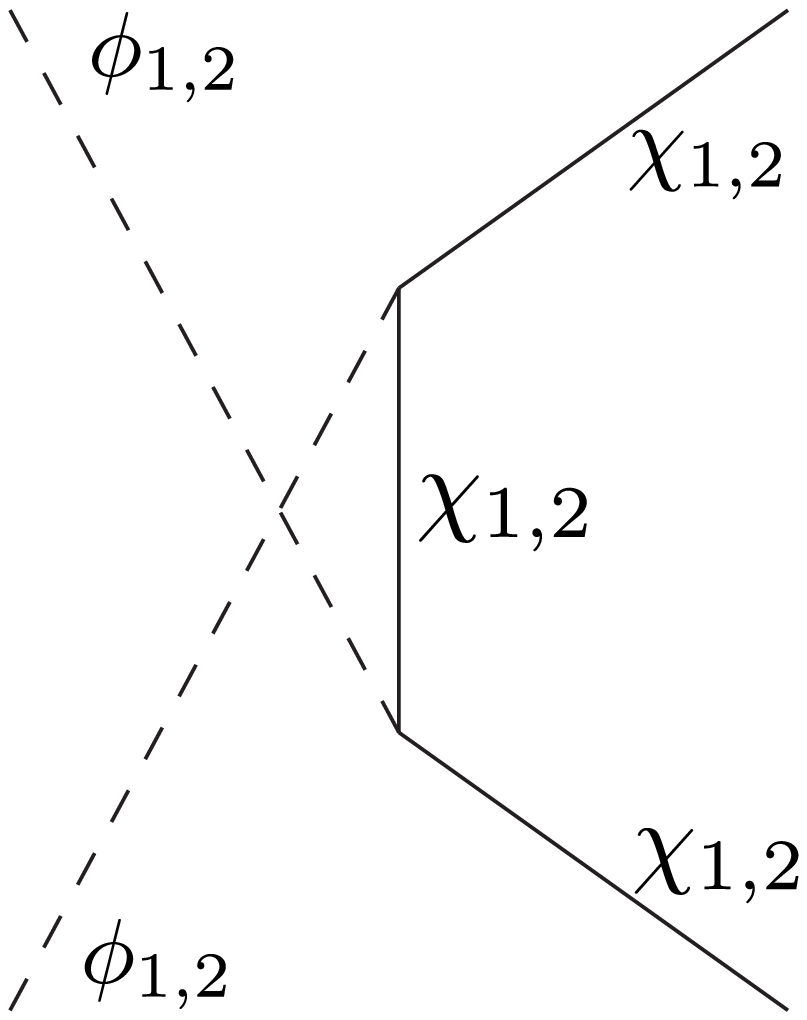}
\caption{$\phi_{1,2} \phi_{1,2} \leftrightarrow \chi_i \chi_i$ diagrams. }
\label{HHDiagrams}
\end{figure}

\begin{figure}
\includegraphics[width=0.2\textwidth]{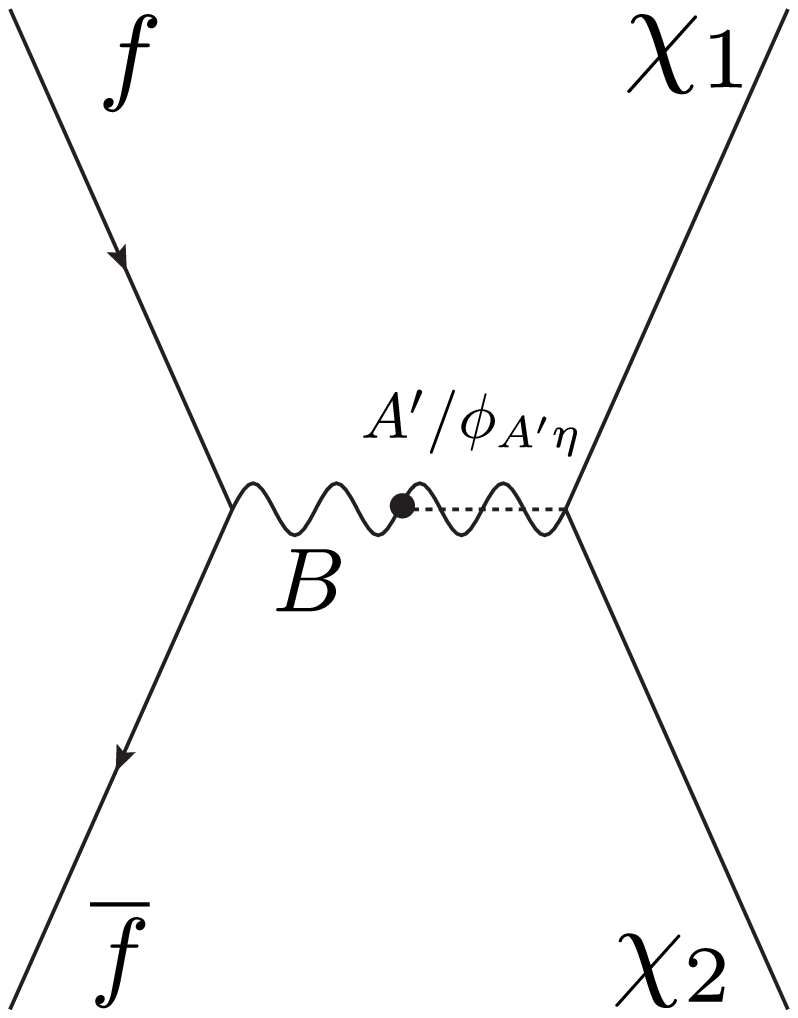}~
\includegraphics[width=0.2\textwidth]{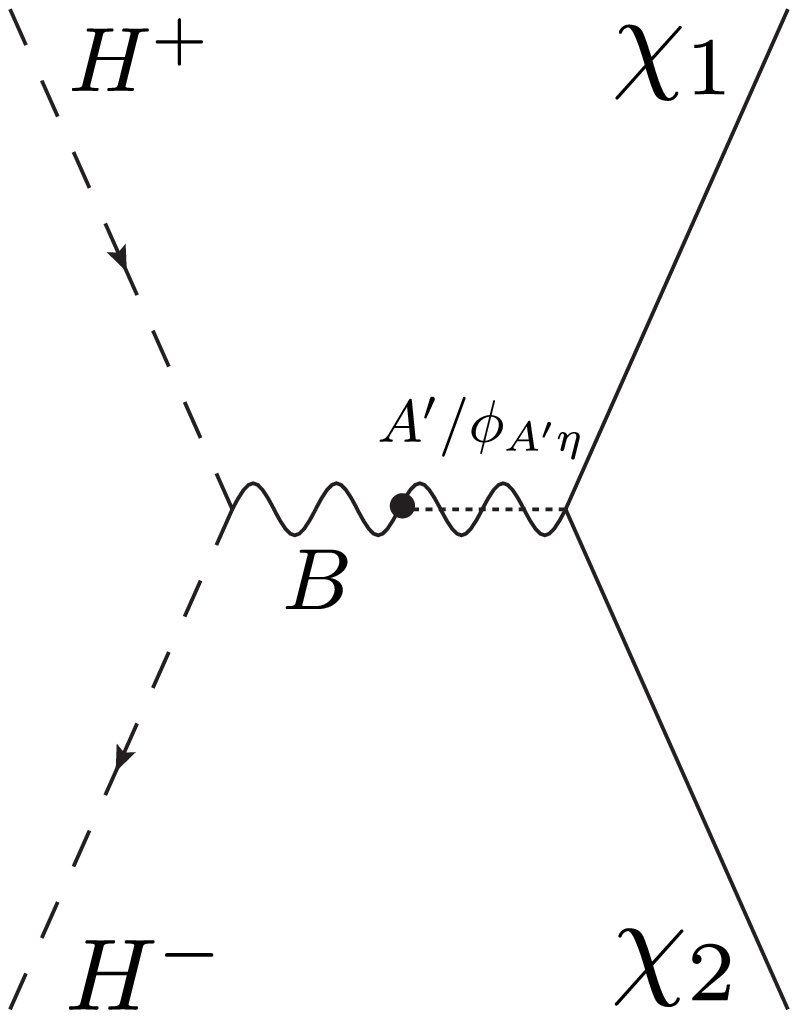}
\caption{$\overline{f} f \leftrightarrow \chi_1 \chi_2$ (left panel) and $H^+ H^-  \leftrightarrow \chi_1 \chi_2$ (right panel) diagrams.} \label{ChiChi2SMSM}
\end{figure}

\section{Phenomenological constraints}
\label{sec:phecon}

The FIMP dark matter interacts so faintly with the SM sector, so it typically lies far beyond the ability of all the direct and indirect detection experiments. For our model discussed in this paper, the dark Higgs sector and the $A^{\prime}$ are constrained by the experimental data.

In this model, there is always a massless Goldstone boson $\phi_{G \eta}$.  The relic of such a particle behaves as a dark radiation, and can contribute to the effective number of neutrino species $N_{\rm eff}$. The SM neutrino-to-photon density ratio is given by
\begin{eqnarray}
\frac{\rho_{\nu}}{\rho_{\gamma}}=\frac{7}{8} N_{\rm eff} \left(\frac{4}{11}\right)^{\frac{4}{3}}.
\end{eqnarray}
If the dark radiation decouples with the thermal bath at some temperature $T_{\rm dr}$ when the effective degree of freedom is $g^*(T_{\rm dec})$, its density respective to the photon density is calculated to be
\begin{eqnarray}
\frac{\rho_{\rm dr}}{\rho_{\gamma}} = N_{\rm dr} \left( \frac{2}{g^*(T_{\rm dec})} \right)^{\frac{4}{3}}.
\end{eqnarray}
If $\phi_{G \eta}$ is the dark radiation,  $N_{\rm dr}=1$, and its correction to $N_{\rm eff}$ is
\begin{eqnarray}
\delta N_{\rm eff} \approx \frac{7}{8} \left( \frac{11}{2 g^*(T_{\rm dec})} \right)^{\frac{4}{3}}.
\end{eqnarray}
Since $\phi_{G \eta}$ only interacts with the Higgs sector and the dark photon sector, which are massive and disappear from the thermal plasma far above the temperature $T>10$ GeV, compelling $\phi_{G \eta}$ to decouple with the plasma when $g^*(T_{\rm dec}) \approx 100$. Therefore,
\begin{eqnarray}
\delta N_{\rm eff} \approx 0.02.
\end{eqnarray}
This is fairly safe within the Planck data \cite{Zyla:2020zbs, Planck:2018vyg, DiValentino:2019dzu}.

The exotic Higgs bosons introduced in this model mix with the SM Higgs boson, and might alter the SM-like Higgs boson's phenomenology. 
According to Ref.~\cite{Falkowski:2015iwa}, when the mixing angle between the SM-like Higgs boson and the exotic Higgs boson X $\sin \theta_{hX} \lesssim 0.2$,  the SM-like Higgs boson can fit all the experimental data safely, so we set this criterion during our scanning processes. Another stringent bound in the Higgs sector is the $h_{\rm SM} \rightarrow \phi_{G \eta} \phi_{G \eta}$, which cannot be prohibited kinematically because of the massless $\phi_{G \eta}$.  The width of the SM Higgs boson decaying into an exotic massless scalar boson $X$ can be estimated as
\begin{eqnarray}
\Gamma_{h \rightarrow XX} = \frac{\lambda_{hX}^2 v_h^2}{32 \pi m_h}.
\end{eqnarray}
$\lambda_{hX}$ is the effective $h$-$h$-$X$-$X$ coupling. We just estimate the bounds $\frac{\lambda_{hX}^2 v_h^2}{32 \pi m_h} \lesssim \Gamma_{h_{\rm SM}}=4.07$ GeV\cite{Zyla:2020zbs, LHCHiggsCrossSectionWorkingGroup:2011wcg, Dittmaier:2012vm, LHCHiggsCrossSectionWorkingGroup:2013rie}. This requires $\lambda_{hX} \lesssim 0.02$, which should be compatible with $\lambda_{s,wh}$ in order of magnitude, so we will set $\lambda_{s,wh} < 0.01$ during our scanning process to avoid this constraint.

Finally, the $A^{\prime}$ in our model might be produced at the LHC through its mixing term (\ref{YExoticMixing}) with the SM $Z/\gamma$ bosons. Ref.~\cite{Zyla:2020zbs} had summarized the detector bounds on such kind of vector bosons. For example, Ref.~\cite{CMS:2019tbu} constrained the $\frac{[\sigma \cdot B] Z^{\prime}}{[\sigma \cdot B] Z} < 10^{-5} \sim 10^{-8}$ depending on different $Z^{\prime}$ masses.  However, compared with the $Z$ production channel, (\ref{YExoticMixing}) at least introduces a $\epsilon^2$ suppression factor, not to mention the suppression due to the large $A^{\prime}$ masses we adopted in this paper. As we will illustrate, we are going to fix $\epsilon = 10^{-4}$ in each of our benchmark points listed in Tab.~\ref{BPs}, which is free from the $Z^{\prime}$ bounds.

\section{Numerical results }
\label{sec:numres}

For all the $2 \leftrightarrow 2$ processes described in Sec.~\ref{FreezeInRates}, there exists at least one s-channel diagram. Sometimes the s-channel mediator becomes on-shell when it is above the threshold to open up a corresponding $1 \leftrightarrow 2$ process. The rigorous manipulation requires to resum the one-loop ``string series'' of the mediator's self-energy diagrams. This modifies its thermal propagators to avoid the infinity, just similar to the familiar manipulation with the Breig-Wignar propagators in the zero-temperature case.  However, due to the invalidation of the Lorentz invariance in the finite temperature, the ``imaginary parts'' of the s-mediators are no longer a constant for us to evaluate conveniently. Since in the freeze-in case, if any $1 \leftrightarrow 2$ processes appears to be non-zero, the ``off-shell'' part of the $2 \leftrightarrow 2$ processes are expected sub-dominant due to the extra couplings, and the``on-shell'' part of the corresponding s-mediator should also be attributed to the $1 \leftrightarrow 2$ processes, thus should be removed to refrain the double-counting. Therefore,  an $2 \leftrightarrow 2$ process is counted only when its corresponding $1 \leftrightarrow 2$ processes disappear.

As the temperature drops, the VEVs of our scalar fields varies. In both the scenario I and II, first-order phase transition might occur at $T_p \sim 1$ TeV scale and emit gravitational waves. We scan randomly through the parameter space $v_s \in [50,  5000]$ GeV, $v_w \in [100,  15000]$ GeV, $\lambda_{sh} \in [0, 0.01]$, $\lambda_{wh} \in [0, 0.01]$, $\lambda_s \in [0, \pi]$, $\lambda_w \in [0, \pi]$, $\lambda_{sw} \in [0, \pi]$, $g_D \in [0.8, 1.2]$ for Scenario I, and $v_s \in [50, 5000]$ GeV, $\lambda_{sh} \in [0, 0.01]$, $\lambda_s \in [0, \pi]$, $g_D \in [0.8, 1.2]$, $v_w t_w \in [1, 10000]$ GeV for Scenario II for the calculations of the $\alpha$ and $\beta$ defined in Eq.~(\ref{alphabetaForGW}) which are calculated at $T_*=T_p$, to estimate the gravitational wave produced from the first-order phase transition. In both scenarios,  the lightest non-SM-like Higgs boson is constrained above 180 GeV to prevent the strong mixing between the exotic Higgs boson and the SM-like Higgs boson. Here the $\lambda_{sh}, \lambda_{wh}<0.01$ criterion are enough to confine the SM-like Higgs boson within the phenomenological constraints, and bounded from below criterion can be checked numerically by both the \texttt{CosmoTransitions} and \texttt{PhaseTracer}. We plot our results in Fig.~\ref{alphabetaFigure}. Among them we adopt two benchmark points for each scenario, which are called BP\underline{~}S1\underline{~}1, BP\underline{~}S1\underline{~}2, BP\underline{~}S2\underline{~}1, BP\underline{~}S2\underline{~}2, with their location in the $\alpha$-$\beta$ plain marked in Fig.~\ref{alphabetaFigure}, and their parameter values assigned according to Tab.~\ref{BPs}.

\begin{figure}
\includegraphics[width=0.48\textwidth]{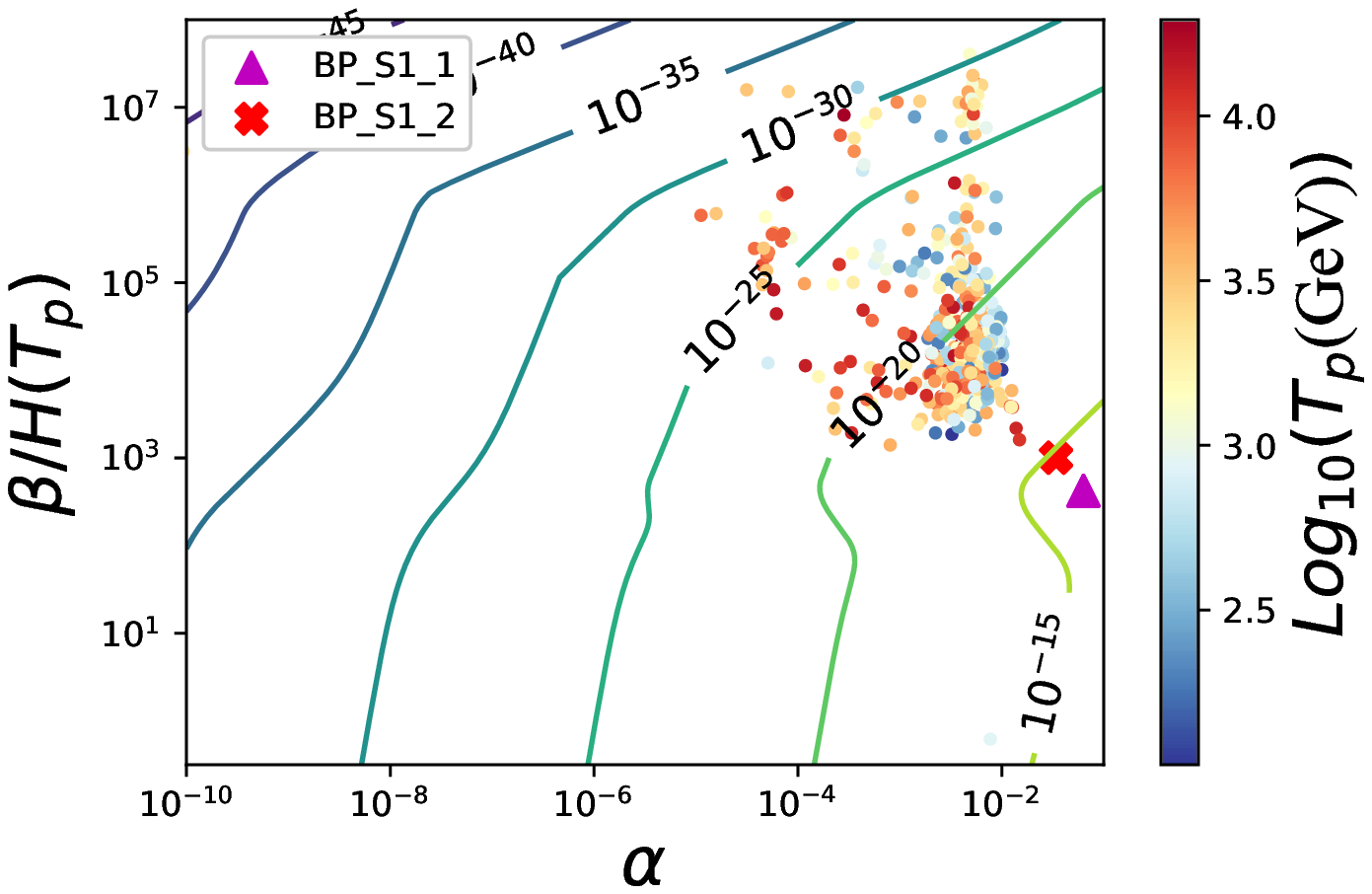}
\includegraphics[width=0.48\textwidth]{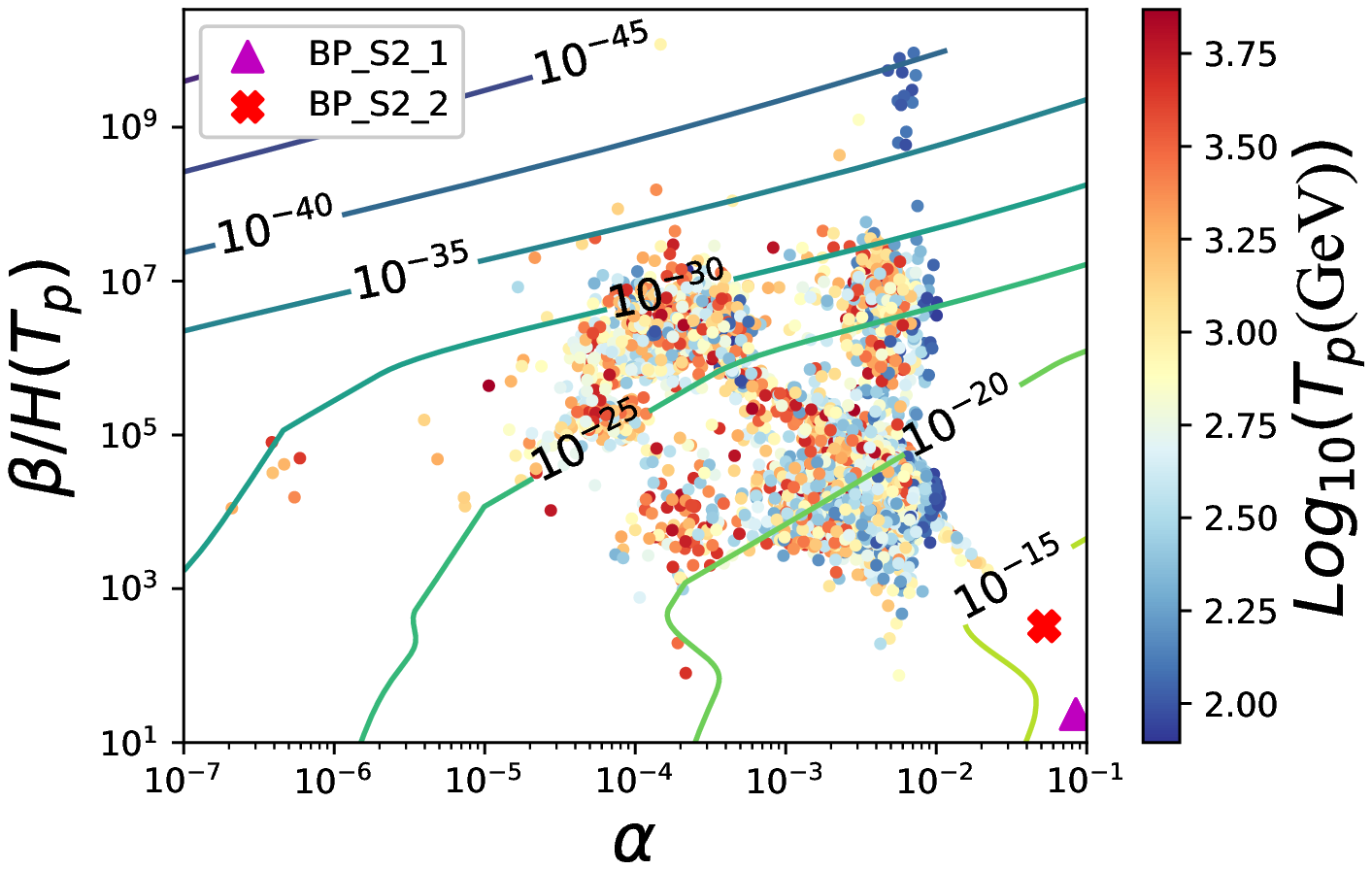}
\caption{The scanned points plotted in the $\alpha$ versus $\beta/H(T_p)$ space for Scenario I (left pannel) and Scenario II (right pannel).  The color bar indicates the percolation temperature $T_p$. If more than one step of first-order phase transition occur, we only adopt the one that generates the largest peak value of the gravitational wave relics $\Omega_{\rm GW, 1PT}(f) h^2$ contributed from the first-order phase transition. The contour is the reference peak value of the $\Omega_{\rm GW, 1PT}(f) h^2$ evaluated at $T_p=500$ GeV.} \label{alphabetaFigure}
\end{figure}

\begin{figure}
\includegraphics[width=0.48\textwidth]{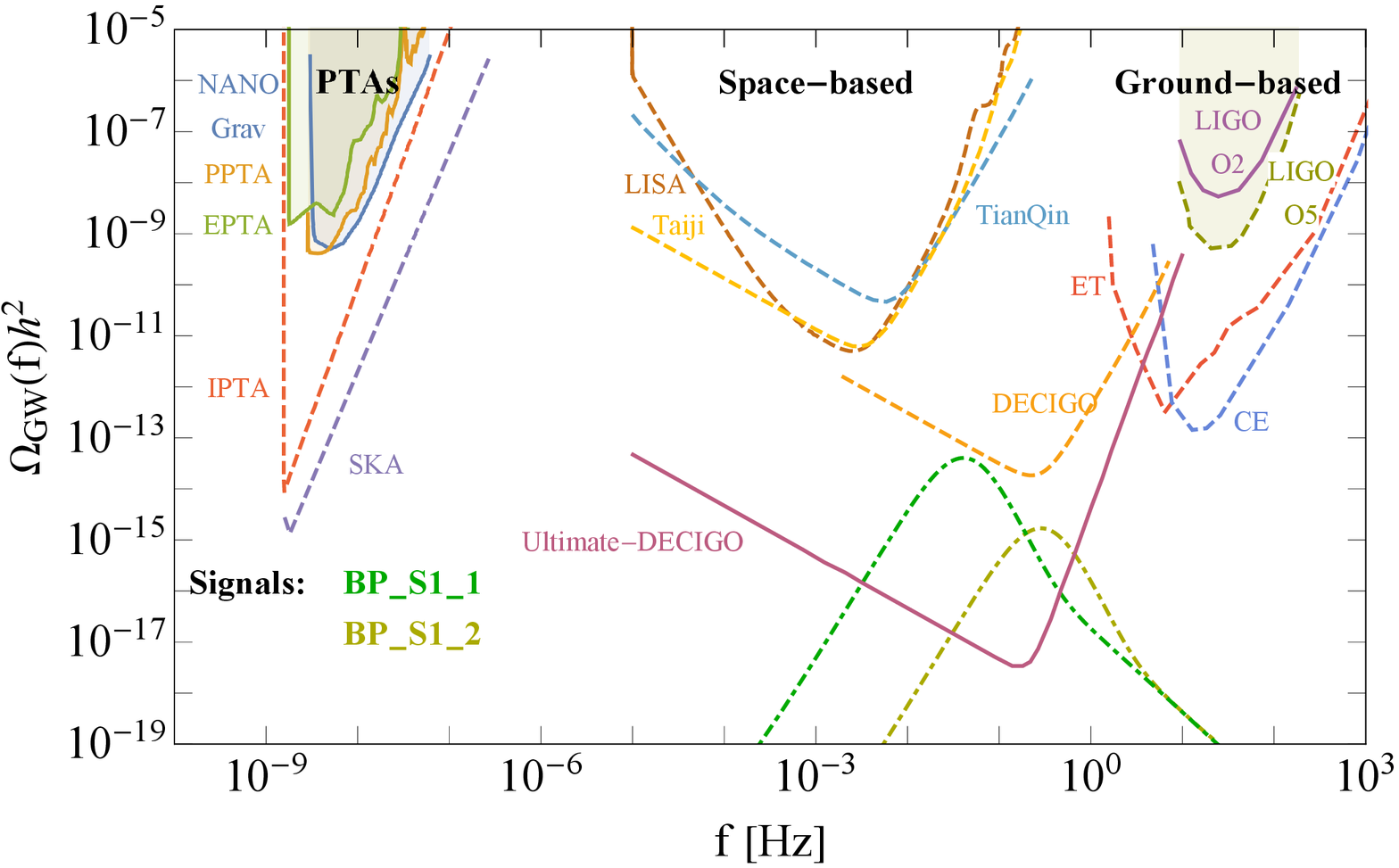}
\includegraphics[width=0.48\textwidth]{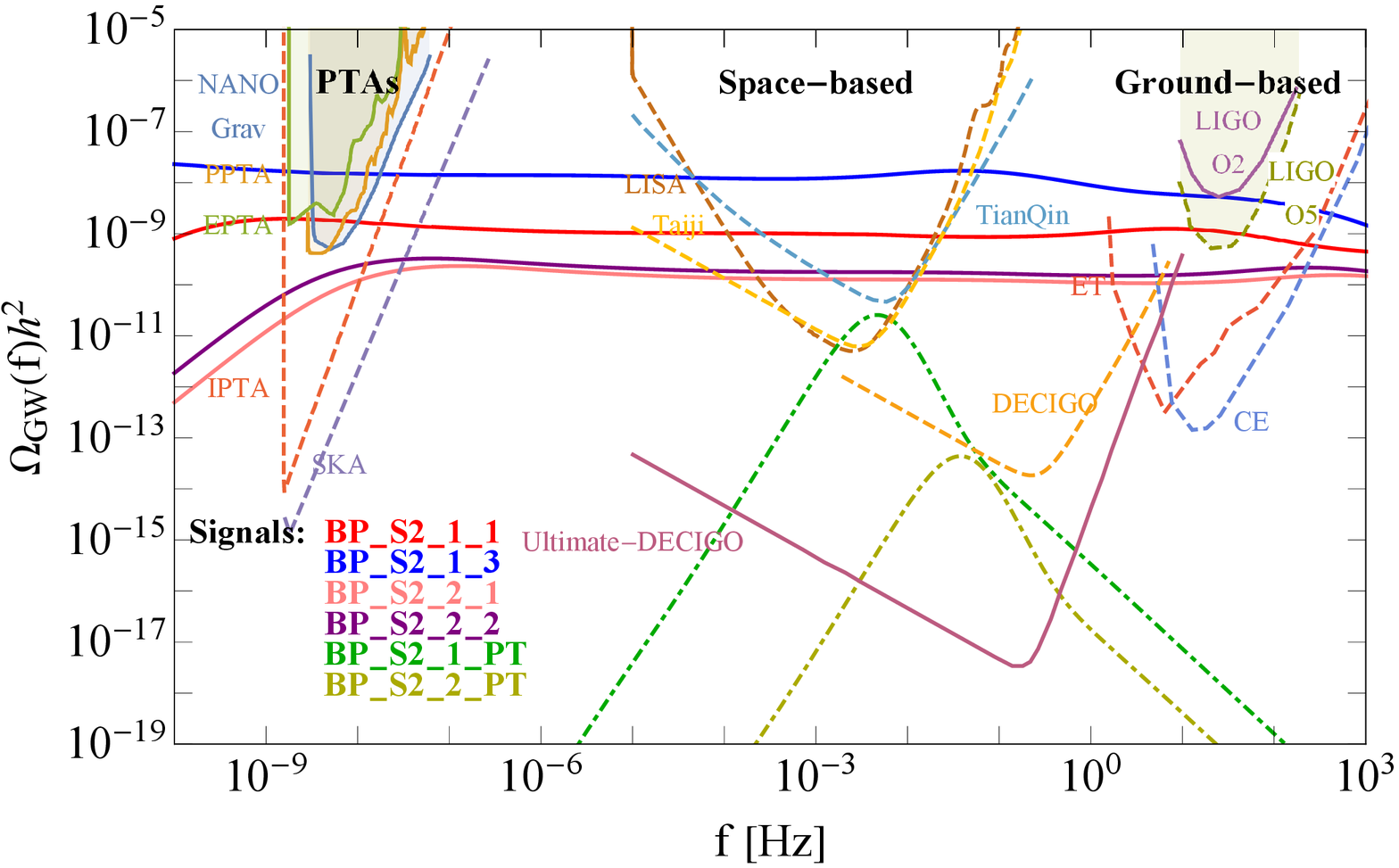}
\caption{Predicted gravitational wave spectrum for each of the (sub-)benchmark points in Scenario I (left panel) and Scenario II (right panel). In the right panel, the gravitational wave spectrum induced by the $v_s$-scale phase transitions are also plotted in the dashed-dotted lines.} \label{GWSpectrum}
\end{figure}

\begin{table}
\begin{tabular}{|c|c|c|c|c|}
\hline
~&\multicolumn{4}{|c|}{Values} \\
\hline
Parameters & BP\underline{~}S1\underline{~}1 & BP\underline{~}S1\underline{~}2 & BP\underline{~}S2\underline{~}1 & BP\underline{~}S2\underline{~}2\\
\hline
$v_s/$GeV & 1303.2 & 2612.7 & 4346.3 & 1418.9\\
$v_w/$GeV & 748.3 & 1744.2 & $\frac{2.667 \times 10^3 g_D}{2 g_{\chi}}$& $\frac{0.30638 \times 10^3 g_D}{2 g_{\chi}}$ \\
$\lambda_{sh}$ & 0.001341 & 0.0003910 & 0.006935 &0.009652 \\
$\lambda_{wh}$ & 0.006998 &  0.006582 &  & \\
$\lambda_s$ & 0.06752 & 0.06061 & 0.007278 & 0.02294 \\
$\lambda_w$ & 1.1991 & 1.7614 & 0.6 & 0.6 \\
$\lambda_{sw}$& 0.4964 & 0.5354 &  & \\
$g_D$ & 0.9688 & 0.9472 & 0.9118 & 1.1102 \\
$T_p/$GeV & 1725.1, 373.8 & 4089.4,  046.1 & 795.9 & 444.4 \\
\hline
\end{tabular}

\begin{tabular}{|c|c|c|c|c|c|c|}
\hline
~&\multicolumn{6}{|c|}{Values} \\
\hline
Parameters & BP\underline{~}S1\underline{~}1\underline{~}1 & BP\underline{~}S1\underline{~}1\underline{~}2 & BP\underline{~}S1\underline{~}1\underline{~}3 & BP\underline{~}S1\underline{~}2\underline{~}1 & BP\underline{~}S1\underline{~}2\underline{~}2 & BP\underline{~}S1\underline{~}2\underline{~}3\\
\hline
$m_{\chi}/$GeV & $2000$ & $2000$ & $2000$ & $2000$ & $2000$ & $2000$ \\
$g_{\chi}$ & $6.76\times10^{-12}$ & $6.09 \times 10^{-12}$ & $1.34 \times 10^{-12}$ & $5.33 \times 10^{-12}$ & $3.41 \times 10^{-12}$ & $4.42 \times 10^{-13}$ \\
$y_{\chi}$ & $0.1g_{\chi}$ & $g_{\chi}$ & $10 g_{\chi}$ & $0.1g_{\chi}$ & $g_{\chi}$ & $10 g_{\chi}$ \\
\hline
\hline
Parameters & BP\underline{~}S2\underline{~}1\underline{~}1 & & BP\underline{~}S2\underline{~}1\underline{~}3 & BP\underline{~}S2\underline{~}2\underline{~}1 & BP\underline{~}S2\underline{~}2\underline{~}2 & \\
\hline
$m_{\chi}/$GeV & $2000$ &  & $2000$ & $2000$ & $2000$ &  \\
$m_{\chi_1}/$GeV & $1853.8$ &  & $-2385.6$ & $1975.9$ & $1759.5$ &   \\
$g_{\chi}$ & $ 5.00 \times10^{-12}$ & & $3.64 \times 10^{-13}$ & $5.64 \times 10^{-12}$ & $4.03 \times 10^{-12}$ &  \\
$y_{\chi}$ & $0.1g_{\chi}$ &  & $3 g_{\chi}$ & $0.1g_{\chi}$ & $g_{\chi}$ & \\
\hline
\end{tabular}
\caption{Benchmark points with their parameters. During our calculations, the mixing parameters between the dark vector boson and the hypercharge boson in (\ref{YExoticMixing}) are set to be a universal $\epsilon=10^{-4}$, and the $V_{sw}$ defined in (\ref{SWMixing}) are assigned to be $3 \times 10^{-5} \frac{v_s}{v_w}$ among all Scenario II benchmark points. The two benchmark points in Scenario I induce two-step phase transitions, thus both the $T_p$ values are listed.} \label{BPs}
\end{table}

In scenario II, $v_w \gg v_s$, and the spontaneously symmetry breaking around the $v_w$-scale can create the cosmic strings, ensuing the gravitational waves relics. Notice that around the $v_s$-scale, the $\Phi_w$ sectors had been integrated out, and the only remained contribution is the $y_w v_w$ within the $A^{\prime}$ mass term from (\ref{mAPrime}). Since $y_w$ and  $v_w$ do not separate in all of the processes during the $v_s$-scale phase transitions, one can regard them as a whole. Therefore, in the first table of Tab.~\ref{BPs}, $v_w$ remained undetermined, and for each BP\underline{~}S2\underline{~}1 and BP\underline{~}S2\underline{~}2, the $v_s$-scale phase transition induced gravitational waves still remain unchanged as $g_\chi$ varies, because the product $y_w v_w$ has been fixed for each of them.

For further comparison, we also plot the TeV-scale phase evolutions for all of our benchmark points in Fig.~\ref{PhaseEvolution}
\begin{figure}
\includegraphics[width=0.48\textwidth]{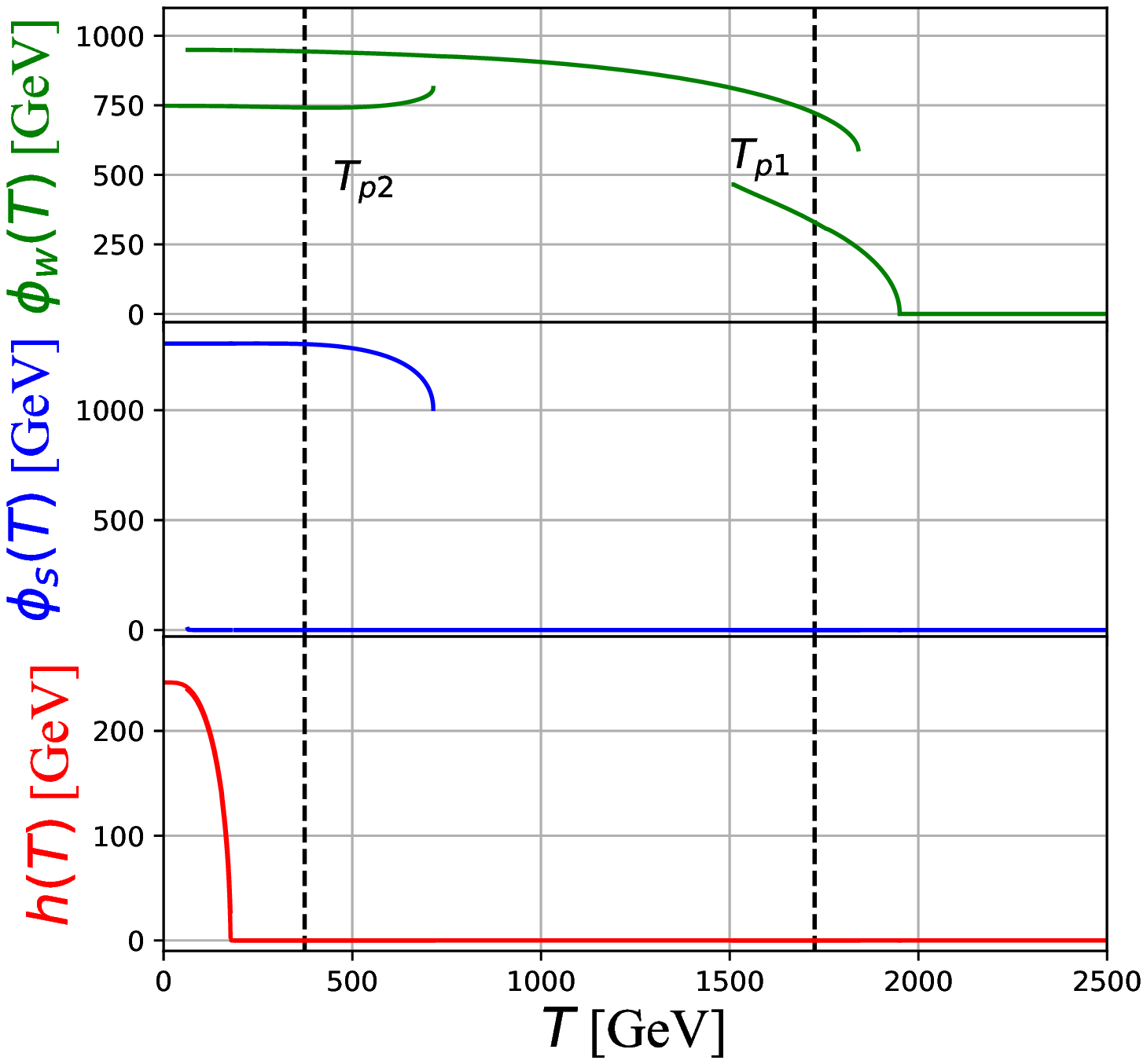}
\includegraphics[width=0.48\textwidth]{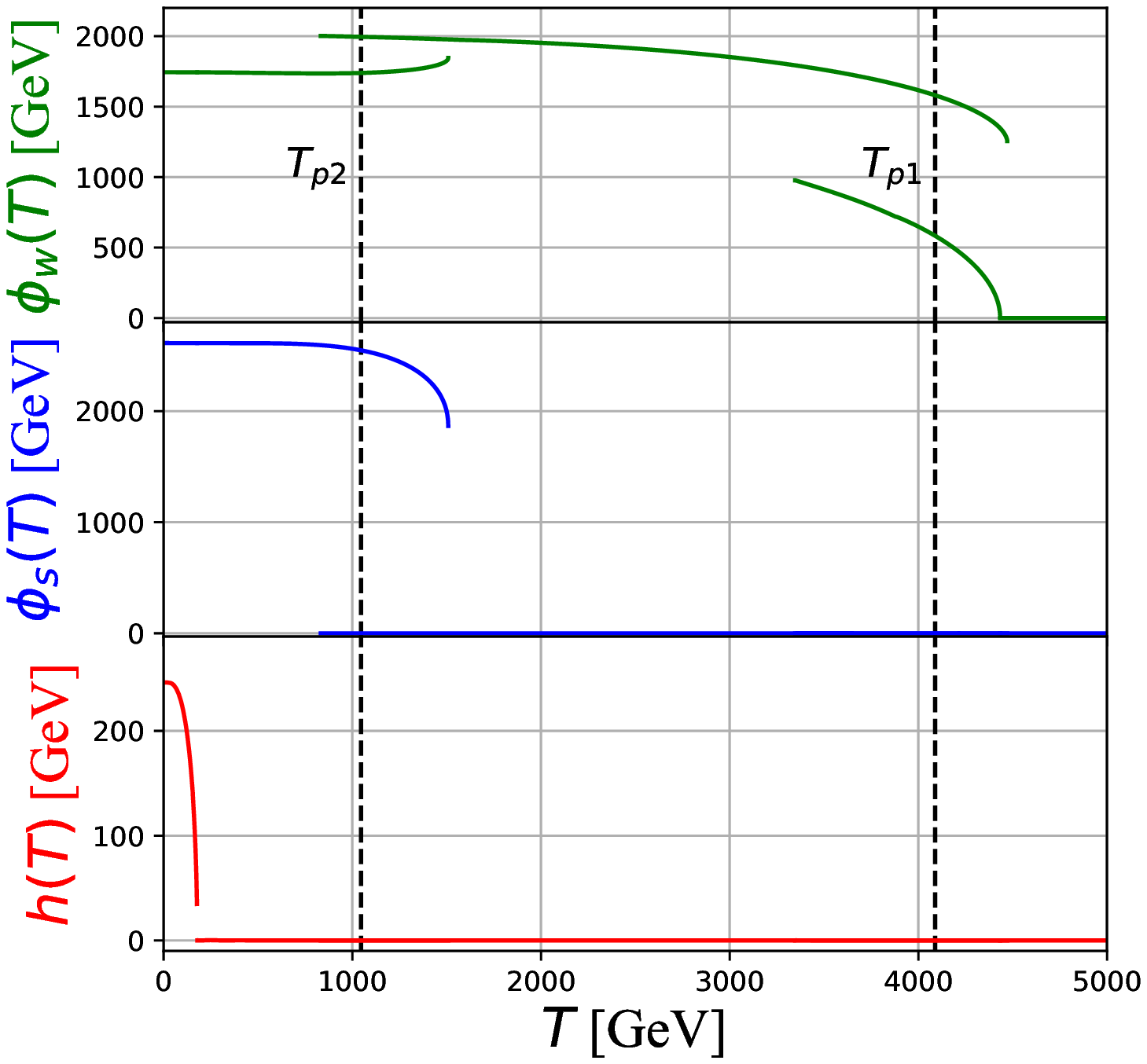}
\includegraphics[width=0.48\textwidth]{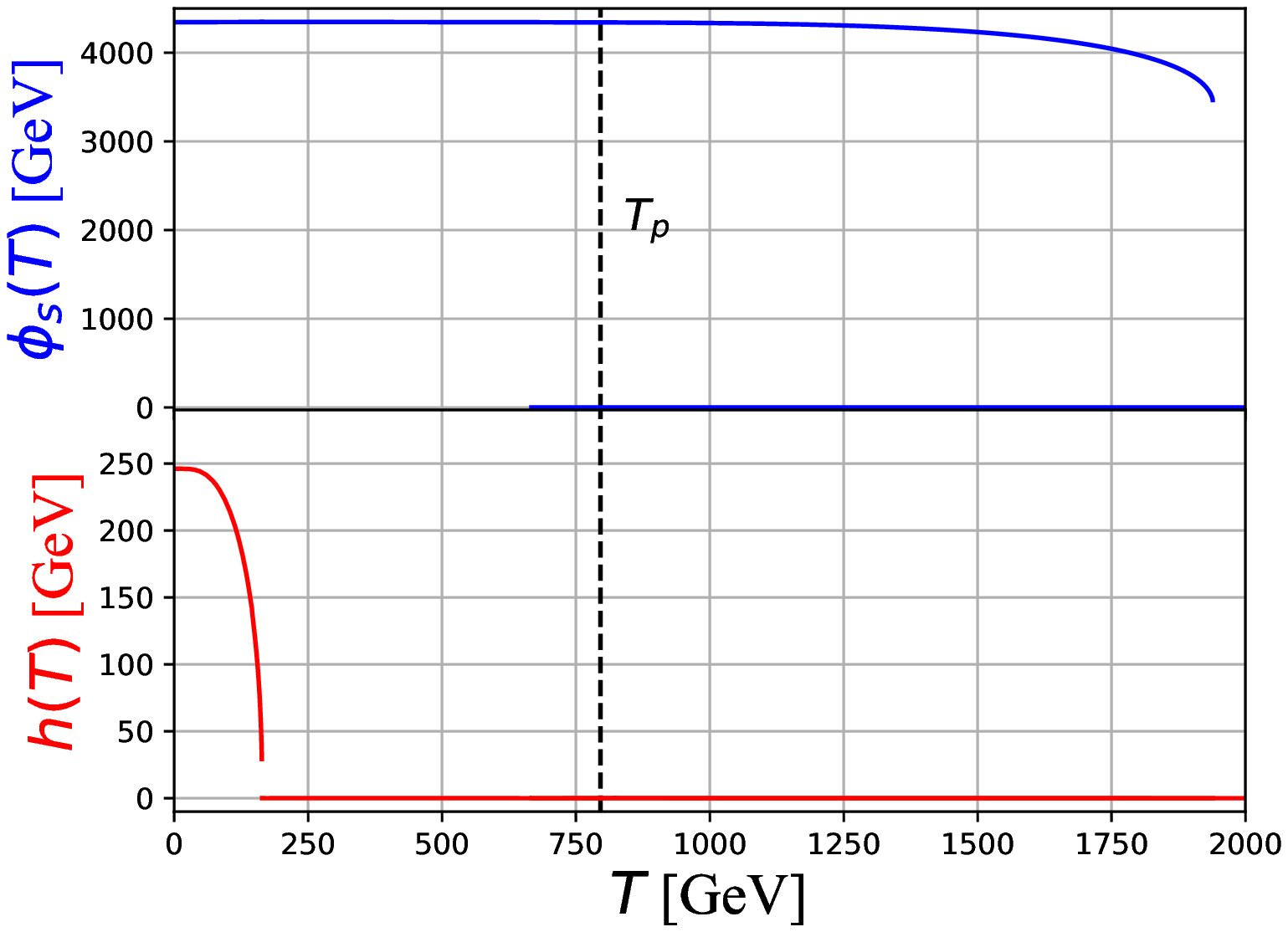}
\includegraphics[width=0.48\textwidth]{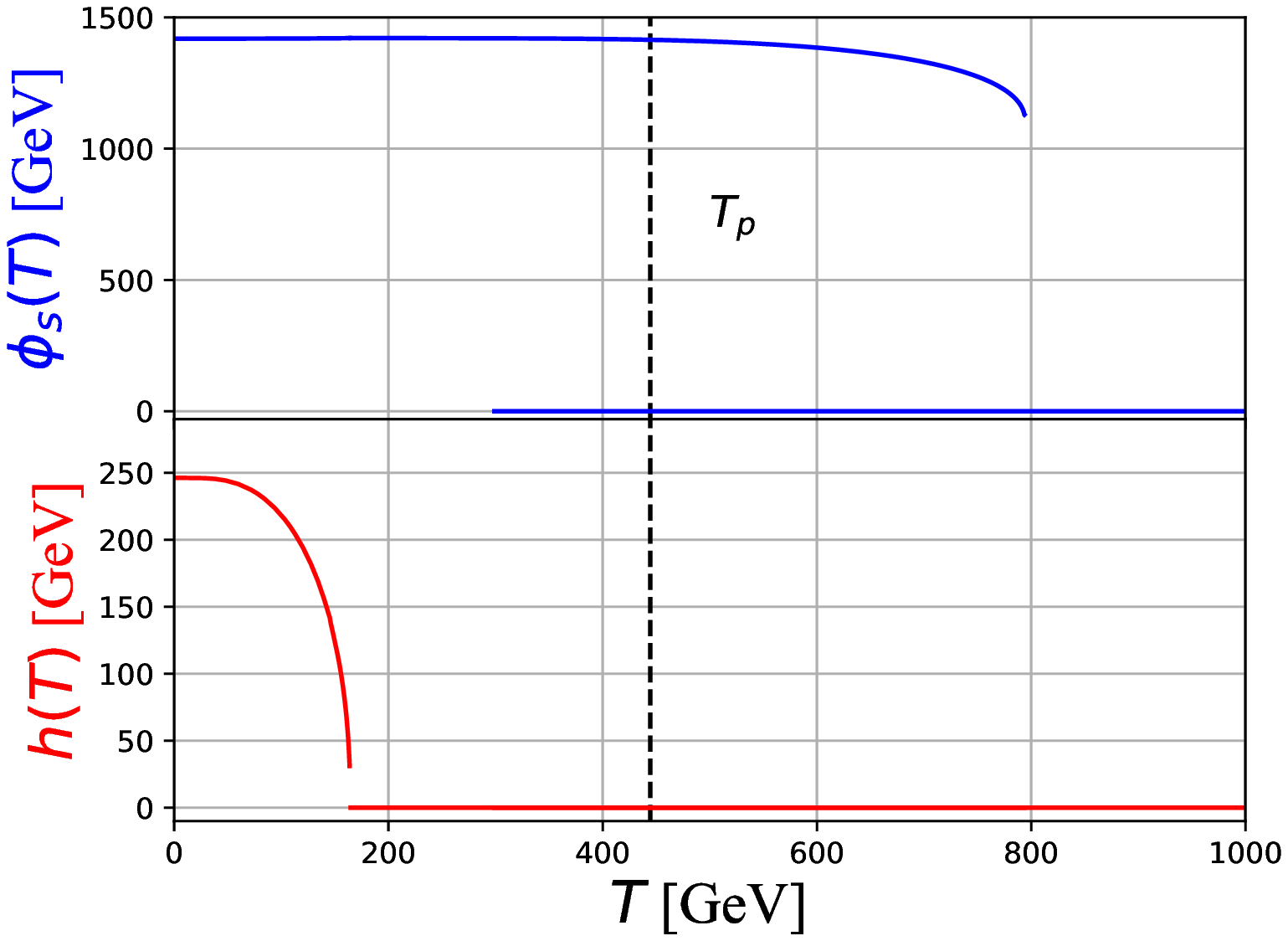}
\caption{Phase evolution figures for BP\underline{~}S1\underline{~}1 (top left), BP\underline{~}S1\underline{~}2 (top right), BP\underline{~}S1\underline{~}2 (bottom left), and BP\underline{~}S1\underline{~}2 (bottom right). Each of the phase evolution figure is marked with the percolation temperature $T_p$. The BP\underline{~}S1\underline{~}1 and BP\underline{~}S1\underline{~}2 (top right) undergo two-step phase transitions, so $T_{p 1}$, $T_{p 2}$ are remarked.} \label{PhaseEvolution}
\end{figure}

However, the cosmic strings induced gravitational wave spectrum shift as $v_w$ moves. In this paper, we adopt three sub-benchmark points  BP\underline{~}SX\underline{~}Y\underline{~}(1,2,3) for each of the BP\underline{~}SX\underline{~}Y,corresponding to the $y_{\chi}=0.1 g_{\chi}$, $y_{\chi}=1 g_{\chi}$, and $y_{\chi}=10 g_{\chi}$ conditions respectively. The exact values assigned to them are selected so that the dark matter relic density $\Omega_{\rm DM}\approx 0.12$\cite{Planck:2018vyg}, and are displayed in the second table of Tab.~\ref{BPs}.  (We will explain later why BP\underline{~}S2\underline{~}1\underline{~}2 and BP\underline{~}S2\underline{~}2\underline{~}3 are missing, and why the $y_{\chi}$ is replaced with $3 g_{\chi}$ rather than $10 g_{\chi}$ in BP\underline{~}S2\underline{~}1\underline{~}2.) In the scenario I, this does not affect the predicted gravitational wave spectrum due to the ignorable cosmic string contributions, however in Scenario II, this will alter the values of $v_w$ to change the $G \mu$ in (\ref{Gmu}). Therefore, we show the predicted gravitational wave spectrum for each of these sub-benchmark points as well as the sensitivities of the proposed gravitational wave experiments in Fig.~\ref{GWSpectrum}.

With the percolation temperature $T_p$ obtained before, we can calculate the evolution of the number density of the dark matter particles.  We adopt the approximation that the phase changes all in the sudden before and after $t_p$.  In scenario I, $\delta m = y_{\chi} v_w$ is tiny compared with $m_{\chi}$, so $m_{\chi}$ is a very good approximation of the dark matter mass. However, in scenario II, evident $\delta m = y_{\chi} v_w$ could even cause $m_{\chi_1}=m_{\chi}-\delta m<0$.  This contradicts with the common acknowledgement that masses are always positive. However, for the fermions, the standard manipulation is to rotate the phase of $\chi_1$ in Eq.~(\ref{ChiMassDiagonalized}) to eliminate this minus sign in the mass terms, with the price of the accumulated evaluation complexity by casting these additional phases to the coupling constants listed in Tab.~\ref{CouplingsTable}.  in this paper, we adopt a simpler but equivalent method to keep all the minus signs of the fermionic mass parameters intact in all of our evaluations.  It is easy to prove that this does not affect the final results, since in the phase space integrations, $m_{\chi_1}$ always appears in its squared values $m_{\chi_1}^2$ to eliminate the minus sign, and in the squared amplitudes, the minus sign in $m_{\chi_1}$ is equivalent with collecting up all the additional phase factors that makes $m_{\chi_1}$ positive.

In this paper, due to our limited computational resources,  we start our calculation when $x=0.05$. This is sufficient because the relic density evolution is not so sensitive to the starting point of $x$ if it is sufficiently small, since the critical freeze-in temperature usually ranges within $0.1 \lesssim x \lesssim 3$. Typically when $x \gtrsim 5$, the freeze-in processes gradually cease, and this is also confirmed by our practical calculations.  Therefore, we require to iterate (\ref{Boltzmann}) until at least when $x>5$.  However, as we have stated, we find it difficult to manipulate the intricate mixings between the various Goldstone bosons after the electroweak phase transition, so we terminate our calculation before $T=200$ GeV to elude the electroweak phase transition. The consistence of these two conditions removes the BP\underline{~}S2\underline{~}2\underline{~}3, in which we estimated that its corresponding $m_{\chi_1} \approx -400$ GeV, and also the BP\underline{~}S2\underline{~}1\underline{~}2, with its corresponding $m_{\chi_1} \approx -540$ GeV.

\begin{figure}
\includegraphics[width=0.48\textwidth]{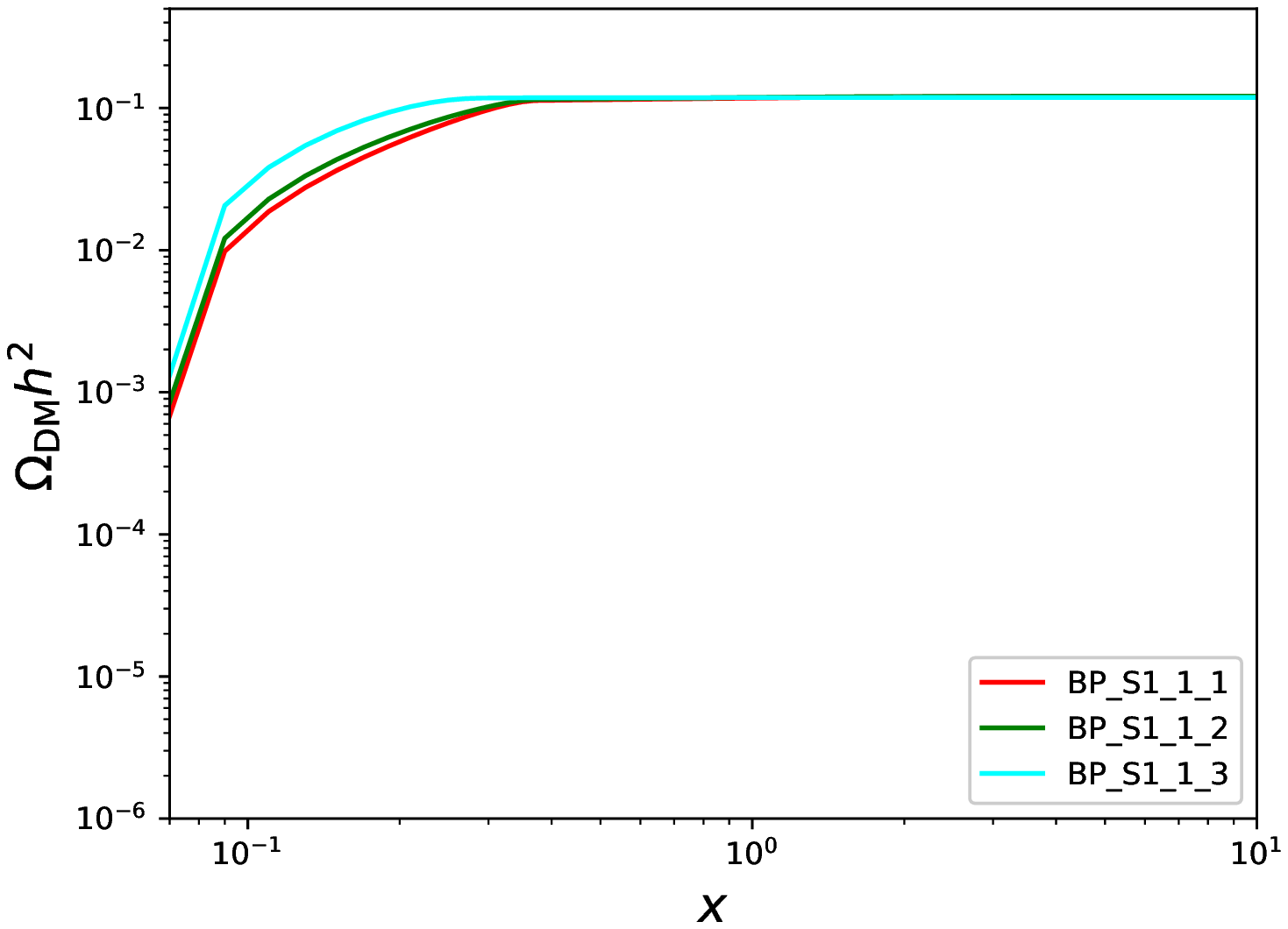}
\includegraphics[width=0.48\textwidth]{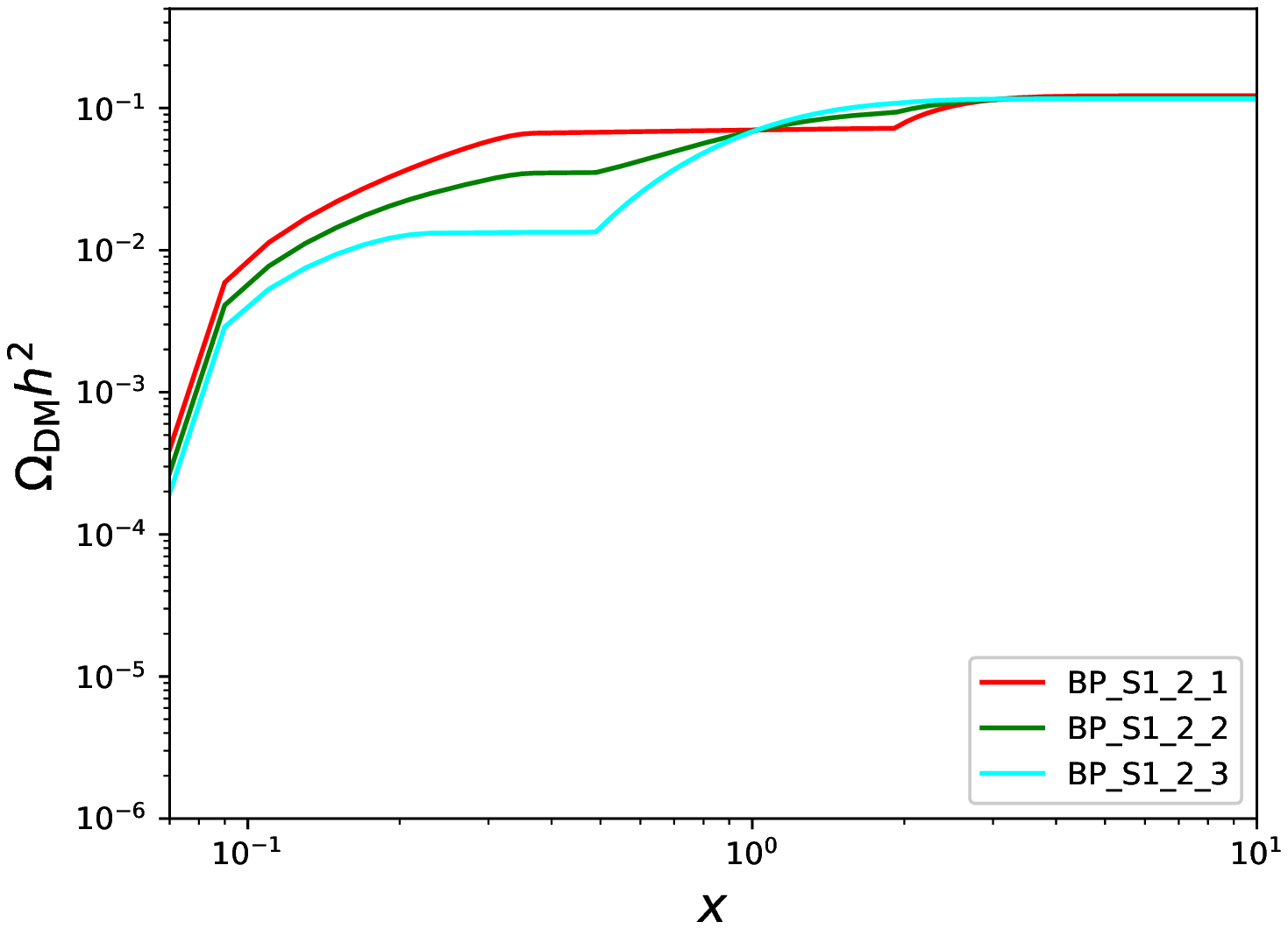}
\caption{The evolution of the dark matter relic in Scenario I.} \label{OmegaS1}
\end{figure}
\begin{figure}
\includegraphics[width=0.48\textwidth]{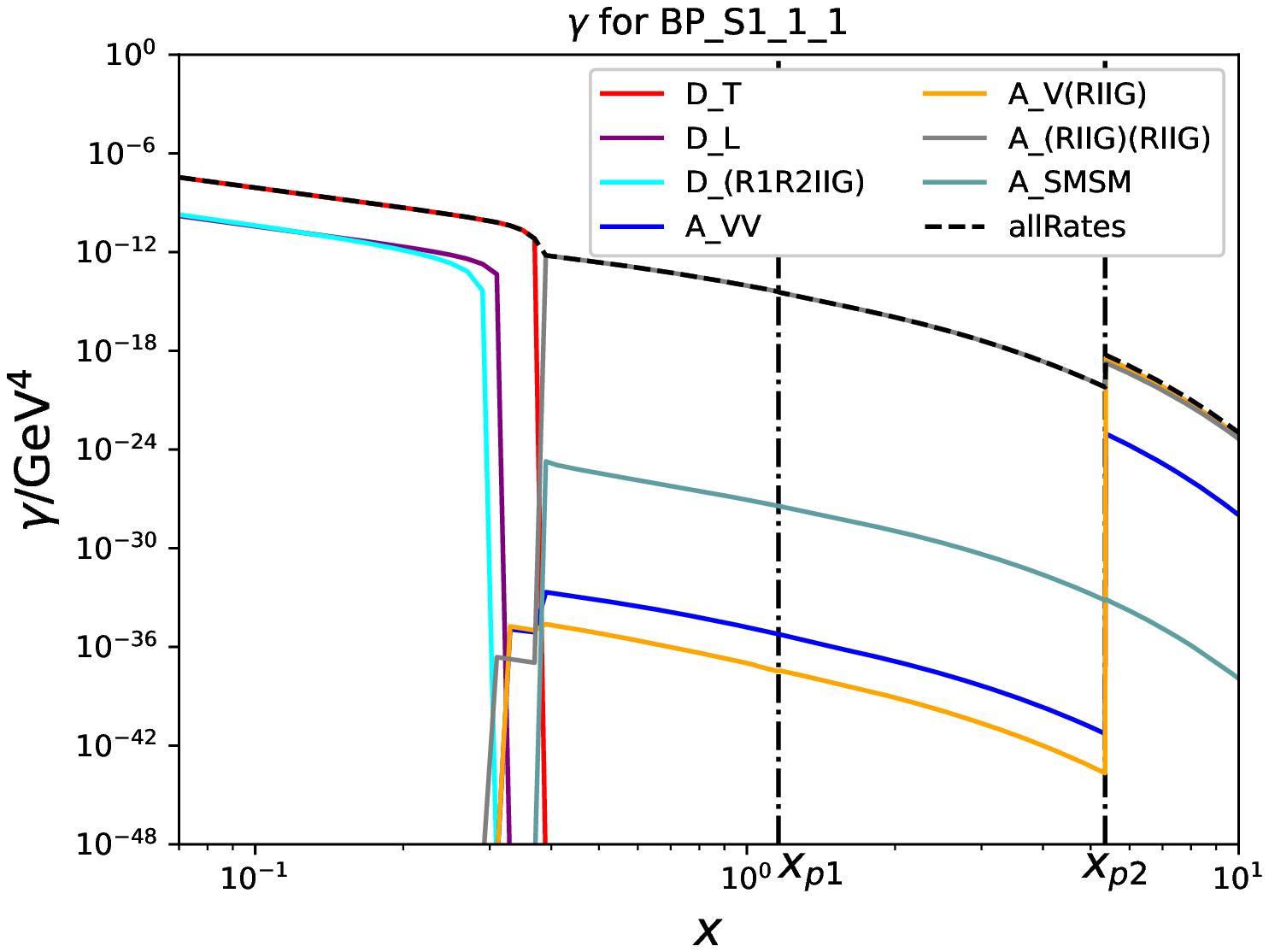}
\includegraphics[width=0.48\textwidth]{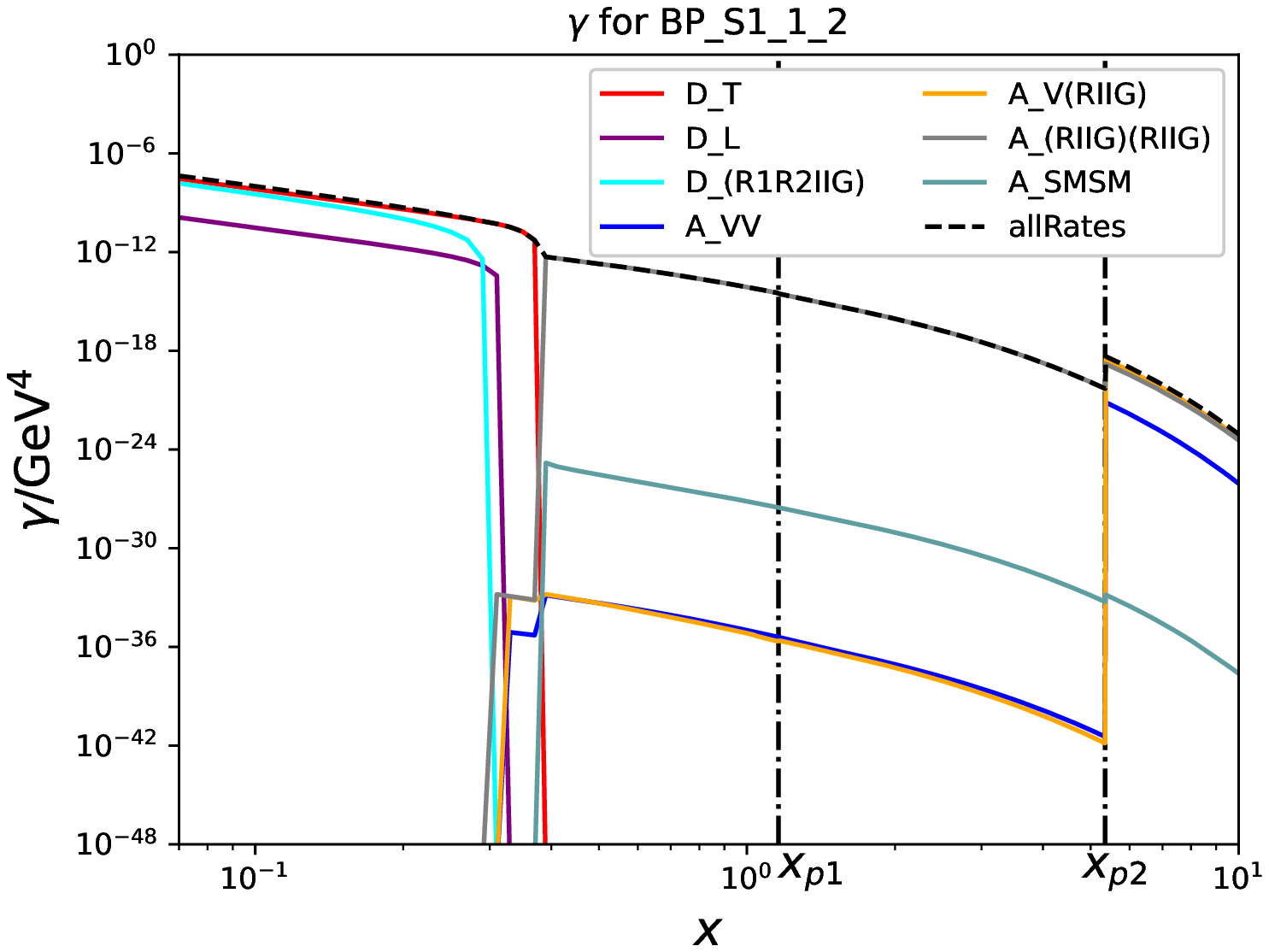}
\includegraphics[width=0.48\textwidth]{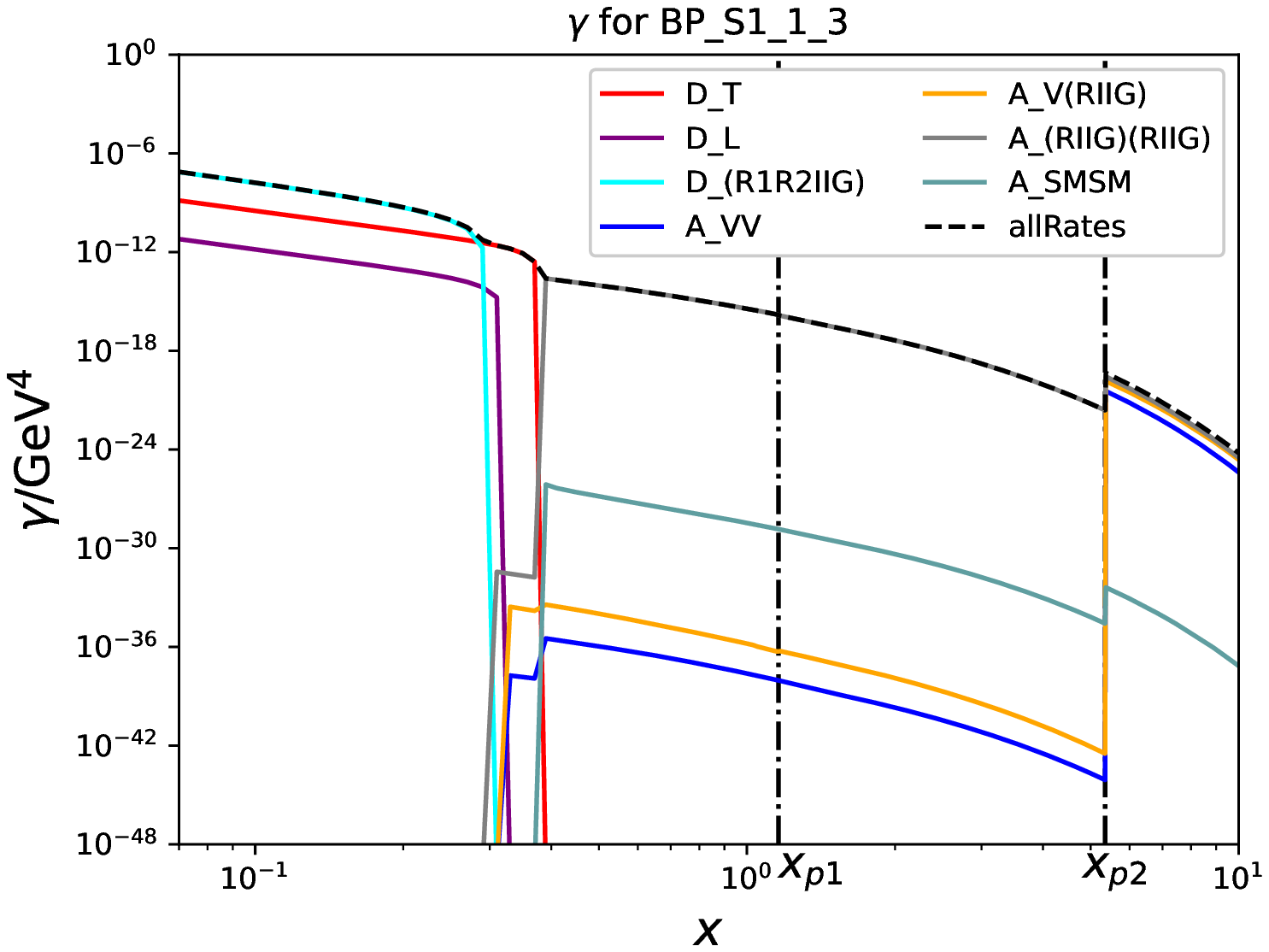}
\includegraphics[width=0.48\textwidth]{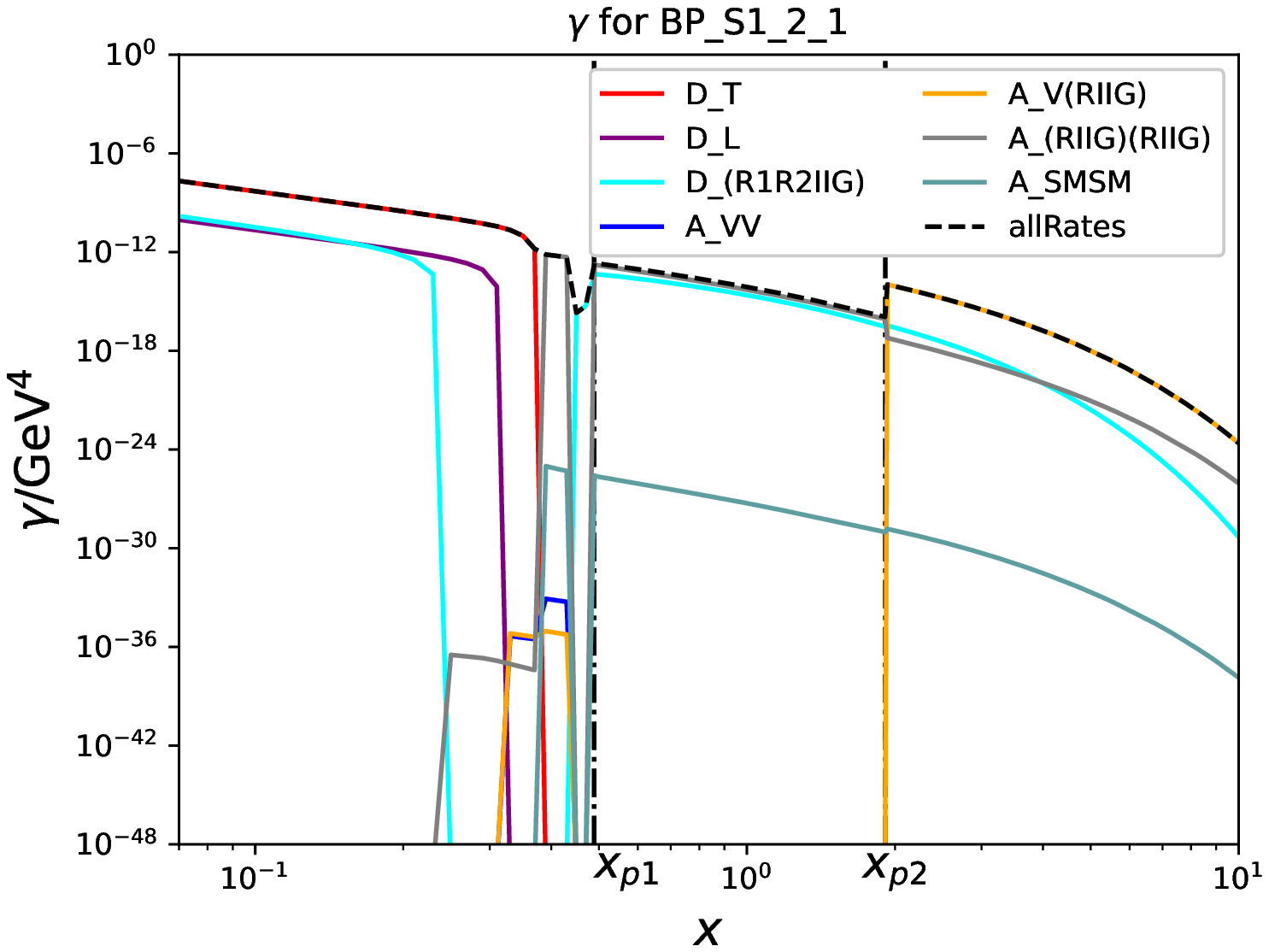}
\includegraphics[width=0.48\textwidth]{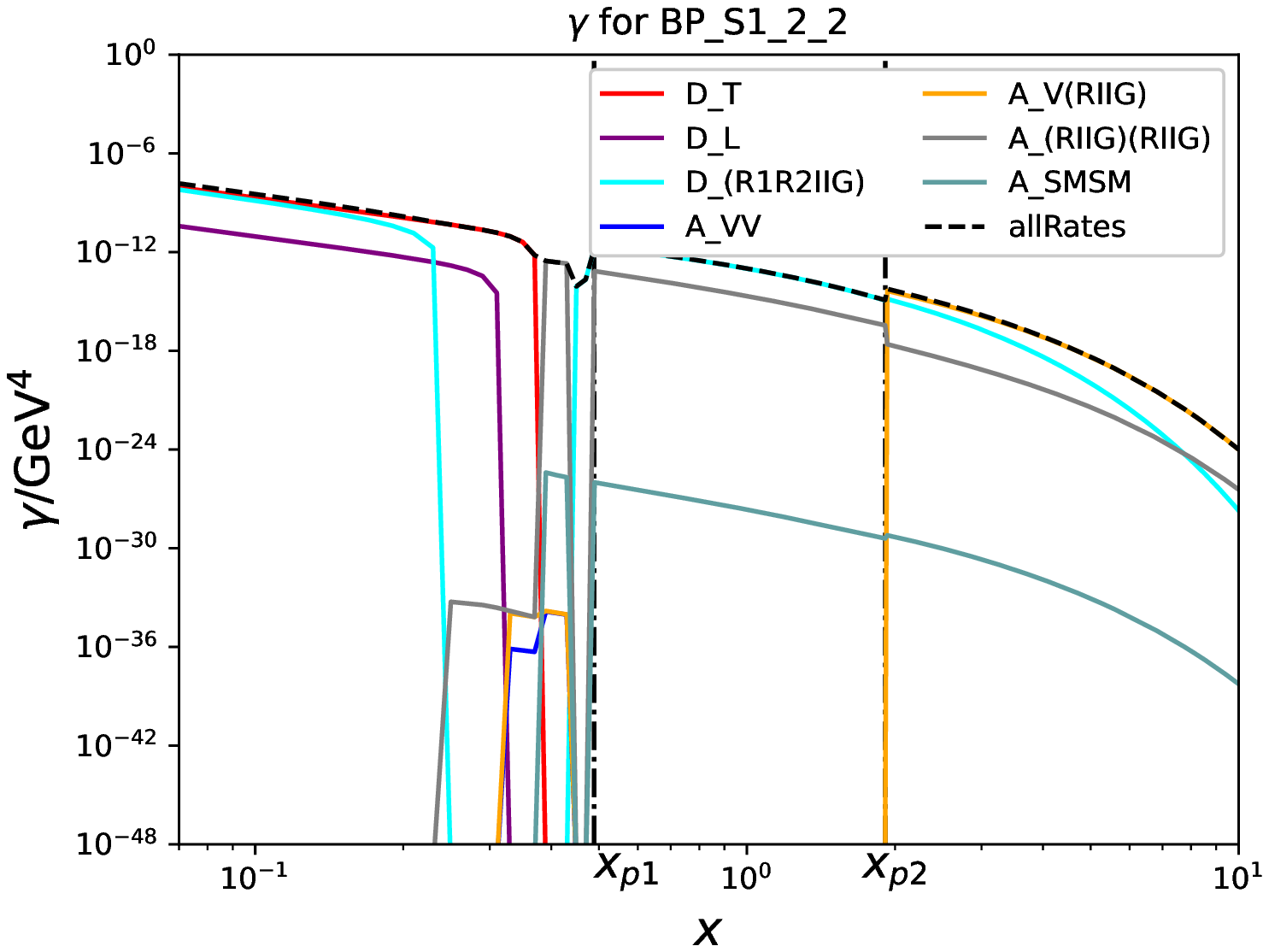}
\includegraphics[width=0.48\textwidth]{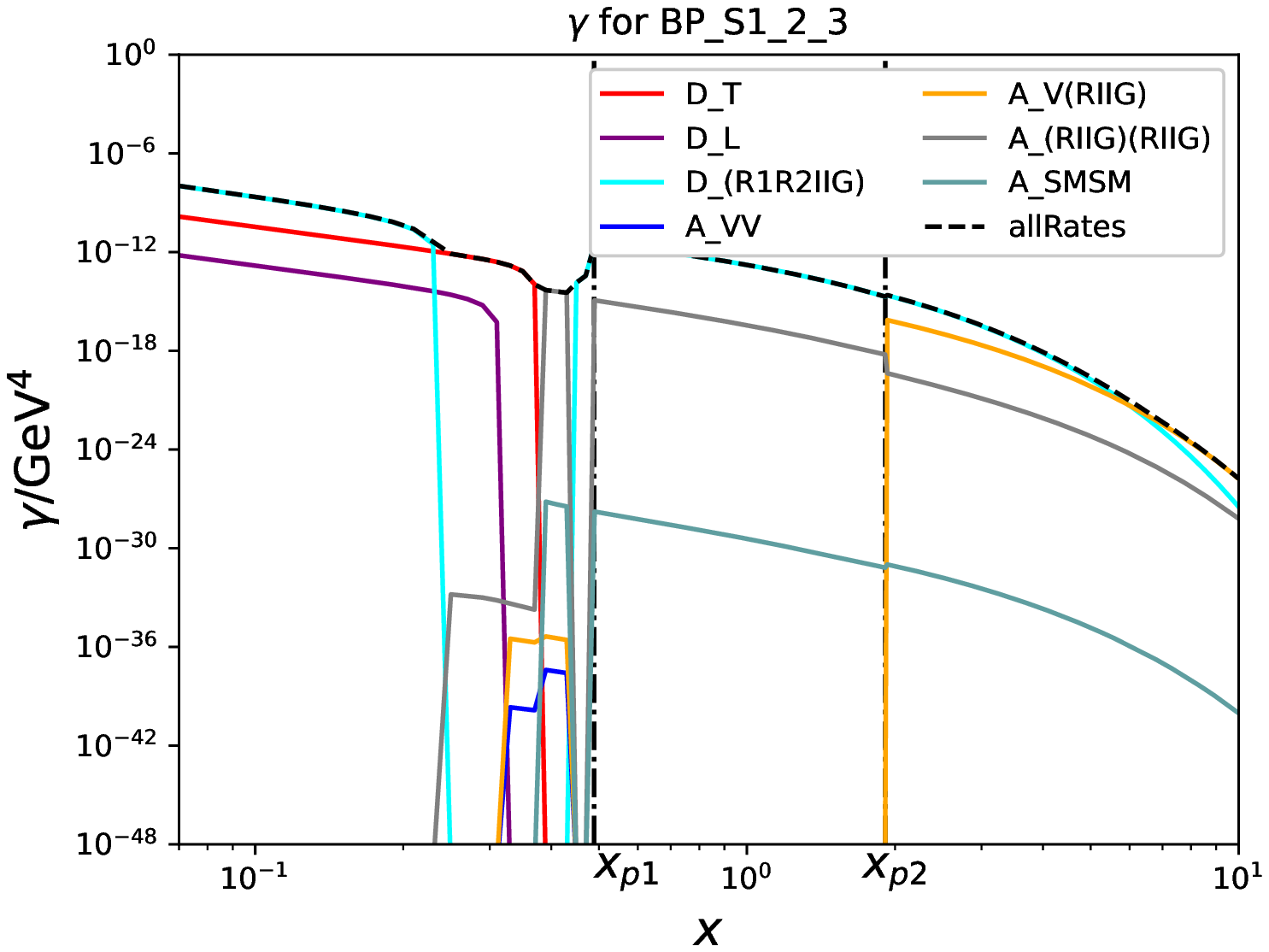}
\caption{Generation rates $\gamma$ as time evolves in Scenario I. Different channels and the total values are plotted.} \label{GammaS1}
\end{figure}

\begin{figure}
\includegraphics[width=0.48\textwidth]{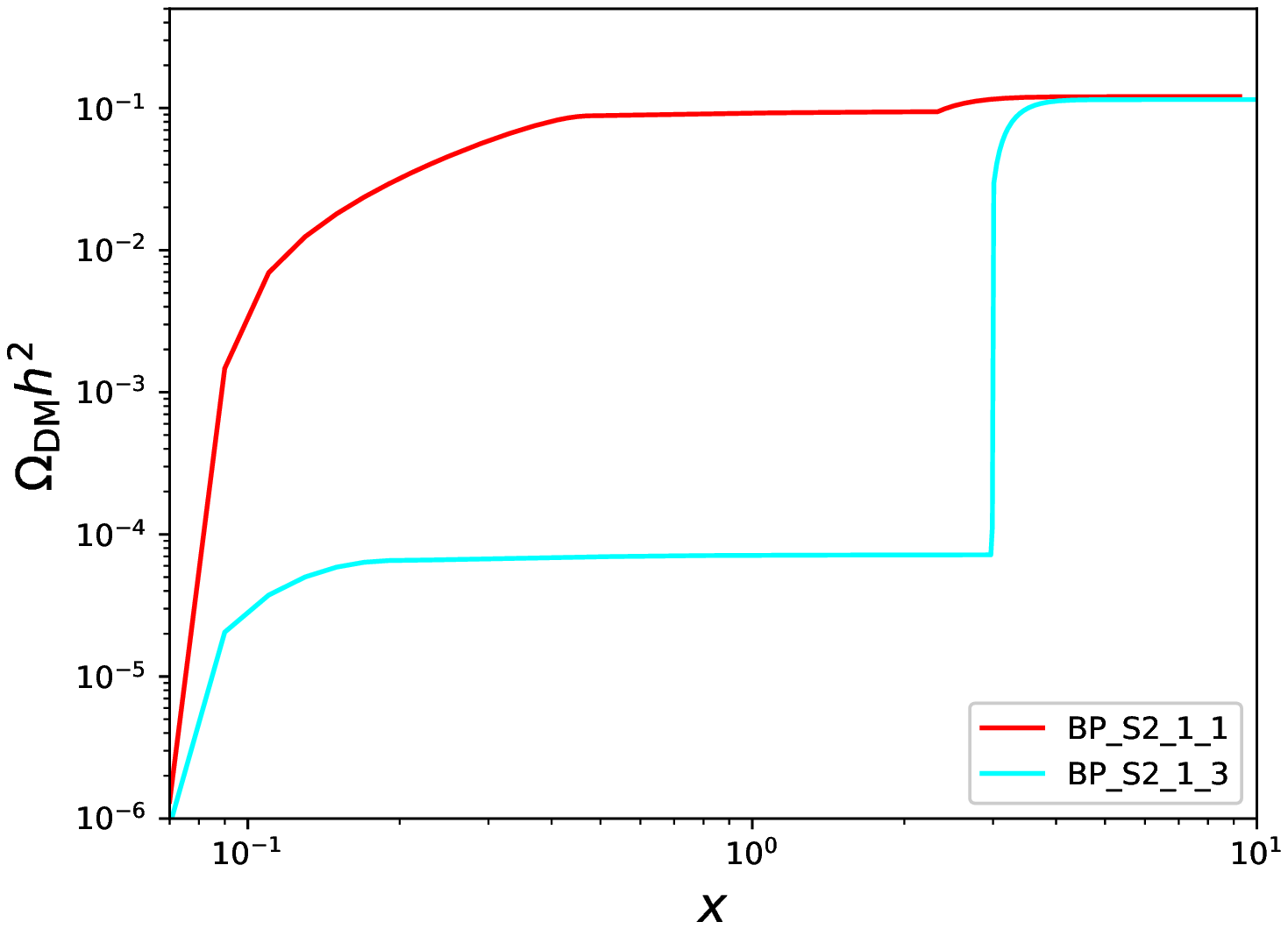}
\includegraphics[width=0.48\textwidth]{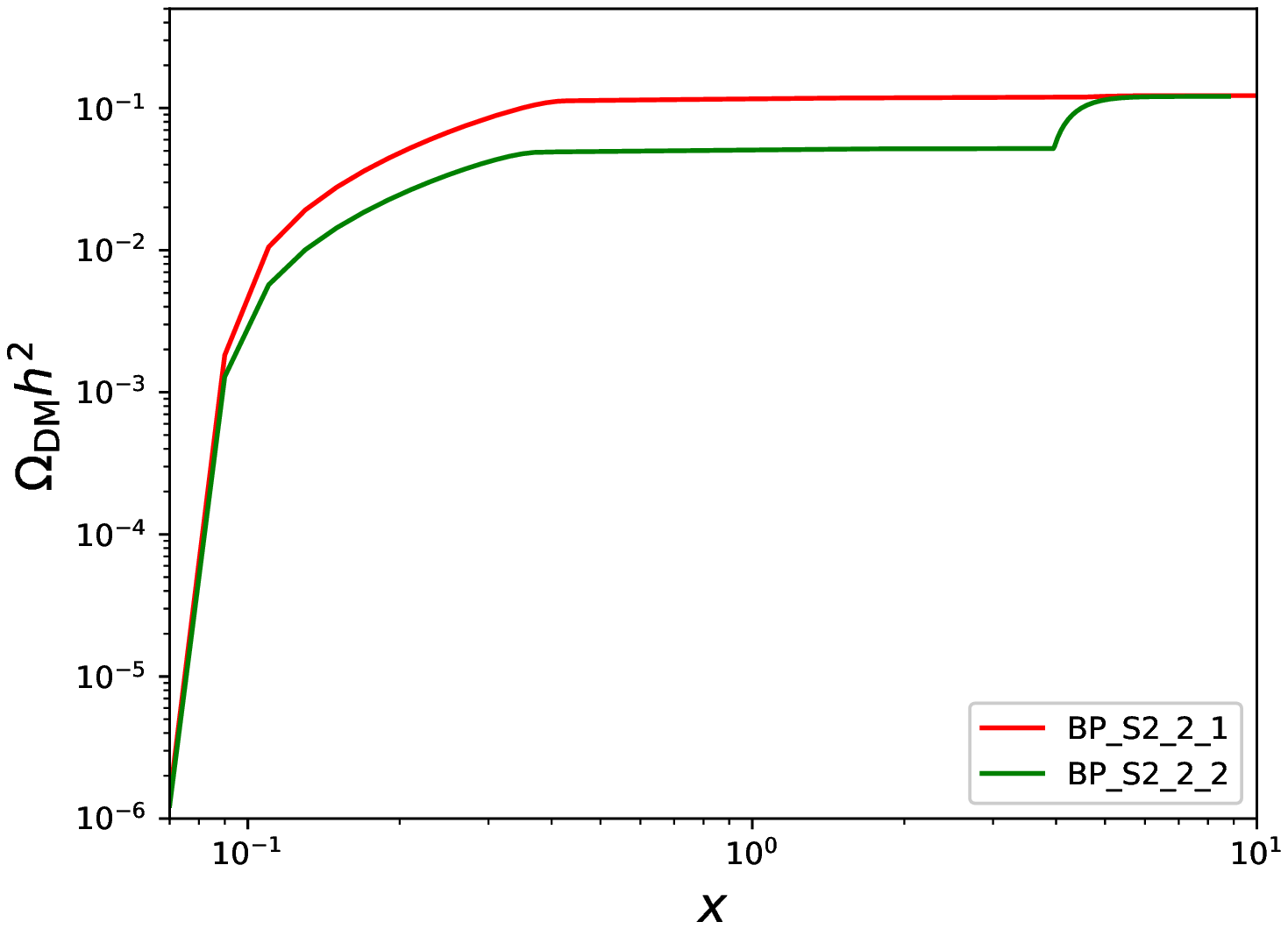}
\caption{The evolution of the dark matter relic in Scenario II.} \label{OmegaS2}
\end{figure}
\begin{figure}
\includegraphics[width=0.48\textwidth]{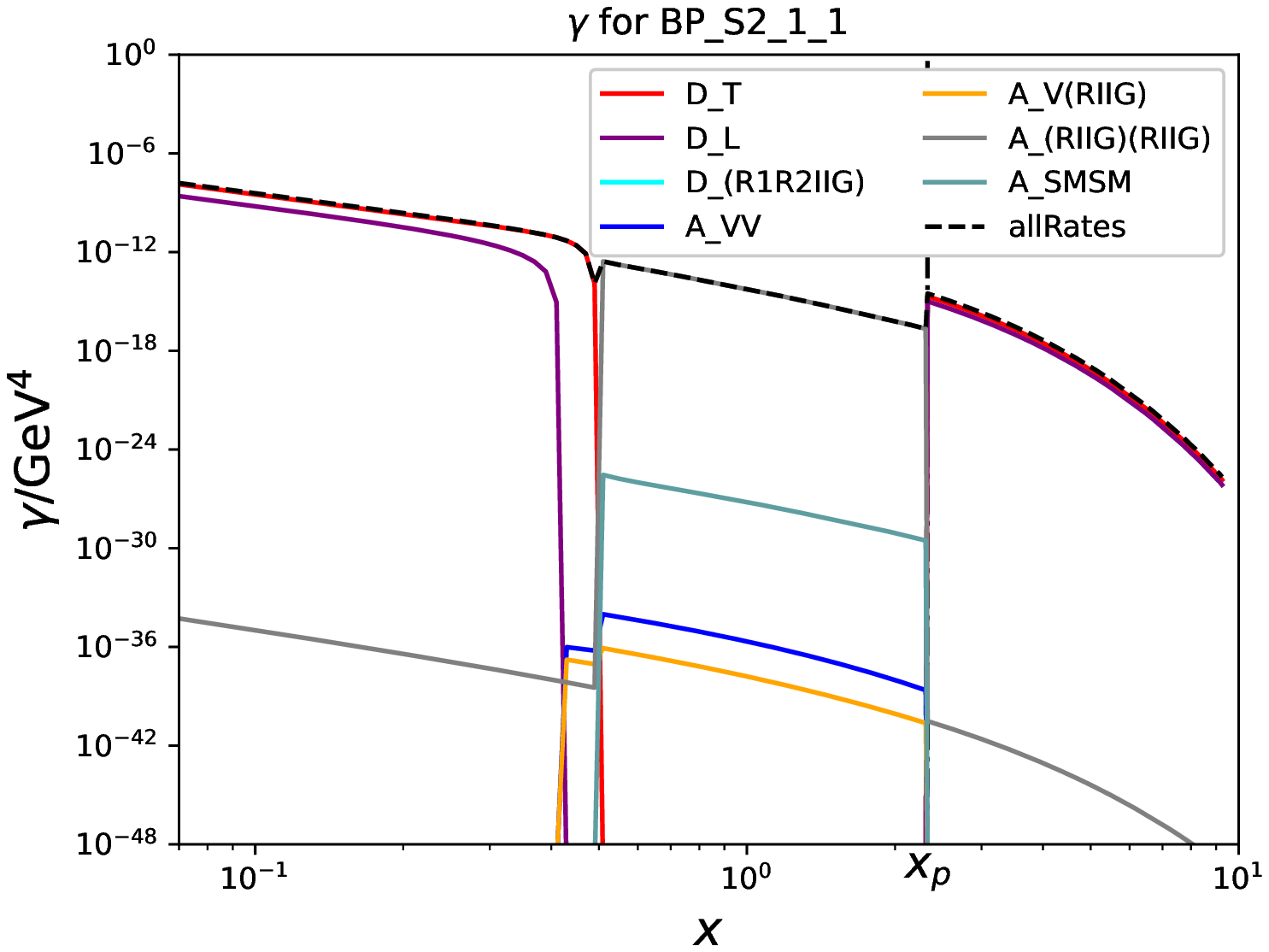}
\includegraphics[width=0.48\textwidth]{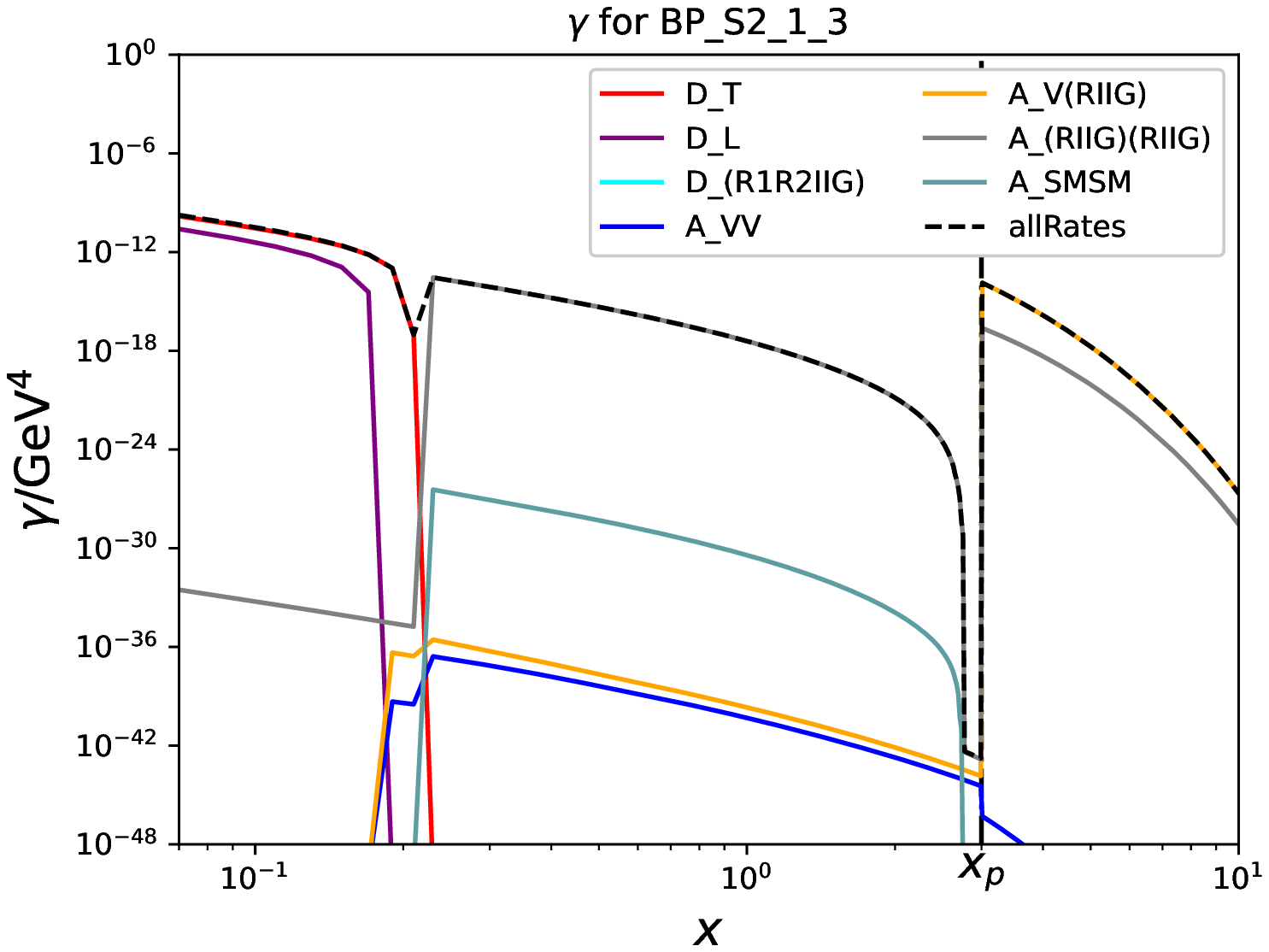}
\includegraphics[width=0.48\textwidth]{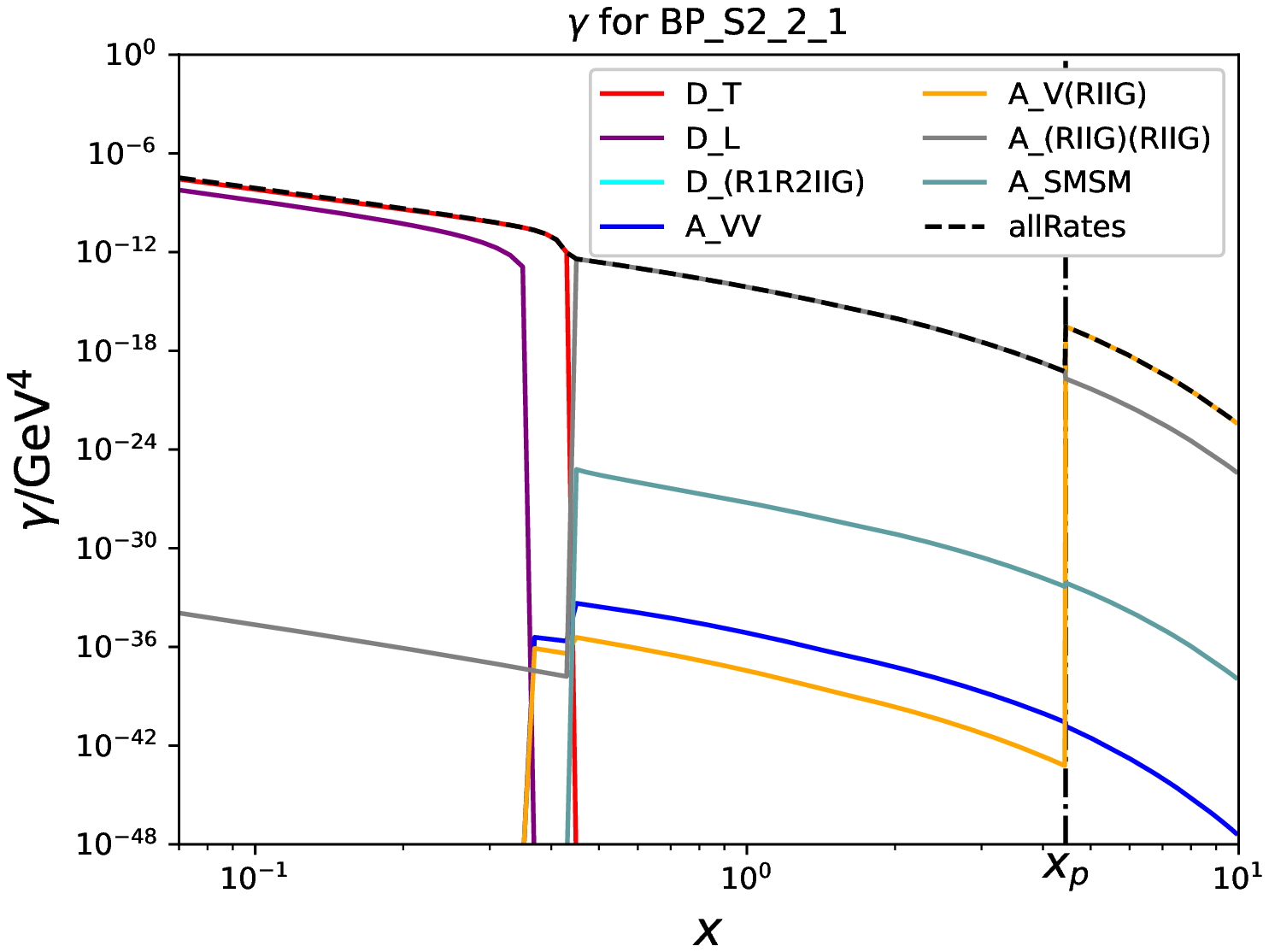}
\includegraphics[width=0.48\textwidth]{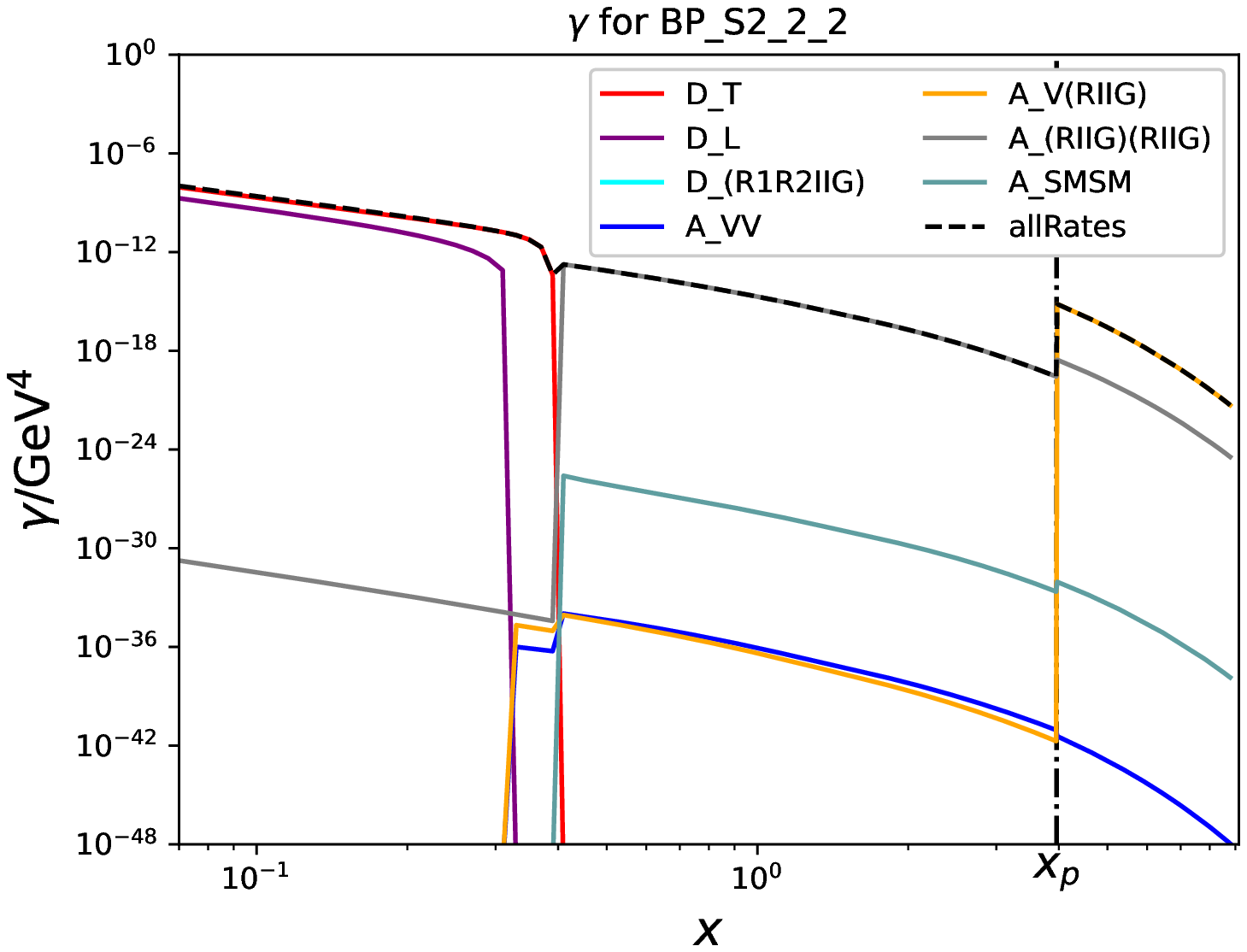}
\caption{Generation rates $\gamma$ as time evolves in Scenario II. Different channels and the total values are plotted.} \label{GammaS2}
\end{figure}

In Fig.~\ref{OmegaS1} and Fig.~\ref{OmegaS2}, we plot the evolutions of the dark matter relic abundance for each of the sub benchmark points in Scenario I and Scenario II respectively.  Fig.~\ref{GammaS1} and Fig.~\ref{GammaS2} are the corresponding evolution of the dark matter generation rates contributing to the right-handed side of (\ref{Boltzmann}). We list the meanings of the channel abbreviations of the $\gamma$ in Tab.~\ref{Meanings}.

\begin{table}
\begin{tabular}{|c|c|}
\hline
Channel Abbreviations & Meaning \\
\hline
D\underline{~}T & Dark matter decayed from the transverse vector boson.\\
\hline
D\underline{~}L & Dark matter decayed from the longitudinal vector boson.  \\
\hline
D\underline{~}(R1R2IIG) & Dark matter decayed from the scalar bosons.  \\
\hline
A\underline{~}VV & Dark matter annihilated from two vector bosons.  \\
\hline
A\underline{~}V(RIIG) & Dark matter annihilated from one vector boson and one scalar boson. \\
\hline
A\underline{~}(RIIG)(RIIG) & Dark matter annihilated from two scalar bosons.  \\
\hline
A\underline{~}SMSM & Dark matter annihilated from two SM particles. \\
\hline
allRates & Summation over all contributions. \\
\hline
\end{tabular}
\caption{Meanings of the abbreviations of the different channels to generate the dark matter.} \label{Meanings}
\end{table}

In Fig.~\ref{GammaS1} and Fig.~\ref{GammaS2} one can clearly see the alternation of the decay and annihilation channels that dominate the total $\gamma$ as the temperature evolves. This is due to the complicated threshold effects as the masses of the vector and scalar boson evolves. During the first-order phase transition, the production rate might jump discontinuously. We marked each of the $x_{p(1,2)}$ corresponding to the percolation temperatures $T_{p(1,2)}$ in all panels of Fig.~\ref{GammaS1}, \ref{GammaS2} for comparing with the phase evolution displayed in Fig.~\ref{PhaseEvolution}. Practically, when we look into the details of the annihilation channels, we found that usually $\chi_1 \chi_2 \rightarrow A^{\prime *} \rightarrow \phi_1 \phi_{A^{\prime} \eta}$ or $\chi_1 \chi_2 \rightarrow A^{\prime *} \rightarrow \phi_1 A^{\prime}_L$ dominate the annihilation channels, because of the larger $A^{\prime}$-$\phi_1$-$\phi_{A^{\prime}}$ coupling constants.

\section{Summary and Future Prospect}
\label{sec:sum}

In this paper, we construct a model where the fermionic dark matter particles feebly interact with the visible sector through the exotic vector boson. In both the dark matter production scenarios under study, it is difficult for the near-future experiments such as LISA, TianQin and Taiji to detect the gravitational wave signal produced by the TeV-scale phase transition. Meanwhile, in the scenario with the spontaneously breakdown of the dark U(1) symmetry by an extremely high scale $v_w$, a significant gravitational wave signal is produced through the cosmic strings decay. This is well within the ability of these experimental proposals.

We enumerated and calculated all the possible $1 \leftrightarrow 2$ and $2 \leftrightarrow 2$ channels to produce the dark matter. The vector boson dispersion relations receive significant and complicated thermal corrections. In place of the rough estimation to treat them as something with a fixed mass for all momentum values, we applied sleeker methodology to separate the transverse, longitudinal and the partly revived Goldstone degree of freedom with different on-shell dispersion relations. The threshold effects induced by the evolution of thermal corrections as temperature drops also significantly alter the dark matter production rates. We calculate these influences and show the results for some benchmark points.

In our paper, we consider the relatively heavy TeV-scale dark matter, and avoid the troublesome mixture between the vector and Goldstone boson sectors for easier evaluation. This is valid in all our benchmark point selections, however for lighter FIMP dark matter with the mass $\lesssim 0.5$ TeV,  or the FIMP dark matter interacting directly with the SM SU(2)$_L\times$U(1)$_Y$ gauge bosons, such effect might be inevitable.  We will concern this situation and do further study on it in the future.

\begin{acknowledgements}
We thank to Junmou Chen, Yang Zhang, Chen Zhang, Hong-Hao Zhang, Zhao-Huan Yu for helpful discussions and communications. Ligong Bian is supported by the National Natural Science Foundation of China under the grants Nos.12075041, 12047564, and the Fundamental Research Funds for the Central Universities of China (No. 2021CDJQY-011 and No. 2020CDJQY-Z003), and Chongqing Natural Science Foundation (Grants No.cstc2020jcyj-msxmX0814). Yi-Lei Tang is supported in part by the National Natural Science Foundation of China under Grants No. 12005312. Part of the calculation was performed on TianHe-2, and we thank for the support of National Supercomputing Center in Guangzhou (NSCC-GZ).
\end{acknowledgements}

\appendix
\clearpage
\section{Thermal masses for SM gauge bosons}
\label{sec:thermass}
When evaluating the effective potential, one has to turn to the imaginary time formalism. The transverse and longitudinal degrees of freedom receives different corrections~\cite{Carrington:1991hz},
\begin{eqnarray}
\Pi _{W^{\pm}}^{L} &=&\frac{11}{6}g^{2}T^{2},~\Pi _{W^{\pm}}^{T}=0  \notag \\
\Pi _{W^{3}}^{L} &=&\frac{11}{6}g^{2}T^{2},~\Pi _{W^{3}}^{T}=0   \\
\Pi _{B}^{L} &=&\frac{11}{6}g'^{2}T^{2} \notag
\end{eqnarray}%
where the script $L$ ($T$) denotes the longitudinal (transversal) mode, and $W$, $B$ denote the $SU(2)_L$ triplet and hyper-charge gauge bosons respectively.

Summed with the VEV-induced terms, them together,  the longitudinally polarized $W$ boson's correction is
\begin{eqnarray}
M_{W^{\pm}_L}^2 = {1 \over 4} g^2 h^2 + \frac{11}{6} g^2 T^2.
\end{eqnarray}
To determine the masses of the longitudinal $Z$ and $A$ one should diagonalize the matrix
\begin{eqnarray}
\frac{1}{4}(h^2)
\begin{pmatrix}
g^2& -g g^\prime \\
-g g^\prime & g^{\prime 2}
\end{pmatrix}
+
\begin{pmatrix}
\frac{11}{6} g^2 T^2 & 0 \\
0 & \frac{11}{6} g^{\prime 2} T^2
\end{pmatrix},
\end{eqnarray}
and adopt the eigenvalues to substitute the $m_{i}^2(h, \phi_s, \phi_w)+c_i(T)$ terms in Eq.~(\ref{DaisyTerms}).

\section{Propagators and on-shell behaviors of the vector boson and the SM fermions} \label{ThermalCorrectionGaugeFermion}

In this paper, we adopt the Goldstone Equivalence Gauge to calculate the interaction rates.  In this gauge, the polarization vector is extended from 4-dimension to 5-dimension by adding up a Goldstone degree of freedom. Here we use $M,N\dots$ to denote the 5-dimensional indices. The previous four numbers $M,N,\dots=0,1,2,3$ indicate the usual Lorentz time and space indices, and the last $M,N,\dots=4$ represents the Goldstone degree of freedom. The full propagator can also be extended into a $5 \times 5$ matrix.  For a particular momentum $k=(k^0, \vec{k})$,  let us define the corresponding $5 \times 5$ project matrix $P_T$, $P_L$, 
\begin{eqnarray} 
& & P_T^{00}=P_T^{0 i}=P_T^{i 0}=P_T^{40}=P_T^{04}=P_T^{4i}=P_T^{i4}=0,  \nonumber \\
& & P_T^{ij}= \delta_{i j}-\frac{k^i k^j}{|\vec{k}|^2}, ~~(i,j=1,2,3)\nonumber \\
& & P_L=\left( \begin{array}{cc}
\frac{k^2}{(n \cdot k)^2}n^{\mu} n^{\nu} & i \frac{m_{A^{\prime}}}{n \cdot k} n^{\mu} \\
-i \frac{m_{A^{\prime}}}{n \cdot k} n^{\nu} & \frac{m_{A^{\prime}}^2}{k^2 + i \epsilon}
\end{array} \right),
\end{eqnarray}
where
\begin{eqnarray}
n^{\mu} = \frac{\sqrt{k^2} \epsilon_{LU}^{\mu}(k)-k^{\mu}}{|\vec{k}|-k^0},
\end{eqnarray}
and $\epsilon_{LU}(k)=(|\vec{k}|, k_0 \frac{\vec{k}}{|\vec{k}|})/\sqrt{k^2}$ is the longitudinal polarization vector in the $R_{\xi}$ gauge. In fact, $P_T$ and $P_L$ are the transverse and longitudinal polarization project matrix, and satisfy
\begin{eqnarray}
P_T^{MN} &=& \sum_{s=\pm} \epsilon^{M*}_s \epsilon^N_s, \nonumber \\
P_L^{MN} &=& \epsilon^{M*}_L \epsilon^N_L, 
\end{eqnarray}
where $\epsilon^M_{\pm}$ and $\epsilon^M_L$ are the transverse and longitudinal polarization vectors respectively.  Here
\begin{eqnarray}
\epsilon^{M}_L(k) = \left(\begin{array}{c}
-\frac{\sqrt{k^2}}{n \cdot k} n^{\mu} \\
i \frac{m_{A^{\prime}}}{\sqrt{k^2}},
\end{array} \right) \label{epsilonL}
\end{eqnarray}
and in the special frameset of $k=(k^0, 0, 0, k^3)$ along the z-axis, $\epsilon_{\pm}^M(k)=\frac{1}{\sqrt{2}}(0, 1, \pm i, 0, 0)$.

In the zero temperature, the Goldstone equivalence gauge propagator of the $A^{\prime}$ is given by
\begin{eqnarray}
D_F^{MN}(k)=\frac{i}{k^2-m_{A^{\prime}}^2+i \epsilon}\left[ P_T+P_L+\frac{k^2-m_{A^{\prime}}^2+i \epsilon}{k^2 + i \epsilon} \left( \begin{array}{cc}
0_{4 \times 4} & 0_{4 \times 1} \\
0_{1 \times 4} & 1
\end{array} \right) \right].
\end{eqnarray}

In the thermal plasma,  adopting the ``$\sigma=\frac{\beta}{2}$-gauge'' and consider the one-loop self-energy diagrams, the full propagator is given by
\begin{eqnarray}
D_{a b}^{\text{full},MN}(k)=U_{ac}(k) \left( \begin{array}{cc}
D_0^{\text{full},MN}(k) & 0 \\
0 & D_0^{\text{full}*,MN}(k)
\end{array} \right)_{cd} U_{db}(k)
\end{eqnarray},
where $a,b,\dots=1,2$, 
\begin{eqnarray}
U(k) = \left( \begin{array}{cc}
\sqrt{1+n_B(k_0)} & \sqrt{n_B(k_0)} \\
\sqrt{n_B(k_0)} & \sqrt{1+n_B(k_0)}
\end{array} \right),
\end{eqnarray}
and
\begin{eqnarray}
D_0^{\text{full},MN}(k)&=&\frac{i}{k^2-m_{A^{\prime}}^2-\Pi_T(k)+i \epsilon} P_T + \frac{i}{k^2-m_{A^{\prime}}^2-\Pi_L(k)+i \epsilon} P_L \nonumber \\
&+& \frac{1}{1-\frac{\Pi_U(k)}{m_A^2}} \frac{i}{k^2+ i \epsilon} \left( \begin{array}{cc}
0_{4 \times 4} & 0_{4 \times 1} \\
0_{1 \times 4} & 1
\end{array} \right). \label{FullPropagatorV}
\end{eqnarray}
$\Pi_{T,L,U}$ are the factors extracted from the one-loop self-energy corrections. In the hard thermal loop (HTL) approximation, they are calculated to be
\begin{eqnarray}
\Pi_L(k) &=& -c_{A^\prime}(T) \frac{k^2}{\vec{k}^2} \left( 1-\frac{k^0}{|\vec{k}|} Q_0 (\frac{k^0}{|\vec{k}|} \right),  \nonumber \\
\Pi_T(k) &=& \frac{1}{2}(2 c_{A^\prime}(T)-\Pi_L(k)),	\label{Pi_TL}
\end{eqnarray}
where
\begin{eqnarray}
Q_0(x) = \frac{1}{2} \ln \frac{x+1}{x-1}.
\end{eqnarray}
Expanded around the minima of the effective potential, $\Pi_U(0)=0$ should be satisfied. Therefore, we can use $\Pi_U(0)$ to estimate $\Pi_U(k)$, so that $\Pi_U(k)=0$.

From (\ref{FullPropagatorV}) one can immediately write down the shifted on-shell equation of both transverse and longitudinal $A^{\prime}$,
\begin{eqnarray}
k^2-m_{A^{\prime}}^2-\Pi_{T,L}(k)=0.
\end{eqnarray}
Each external leg of both transverse and longitudinal vector bosons should be multiplied with the ``renormalization factor'' $\sqrt{Z_{T,L}(k)}$, with their definitions given by
\begin{eqnarray}
Z_{T,L}(k)=\frac{2 k_0}{2 k_0-\frac{\partial \Pi_{T,L}(k)}{\partial k_0}}. \label{ZAPrime}
\end{eqnarray}
A fraction of Goldstone degree of freedom has also been spewed out by the longitudinal polarization vector boson. Calculating $D_0^{\text{full}, 44}(k)$ from (\ref{FullPropagatorV}), one acquires
\begin{eqnarray}
\Delta_{\text{GS}}^F(k)=\frac{k^2-\Pi_L(k)+i \epsilon}{k^2-m_{A^{\prime}}^2-\Pi_L(k)+i \epsilon} \frac{i}{k^2+i \epsilon}. \label{GSPropagator}
\end{eqnarray}
Besides the $k^2-m_{A^{\prime}}^2-\Pi_L(k)$ pole corresponding to the longitudinal polarization vector boson, (\ref{GSPropagator}) includes a branch cut along $k^0 \in (-\vec{k}, \vec{k})$ axis. The imaginary part of $\Delta_{\text{GS}}^F(k)$ peaks near $k^0 =\pm \vec{k}$, so as an approximation we can regard these two peaks as two massless Goldstone fractions. Their ``renormalization factor'' is given by
\begin{eqnarray}
Z_{\text{GS}}=-\frac{2 R(\gamma, \alpha)}{\pi},	\label{ZGS}
\end{eqnarray}
where
\begin{eqnarray}
R(\gamma, \alpha) = \int_0^{|\vec{k}|+\delta} -|\vec{k}| \text{Im}[i \Delta_{\text{GS}}^F(k^0, \vec{k})] dk^0,
\end{eqnarray}
and $\gamma = \frac{c_{A^{\prime}}(T)}{\vec{k}^2}$, $\alpha= \frac{m_{A^{\prime}}^2}{\vec{k}^2}$.

For the leptons and quarks, we only consider the processes above the electroweak phase transition critical temperature, so the thermal corrected dispersion relation of a fermion is given by
\begin{eqnarray}
\Delta_{\pm f}^{-1} (k) = \left( k^0 \mp |\vec{k}_1| - \frac{m_f^2}{2 | \vec{k}_1|} \left[ \left( 1 \mp \frac{k^0}{|\vec{k}|} \right) \ln \frac{k^0+|\vec{k}|}{k^0-|\vec{k}|} \pm 2  \right] \right)=0, \label{FDispersion}
\end{eqnarray}
where $f$ indicates any left- or right-handed leptons and quarks, so
\begin{eqnarray}
m_{e_L, \nu_{eL}, \mu_L, \nu_{\mu L}, \tau_L, \nu_{\tau L}}&=&\frac{g_1^2 + 3 g_2^2}{32} T^2, \nonumber \\
m_{e_R, \mu_R, \tau_R} &=& \frac{g_1^2 T^2}{8}, \nonumber \\
m_{u_L, c_L, t_L, d_L, s_L, b_L} &=& \frac{(g_1/3)^2 + 3 g_2^2}{32} T^2 + \frac{g_s^2 T^2}{6}, \nonumber \\
m_{u_R, c_R, t_R}&=& \frac{(2 g_1/3)^2 T^2}{8} + \frac{g_s^2 T^2}{6},  \nonumber \\
m_{d_R, s_R, b_R}&=& \frac{( g_1/3)^2 T^2}{8} + \frac{g_s^2 T^2}{6}. \label{LeptonThermalMasses}
\end{eqnarray}

There are four solutions of the equations in (\ref{FDispersion}), indicating one particle, one hole, and one anti-particle as well as one anti-hole. The ``renormalization factors'' become
\begin{eqnarray}
Z_f(k) = \frac{(k^0)^2-\vec{k}^2}{2 m_f^2}.
\end{eqnarray}

\section{From self-energy diagrams to $1 \leftrightarrow 2$, $2 \leftrightarrow 2$ interaction rates}\label{intrate}

In the literature, the evaluation of the changing rate of some particle number density in the finite temperature environment relies on the imaginary part of the self-energy diagrams. In this paper, for the sake of intuitiveness we depend on the tree-level diagrams, just as everybody learned from the quantum field theory textbooks in the zero-temperature case.This appendix section aims at illustrating the equivalence between these two methods.

It is now convenient to rely on the $\sigma=0$ gauge to calculate the imaginary part of the self-energy diagrams. Following the usual literature, let us denote $1$ and $2$ as the two types of vertices. The production rate of one of the dark matter particle, e.g., $\chi_1$ is extracted from $\Pi^<(k)$. This is calculated from the self-energy diagrams in which the leftmost vertex is in type $2$ and the rightmost vertex is in type $1$. Since both the $\chi_1$ and $\chi_2$ are far from thermal equilibrium with the plasma, it is convenient to appoint $T=0$ in all the $\chi_{1,2}$ propagators only, while keeping all the other temperature terms normal in other particle propagators. Therefore,
\begin{eqnarray}
D_{\chi_i 11}^F(k)&=&D_{\chi_i 22}^{F*}(k) = \frac{i(k\!\!\!/+m_{\chi_i})}{k^2-m_{\chi_i}^2+i \epsilon}, \nonumber \\
D_{\chi_i 12}^F(k)&=&\theta(-k_0) 2 \pi \delta(k^2-m^2), \nonumber \\
D_{\chi_i 21}^F(k)&=&\theta(k_0) 2 \pi \delta(k^2-m^2).
\end{eqnarray}

\begin{figure}
\includegraphics[width=0.4\textwidth]{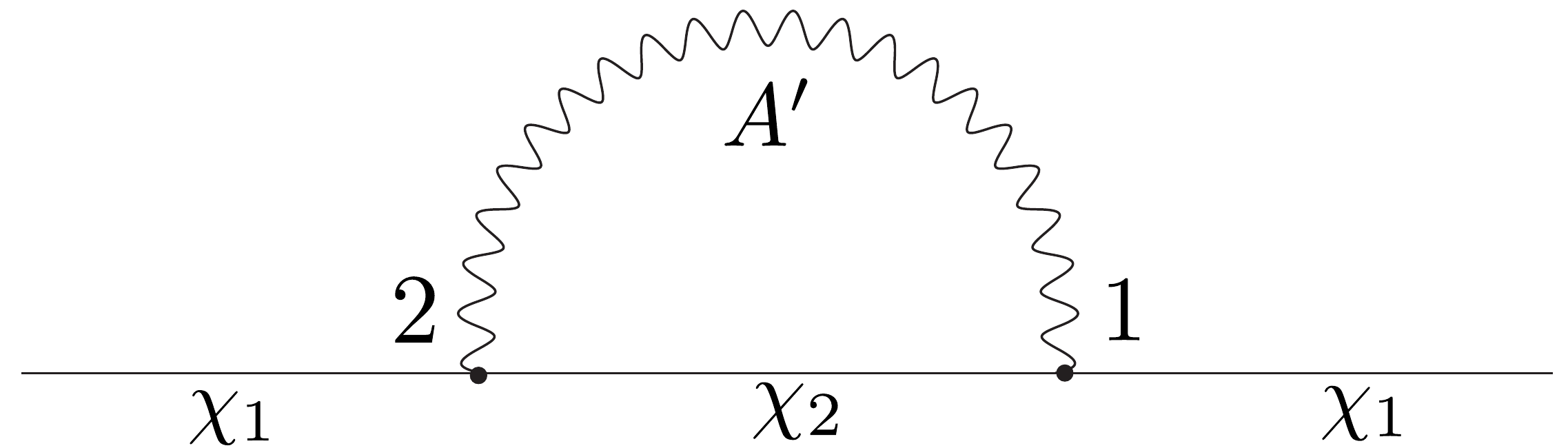}
\caption{One-loop self-energy diagram inducing the $1 \leftrightarrow 2$ contributions to the production rates.}
\label{OneLoop}
\end{figure}
\begin{figure}
\includegraphics[width=0.4\textwidth]{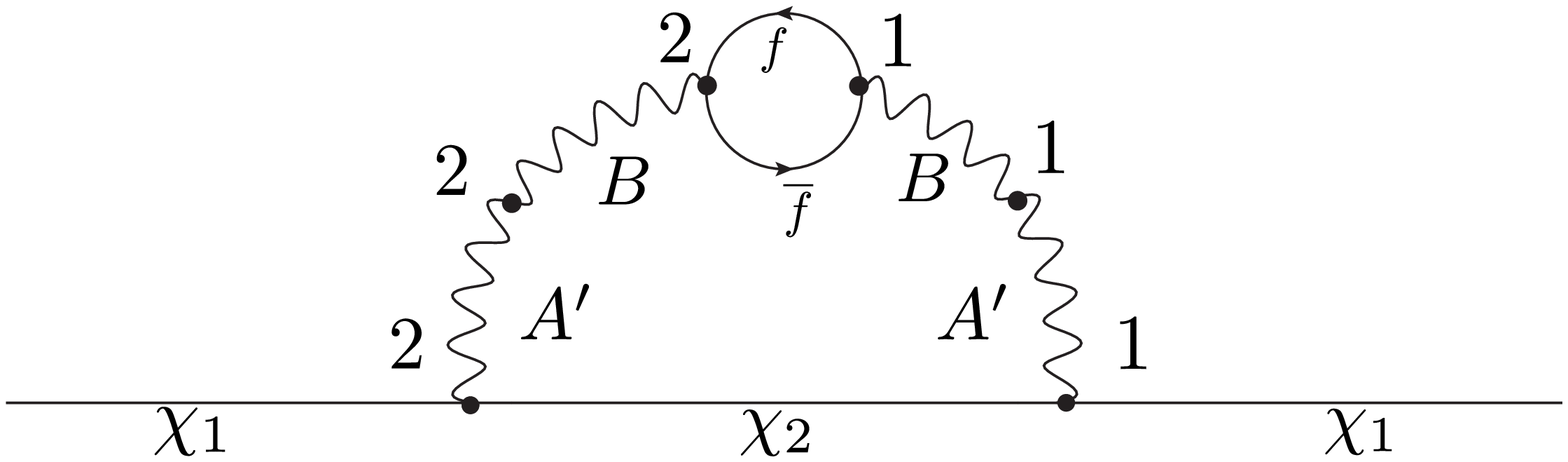}
\includegraphics[width=0.4\textwidth]{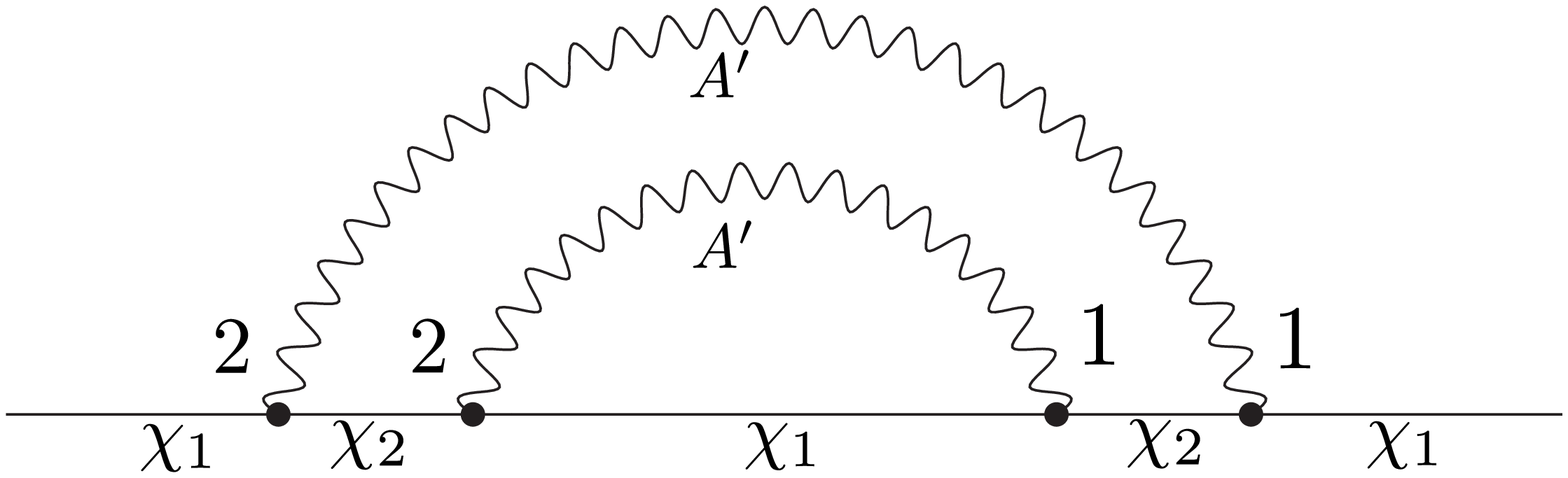}
\caption{Two-loop self-energy diagram inducing the $2 \leftrightarrow 2$ contributions to the production rates.}
\label{TwoLoop}
\end{figure}
\begin{figure}
\includegraphics[width=0.4\textwidth]{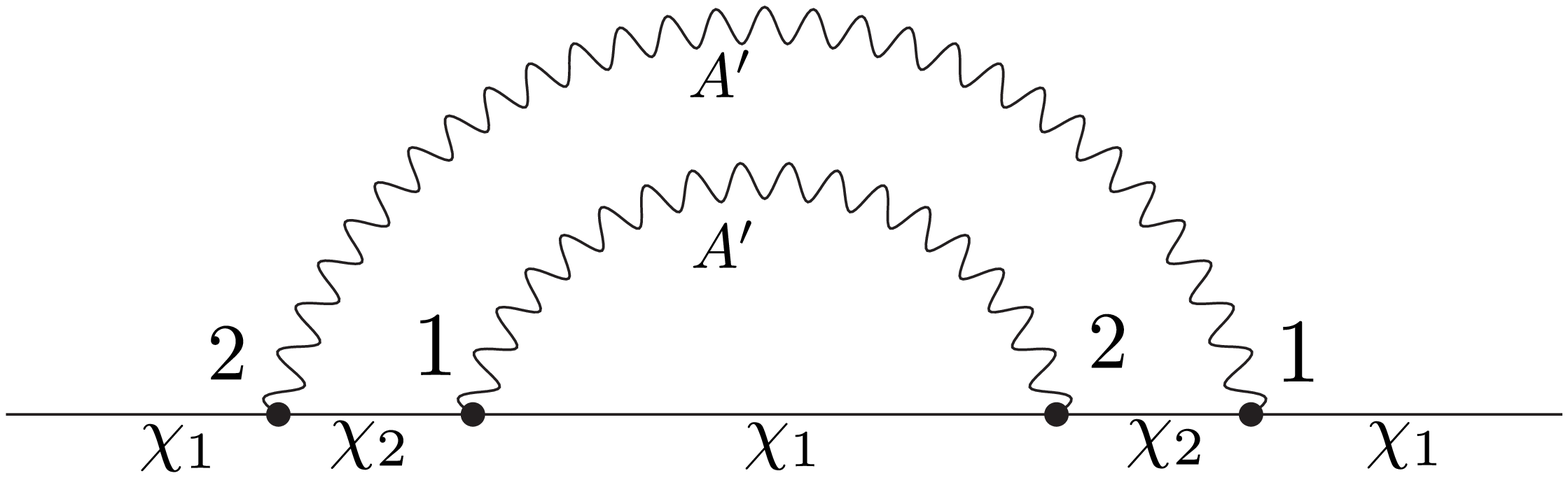}
\caption{Two-loop self-energy diagram inducing the $2 \leftrightarrow 2$ contributions to the production rates.}
\label{TwoLoop_Anomaly}
\end{figure}

The $1 \leftrightarrow 2$ processes are extracted from the one-loop diagrams. One of the example is listed in Fig.~\ref{OneLoop}.  The $D_{21}^F$ propagators connecting the type-1 and type-2 vertices are finally reduced to the on-shell phase-space integrals. Therefore, the one loop self-energy diagram in Fig.~\ref{OneLoop} is cut into the tree-level $1 \leftrightarrow 2$ diagrams which is compatible with our previous evaluations.

The $2 \leftrightarrow 2$ processes arise from the two-loop diagrams.  We show two examples in Fig.~\ref{TwoLoop}. Cutting the propagators between the type-1 and type-2 vertices induce the $2 \leftrightarrow 2$ diagrams. The left panel entails the s-channel process,  and the right panel gives the t-channel processes.

Other arrangements of the vertex types are possible. In other cases, the two-loop diagrams can be cut into more than one processes. For example, Fig.~\ref{TwoLoop_Anomaly} results in four parts if we cut all the connections between the type-1 and type-2 vertices, indicating successive real processes.  However, when all the propagators connecting the type-1 and type-2 vertices become on-shell,  the one-loop induced $1 \leftrightarrow 2$ contributions will also arise with dominate result due to the lower perturbative orders. Therefore, we can safely omit all high-order two-loop processes when there exist non-zero one-loop contributions inducing the $1 \leftrightarrow 2$ processes. This is actually what we did in our previous operations.

\section{Detailed Formulas Evaluating the Freeze-in Processes} \label{DetailsFreezeIn}
\subsection{$1 \leftrightarrow 2$ Processes}

The $1 \leftrightarrow 2$ processes include the $A^{\prime} \leftrightarrow \chi_1 \chi_2$ and the $\phi_{1,2} \leftrightarrow \chi_{i}\chi_{i}$(i=1,2) processes. The longitudinal polarization of $A^{\prime}$ involves the accessory $\phi_{A^{\prime} \eta} \leftrightarrow \chi_1 \chi_2$ terms. Before $\phi_{s,w}$ acquire the VEVs, $\phi_{s\eta, w\eta}$ become massive and might decay into $\chi_1 \chi_2$. All the possible diagrams are listed in Fig.~\ref{OneTwoDiagrams}.

Let us calculate the $A^{\prime} \leftrightarrow \chi_{1} \chi_{2}$ processes at first. In the practical evaluation, without loss of generality we rotate to the coordination that the $A^{\prime}$ momentum $p_{A^{\prime}}^{\mu} =(E_{A^{\prime}}, 0, 0, p_{A^{\prime}})$.  The polarization vectors are extended to 5-dimensional vectors, with the additional element assigned to the Goldstone degree of freedom, as illustrated in Appendix~\ref{ThermalCorrectionGaugeFermion}. The transverse polarization vector $\epsilon_{\pm}^{M}(k)=(\epsilon_{\pm}^{\mu}, 0)$, and $\epsilon_{\pm}^{\mu}(k)=(0, 1, \pm i, 0)$. The inward longitudinal polarization vector $\epsilon_{L}^M(k)$ is given by (\ref{epsilonL}).

The 4-dimensional vector part of the matrix element is given by
\begin{eqnarray}
i \mathcal{M}_{r,s,V}^{\mu}= t_{\chi} g_D (p_{A^{\prime}}) \overline{u}_r(p_{\chi_1}) \gamma^{\mu} v_s(p_{\chi_2}).
\end{eqnarray}
The Goldstone part of the matrix element is
\begin{eqnarray}
i \mathcal{M}_{r,s,V}^{4} = i y_{A^{\prime} \eta} \overline{u}_r(p_{\chi_1}) v_s(p_{\chi_2}).
\end{eqnarray}
Here, $p_{\chi_1,\chi_2}$ are the momentum of the dark matter particles $\chi_{1,2}$ which satisfy the energy-momentum conservation laws $p_{\chi_1}+p_{\chi_2}=p_{A^{\prime}}$ . Considering the statistic and renormalization factors, the summed squared matrix elements for a particular polarization vector $\epsilon_{t}^M(p_{A^{\prime}})$ where $t=\pm,L$ are given by
\begin{eqnarray}
A_{A^{\prime},t}=\sum_{r,s,m,n} \mathcal{M}_{r,s,V}^M \mathcal{M}_{r,s,V}^{*N} \epsilon_{t M}(p_{A^{\prime}})^* \epsilon_{t N}(p_{A^{\prime}}) f_B(\frac{p_{A^{\prime}}^0}{T}) Z_{t},
\end{eqnarray}
where the index $t$ does not undergo an Einstein summation. The definition of $Z_{t}$ is given by Eq.~(\ref{ZAPrime}), where $T$ in Eq.~(\ref{ZAPrime}) indicates both ``$+$'' and ``$-$''.  Besides the $f_B$ appeared above, we give both the definitions of the fermionic and bosonic statistic factors $f_{F,B}$,
\begin{eqnarray}
f_F(x) &=& \frac{1}{e^x+1}, \nonumber \\
f_B(x) &=& \frac{1}{e^x-1}.
\end{eqnarray}

The thermally averaged interaction rate can be the expressed as
\begin{eqnarray}
\gamma_{A^{\prime},t} &=& \int \frac{d^3 \vec{p}_{A^{\prime}} d^3 \vec{p}_{\chi_1} d^3 \vec{p}_{\chi_2}}{(2 \pi)^9 (2 p_{A^{\prime}}^0)  (2 p_{\chi_1}^0)  (2 p_{\chi_2}^0)} A_{A^{\prime},t} (2 \pi)^4 \delta^4(p_{A^{\prime}}-p_{\chi_1}-p_{\chi_2}) \nonumber \\
&=& \int \frac{d^3 \vec{p}_{A^{\prime}} d^3 \vec{p}_{\chi_1}}{(2 \pi)^9 (2 p_{A^{\prime}}^0)  (2 p_{\chi_1}^0)  (2 p_{\chi_2}^0)} A_{A^{\prime},t} (2 \pi)^4 \delta(p_{A^{\prime}}^0-p_{\chi_1}^0-p_{\chi_2}^0). \label{gammaAPrime}
\end{eqnarray}
Notice that the $p_{A^{\prime}}^0$ appeared in (\ref{gammaAPrime}) can be extracted from solving the dispersion equations
\begin{eqnarray}
p_{A^{\prime}}^2 - m_{A^{\prime}}^2 - \Pi_{T,L}(p_{A^{\prime}})=0,
\end{eqnarray}
where $\Pi_{T,L}$ are given by (\ref{Pi_TL}), and one should adopt $\Pi_T$ for $t=\pm$, $\Pi_L$ for $t=L$.

To integrate out the $d^3 \vec{p}_{\chi_1}$, one can boost into the frame where $A^{\prime}$ is at rest. Define $\beta=\frac{|\vec{p}_{A^{\prime}}|}{p_{A^{\prime}}^0}$, $\gamma=\frac{1}{\sqrt{1-\beta^2}}$, we have
\begin{eqnarray}
p_{\chi_1 A^{\prime}}^0 &=& \gamma(p_{\chi_1}^0-\beta |\vec{p}_{\chi_1}| \cos \theta_{\chi_1}), \\
|\vec{p}_{\chi_1 A^{\prime}}| \cos \theta_{\chi_1 A^{\prime}} &=& \gamma (|\vec{p}_{\chi_1}| \cos \theta_{\chi_1} - \beta p_{\chi_1}^0), \\
|\vec{p}_{\chi_1 A^{\prime}}| \sin \theta_{\chi_1 A^{\prime}} &=& |\vec{p}_{\chi_1}| \sin \theta_{\chi_1} ,
\end{eqnarray}
where $\theta_{\chi_1}$ is the angle between $\vec{p}_{\chi_1}$ and $\vec{p}_{A^{\prime}}$, $p_{\chi_1 A^{\prime}}=(p_{\chi_1 A^{\prime}}^0, \vec{p}_{\chi_1 A^{\prime}})$ is the $\chi_1$ momentum in the $A^{\prime}$'s rest frame, and $\theta_{\chi_1 A^{\prime}}$ is the corresponding direction after the boost. Since we ignore the thermal corrections on $\chi_{1,2}$, so their dispersion relations are not corrected, then
\begin{eqnarray}
\int \frac{d^3 \vec{p}_{\chi_1}}{(2 \pi)^6 (2 p_{\chi_1}^0)(2 p_{\chi_2}^0)} (2 \pi)^4 \delta(p_{A^{\prime}}^0-p_{\chi_1}^0-p_{\chi_2}^0) = \int \frac{\sin \theta_{\chi_1 A^{\prime}} d\theta_{\chi_1 A^{\prime}} d \phi_{\chi_1}}{4 \pi} \frac{2 |\vec{p}_{\chi_1 A^{\prime}}|}{\sqrt{p_{A^{\prime}}^2}}, \label{PhaseSpaceChi}
\end{eqnarray}
where $\phi_{\chi_1}$ is the azimuth angle of $\vec{p}_{\chi_1}$ relative to the $\vec{p}_{A^{\prime}}$ vector direction, and this angle remains unchanged after the boost.

Substituting Eq.~(\ref{PhaseSpaceChi}) into Eq.~(\ref{gammaAPrime}), we can collect all the elements to calculate the $\gamma_{A^{\prime},t}$.

In the finite temperature case, the remained ``free'' Goldstone degree of freedom, which corresponds to the tachyonic branching cut in the $A^{\prime}$ propagator,  can be approximated to be a massless particle-like object. The direct production of $\chi_1 \chi_2$ from an ``on-shell'' Goldstone fraction is kinematically forbidden, so we do not need to calculate this channel.

$\phi_{1,2}$ can also decay into a pair of $\chi_1 \chi_1$ or $\chi_2 \chi_2$. The dispersion relation of a scalar particle is simpler than the vector boson, and can be straightforwardly written as
\begin{eqnarray}
p_{\phi_i}^2=m_i^2,
\end{eqnarray}
where $m_i$ are the masses of the two scalar eigenstates defined in (\ref{MassEigenScalar}). The matrix element is given by
\begin{eqnarray}
i \mathcal{M}_{\phi_i, \chi_j} = i y_{ji}  \overline{u}_r(p_{\chi_j 1}) v_s(p_{\chi_j 2}),
\end{eqnarray}
where $i,j=1,2$, $p_{\chi_j 1,2}$ are the momenta of the first and second $\chi_j$ particle. Then the summed squared matrix elements are given by
\begin{eqnarray}
A_{\phi_1, \chi_j} = \frac{1}{2} \sum_{r,s} \mathcal{M}_{\phi_i, \chi_j} \mathcal{M}_{\phi_i, \chi_j}^* f_B(\frac{p_{\phi_i}^0}{T}).
\end{eqnarray}
The additional $\frac{1}{2}$ is the interchanging factor of the identical particles. 

The phase space integral can be performed with numerical algorithms, however, just as when we calculate the dark matter freeze-out processes, applying the Maxwell distribution $e^{-\frac{p_{\phi_i}}{T}}$ to estimate the $f_B(\frac{p_{\phi_i}^0}{T})$ can significantly simplify the phase space integral,
\begin{eqnarray}
\gamma_{\phi_i, \chi_j} &=& \int \frac{d^3 \vec{p}_{\phi_i} d^3 \vec{p}_{\chi_j 1} d^3 \vec{p}_{\chi_j 2}}{(2 \pi)^9 (2 p_{\phi_i}^0)  (2 p_{\chi_j 1}^0)  (2 p_{\chi_j 2}^0)} A_{\phi_i \chi_j} (2 \pi)^4 \delta^4(p_{\phi_i}-p_{\chi_j 1}-p_{\chi_j 2}) \nonumber \\
&\approx& \frac{T^3}{2 \pi^2} z_{\phi_i}^2 K_1(z_{\phi_i}) \Gamma_{\phi_i, \chi_j},  \label{gammaPhiI}
\end{eqnarray}
where $z_{\phi_i} = \frac{m_i}{T}$, $K_1$ is a Bessel function, and
\begin{eqnarray}
\Gamma_{\phi_i, \chi_j} = \frac{\sqrt{\frac{m_{i}^2}{4}-m_{\chi_j}^2}}{8 \pi m_{i}^2} A_{\phi_1, \chi_j}.
\end{eqnarray}

When all of the $\phi_{s,w}=0$, both $\phi_{s\eta, w\eta}$ can become massive to open the $\phi_{s\eta, w\eta} \rightarrow \chi_1 \chi_2$ channel. The evaluations of the corresponding $\gamma_{\phi_{s\eta, w\eta}}$ are very similar to $\gamma_{\phi_i, \chi_j}$, so we do not show the details here.

\subsection{$2 \leftrightarrow 2$ processes without the SM external states}

The $2 \leftrightarrow 2$ processes without the SM external states include $A^{\prime} A^{\prime} \leftrightarrow \chi_i \chi_i$, $A^{\prime} \phi_{A^{\prime}\eta, G\eta} \leftrightarrow \chi_i \chi_i$, $\phi_{A^{\prime}\eta, G\eta} \phi_{A^{\prime}\eta, G\eta} \leftrightarrow \chi_i \chi_i$, $\phi_{1,2} \phi_{1,2} \leftrightarrow \chi_i \chi_i$, $\phi_{1,2} A^{\prime} \leftrightarrow \chi_1 \chi_2$, $\phi_{1,2} \phi_{A^{\prime}\eta, G\eta} \leftrightarrow \chi_1 \chi_2$ channels. We will enumerate their contributions to the production rate of the dark matter.

For the $A^{\prime} A^{\prime} \leftrightarrow \chi_1 \chi_2$ channel, the Feynmann diagrams are listed in Fig.~\ref{AADiagrams}, \ref{GADiagrams}, and \ref{GGDiagrams}. The pure vector part of the matrix elements are given by
\begin{eqnarray}
i \mathcal{M}_{A^{\prime} A^{\prime}, \chi_i rs}^{M N} = i \mathcal{M}_{A^{\prime} A^{\prime}, s, \chi_i rs}^{M N} +i \mathcal{M}_{A^{\prime} A^{\prime}, t, \chi_i rs}^{M N} +i \mathcal{M}_{A^{\prime} A^{\prime}, u, \chi_i rs}^{M N},
\end{eqnarray}
where
\begin{eqnarray}
i \mathcal{M}_{A^{\prime} A^{\prime}, s, \chi_i rs}^{\mu \nu} &=& -g^{\mu \nu} \overline{u}_r(p_{\chi_i 1}) v_s(p_{\chi_i 2}) \sum_{j=1,2} \frac{i y_{ij} G_{A^{\prime} A^{\prime} j} }{(p_{\chi_i 1}+p_{\chi_i 2})^2-m_{\phi_j}^2}, \\
i \mathcal{M}_{A^{\prime} A^{\prime}, t, \chi_i rs}^{\mu \nu} &=& - \overline{u}_r(p_{\chi_i 1}) \gamma^{\mu} \frac{g_{\chi}^2 i}{p\!\!\!/_{\chi_i 1}-p\!\!\!/_{A^{\prime} 1}-m_{\chi_{(3-i)}}} \gamma^{\nu} v_s(p_{\chi_i 2}), \\
i \mathcal{M}_{A^{\prime} A^{\prime}, u, \chi_i rs}^{\mu \nu} &=& - \overline{u}_r(p_{\chi_i 1}) \gamma^{\nu} \frac{g_{\chi}^2 i}{p\!\!\!/_{\chi_i 1}-p\!\!\!/_{A^{\prime} 2}-m_{\chi_{(3-i)}}} \gamma^{\mu} v_s(p_{\chi_i 2}).
\end{eqnarray}
Here, of course $p_{A^{\prime} (1,2)}$ and $p_{\chi_i (1,2)}$ are the momentum of the external particles respectively, and they satisfy $p_{A^{\prime} 1} + p_{A^{\prime} 2} = p_{\chi_i 1} + p_{\chi_i 2}$.

The Goldstone-vector part of the matrix element characterized by Fig.~\ref{GADiagrams} is given by
\begin{eqnarray}
i \mathcal{M}_{A^{\prime} A^{\prime}, \chi_i rs}^{4 \mu} = i \mathcal{M}_{A^{\prime} A^{\prime}, s, \chi_i rs}^{4 \mu} +i \mathcal{M}_{A^{\prime} A^{\prime}, t, \chi_i rs}^{4 \mu} +i \mathcal{M}_{A^{\prime} A^{\prime}, u, \chi_i rs}^{4 \mu},
\end{eqnarray}
where
\begin{eqnarray}
i \mathcal{M}_{A^{\prime} A^{\prime}, s, \chi_i rs}^{4 \mu} &=& (p_{\chi_i 1}+p_{\chi_i 2} + p_{A^{\prime} 1})^{\mu} \overline{u}_r(p_{\chi_i 1}) v_s(p_{\chi_i 2}) \sum_{j=1,2} \frac{y_{ij} g_{j A^{\prime}}  }{(p_{\chi_i 1}+p_{\chi_i 2})^2-m_{\phi_j}^2}, \\
i \mathcal{M}_{A^{\prime} A^{\prime}, t, \chi_i rs}^{4 \mu} &=& (\delta_{1 i}-\delta_{2 i}) \overline{u}_r(p_{\chi_i 1}) \frac{g_{\chi} y_{A^{\prime}}}{p\!\!\!/_{\chi_i 1}-p\!\!\!/_{A^{\prime} 1}-m_{\chi_{(3-i)}}} \gamma^{\mu} v_s(p_{\chi_i 2}), \\
i \mathcal{M}_{A^{\prime} A^{\prime}, u, \chi_i rs}^{4 \mu} &=&  -(\delta_{1 i}-\delta_{2 i}) \overline{u}_r(p_{\chi_i 1}) \gamma^{\mu}  \frac{g_{\chi} y_{A^{\prime}} }{p\!\!\!/_{\chi_i 1}-p\!\!\!/_{A^{\prime} 2}-m_{\chi_{(3-i)}}} v_s(p_{\chi_i 2}).
\end{eqnarray}

The evaluation of $i \mathcal{M}_{A^{\prime} A^{\prime}, \chi_i rs}^{\mu 4} $ is similar to $i \mathcal{M}_{A^{\prime} A^{\prime}, \chi_i rs}^{4 \mu}$, except the interchanging between $p_{A^{\prime} 1}$ and $p_{A^{\prime} 2}$.

The Goldstone-Goldstone part of the matrix element characterized by Fig.~\ref{GGDiagrams} is given by
\begin{eqnarray}
i \mathcal{M}_{A^{\prime} A^{\prime}, \chi_i rs}^{4 4} = i \mathcal{M}_{A^{\prime} A^{\prime}, s, \chi_i rs}^{4 4} +i \mathcal{M}_{A^{\prime} A^{\prime}, t, \chi_i rs}^{4 4} +i \mathcal{M}_{A^{\prime} A^{\prime}, u, \chi_i rs}^{4 4},
\end{eqnarray}
where
\begin{eqnarray}
i \mathcal{M}_{A^{\prime} A^{\prime}, s, \chi_i rs}^{4 4} &=& -\overline{u}_r(p_{\chi_i 1}) v_s(p_{\chi_i 2}) \sum_{j=1,2} \frac{i y_j A_{j A^{\prime} A^{\prime}} }{(p_{\chi_i 1}+p_{\chi_i 2})^2-m_{\phi_j}^2},  \\
i \mathcal{M}_{A^{\prime} A^{\prime}, t, \chi_i rs}^{4 4} &=& -\overline{u}_r(p_{\chi_i 1}) \frac{y_{A^{\prime}}^2 i}{p\!\!\!/_{\chi_i 1}-p\!\!\!/_{A^{\prime} 1}-m_{\chi_{(3-i)}}} v_s(p_{\chi_i 2}), \\
i \mathcal{M}_{A^{\prime} A^{\prime}, u, \chi_i rs}^{4 4} &=& -\overline{u}_r(p_{\chi_i 1}) \frac{y_{A^{\prime}}^2 i}{p\!\!\!/_{\chi_i 1}-p\!\!\!/_{A^{\prime} 2}-m_{\chi_{(3-i)}}} v_s(p_{\chi_i 2}).
\end{eqnarray}

Therefore, the summed squared matrix elements are
\begin{eqnarray}
A_{A^{\prime} A^{\prime} \rightarrow \chi_i \chi_i, t_1 t_2 } &=& \frac{1}{2} \sum_{r,s,M,N,O,P}  \epsilon_{t_1 M}^*(p_{A^{\prime} 1}) \epsilon_{t_2 N}^*(p_{A^{\prime} 2}) \epsilon_{t_1 O}(p_{A^{\prime} 1}) \epsilon_{t_2 P}(p_{A^{\prime} 2}) \nonumber \\
& & \mathcal{M}_{A^{\prime} A^{\prime}, \chi_i r s}^{MN} \mathcal{M}_{A^{\prime} A^{\prime}, \chi_i r s}^{OP*} f_B(\frac{p_{A^{\prime} 1}^0}{T}) f_B(\frac{p_{A^{\prime} 2}^0}{T}) Z_{t_1} Z_{t_2},
\end{eqnarray}
where $t_1$, $t_2=\pm,L$ denote the polarization of the vector bosons. Then the corresponding production rates are given by
\begin{eqnarray}
\gamma_{A^{\prime} A^{\prime}, i t_1 t_2} = & & \int \frac{d^3 \vec{p}_{A^{\prime} 1} d^3 \vec{p}_{A^{\prime} 2} d^3 \vec{p}_{\chi_1} d^3 \vec{p}_{\chi_2}}{(2 \pi)^{12} (2 p_{A^{\prime} 1}^0) (2 p_{A^{\prime} 2}^0)  (2 p_{\chi_i 1}^0)  (2 p_{\chi_i 2}^0)} A_{A^{\prime} A^{\prime} \rightarrow \chi_i \chi_i, t_1 t_2 }  \nonumber \\
& \times & (2 \pi)^4 \delta^4(p_{A^{\prime} 1} + p_{A^{\prime} 2} - p_{\chi_i 1}-p_{\chi_i 2}).  \label{gammaAPrimeAPrime}
\end{eqnarray}

The $A^{\prime} \phi_{A^{\prime} \eta} \leftrightarrow \chi_1 \chi_2$ processes involve the Fig.~\ref{GADiagrams}, \ref{GGDiagrams}. Here, $\phi_{A^{\prime}}$ is the fraction of Goldstone boson vomited out by the longitudinal polarization of $A^{\prime}$. Rigorously speaking this is a ``branch cut'' rather than a ``particle''. However, as illustrated in Appendix.~\ref{ThermalCorrectionGaugeFermion}, it can be estimated as a massless particle. The matrix elements are given by
\begin{eqnarray}
i \mathcal{M}_{A^{\prime} \phi_{A^{\prime} \eta}, \chi_i rs}^{M} = i \mathcal{M}_{A^{\prime} \phi_{A^{\prime} \eta}, s, \chi_i rs}^{M} +i \mathcal{M}_{A^{\prime} \phi_{A^{\prime} \eta}, t, \chi_i rs}^{M} +i \mathcal{M}_{A^{\prime} \phi_{A^{\prime} \eta}, u, \chi_i rs}^{M}, \label{MAPrimePhiAPrime}
\end{eqnarray}
where
\begin{eqnarray}
i \mathcal{M}_{A^{\prime}  \phi_{A^{\prime} \eta}, s, \chi_i rs}^{\mu} &=& (p_{\chi_i 1}+p_{\chi_i 2} + p_{\phi_{A^{\prime}}})^{\mu} \overline{u}_r(p_{\chi_i 1}) v_s(p_{\chi_i 2}) \sum_{j=1,2} \frac{y_{ij} g_{j A^{\prime}}  }{(p_{\chi_i 1}+p_{\chi_i 2})^2-m_{\phi_j}^2}, \\
i \mathcal{M}_{A^{\prime}  \phi_{A^{\prime} \eta}, t, \chi_i rs}^{\mu} &=& (\delta_{1 i}-\delta_{2 i}) \overline{u}_r(p_{\chi_i 1}) \frac{g_{\chi} y_{A^{\prime}} }{p\!\!\!/_{\chi_i 1}-p\!\!\!/_{\phi_{A^{\prime} \eta}}-m_{\chi_{(3-i)}}} \gamma^{\mu} v_s(p_{\chi_i 2}), \\
i \mathcal{M}_{A^{\prime} \phi_{A^{\prime} \eta}, u, \chi_i rs}^{\mu} &=&  -(\delta_{1 i}-\delta_{2 i}) \overline{u}_r(p_{\chi_i 1}) \gamma^{\mu}  \frac{g_{\chi} y_{A^{\prime}} }{p\!\!\!/_{\chi_i 1}-p\!\!\!/_{A^{\prime}}-m_{\chi_{(3-i)}}} v_s(p_{\chi_i 2}),
\end{eqnarray}
and
\begin{eqnarray}
i \mathcal{M}_{A^{\prime}  \phi_{A^{\prime} \eta}, s, \chi_i rs}^{4} &=& -\overline{u}_r(p_{\chi_i 1}) v_s(p_{\chi_i 2}) \sum_{j=1,2} \frac{y_j A_{j A^{\prime} A^{\prime}} i}{(p_{\chi_i 1}+p_{\chi_i 2})^2-m_{\phi_j}^2},  \\
i \mathcal{M}_{A^{\prime}  \phi_{A^{\prime} \eta}, t, \chi_i rs}^{4} &=& -\overline{u}_r(p_{\chi_i 1}) \frac{y_{A^{\prime}}^2 i}{p\!\!\!/_{\chi_i 1}-p\!\!\!/_{\phi_{A^{\prime} \eta}}-m_{\chi_{(3-i)}}} v_s(p_{\chi_i 2}), \\
i \mathcal{M}_{A^{\prime}  \phi_{A^{\prime} \eta}, u, \chi_i rs}^{4} &=&  -\overline{u}_r(p_{\chi_i 1}) \frac{y_{A^{\prime}}^2 i}{p\!\!\!/_{\chi_i 1}-p\!\!\!/_{A^{\prime} \eta}-m_{\chi_{(3-i)}}} v_s(p_{\chi_i 2}).
\end{eqnarray}
Here, $p_{\chi_i(1,2)}$, $p_{A^{\prime}}$ and $p_{\phi_{A^{\prime} \eta}}$ are the momentum of the corresponding particles, and notice we adopt the estimation of $p_{\phi_{A^{\prime} \eta}}^2=0$. The summed squared matrix elements are therefore
\begin{eqnarray}
A_{A^{\prime} \phi_{A^{\prime} \eta} \rightarrow \chi_i \chi_i, t_1 t_2 } &=& \frac{1}{2} \sum_{r,s,M,N}  \epsilon_{t_1 M}^*(p_{A^{\prime}}) \epsilon_{t_1 N}(p_{A^{\prime}}) \mathcal{M}_{A^{\prime} \phi_{A^{\prime} \eta}, \chi_i r s}^{M} \mathcal{M}_{A^{\prime} \phi_{A^{\prime} \eta} \chi_i r s}^{N*} \nonumber \\
& & f_B(\frac{p_{A^{\prime}}^0}{T}) f_B(\frac{p_{\phi_{A^{\prime} \eta}}^0}{T}) Z_{t} Z_{\text{GS}}. \label{AAPrimePhiAPrimeEta}
\end{eqnarray}
The definition of $Z_{\text{GS}}$ is given by (\ref{ZGS}). The production rates then becomes
\begin{eqnarray}
\gamma_{A^{\prime} \phi_{A^{\prime} \eta}, i t} = & & \int \frac{d^3 \vec{p}_{\phi_{A^{\prime} \eta}} d^3 \vec{p}_{A^{\prime}} d^3 \vec{p}_{\chi_1} d^3 \vec{p}_{\chi_2}}{(2 \pi)^{12} (2 p_{\phi_{A^{\prime} \eta}}^0) (2 p_{A^{\prime}}^0)  (2 p_{\chi_i 1}^0)  (2 p_{\chi_i 2}^0)} A_{A^{\prime} \phi_{A^{\prime} \eta} \rightarrow \chi_i \chi_i, t }  \nonumber \\
& \times & (2 \pi)^4 \delta^4(p_{\phi_{A^{\prime} \eta}} + p_{A^{\prime}} - p_{\chi_i 1}-p_{\chi_i 2}).  \label{gammaAPrimePhiAPrime}
\end{eqnarray}

The $\phi_{A^{\prime} \eta} \phi_{A^{\prime} \eta} \leftrightarrow \chi_i \chi_i$ only corresponds to Fig.~\ref{GGDiagrams}. The matrix elements are given by
\begin{eqnarray}
i \mathcal{M}_{\phi_{A^{\prime} \eta} \phi_{A^{\prime} \eta}, \chi_i rs} = i \mathcal{M}_{\phi_{A^{\prime} \eta} \phi_{A^{\prime} \eta}, s, \chi_i rs} +i \mathcal{M}_{\phi_{A^{\prime} \eta} \phi_{A^{\prime} \eta}, t, \chi_i rs} +i \mathcal{M}_{\phi_{A^{\prime} \eta} \phi_{A^{\prime} \eta}, u, \chi_i rs},
\end{eqnarray}
where
\begin{eqnarray}
i \mathcal{M}_{\phi_{A^{\prime}\eta}  \phi_{A^{\prime}\eta}, s, \chi_i rs} &=& -\overline{u}_r(p_{\chi_i 1}) v_s(p_{\chi_i 2}) \sum_{j=1,2} \frac{i y_j A_{j A^{\prime} A^{\prime}} }{(p_{\chi_i 1}+p_{\chi_i 2})^2-m_{\phi_j}^2},  \\
i \mathcal{M}_{\phi_{A^{\prime}\eta}  \phi_{A^{\prime}\eta}, t, \chi_i rs} &=& -\overline{u}_r(p_{\chi_i 1}) \frac{y_{A^{\prime} \eta}^2 i}{p\!\!\!/_{\chi_i 1}-p\!\!\!/_{\phi_{A^{\prime} \eta} 1}-m_{\chi_{(3-i)}}} v_s(p_{\chi_i 2}), \\
i \mathcal{M}_{\phi_{A^{\prime}\eta}  \phi_{A^{\prime}\eta}, u, \chi_i rs} &=& - \overline{u}_r(p_{\chi_i 1}) \frac{y_{A^{\prime} \eta}^2 i}{p\!\!\!/_{\chi_i 1}-p\!\!\!/_{\phi_{A^{\prime} \eta} 2}-m_{\chi_{(3-i)}}} v_s(p_{\chi_i 2}).
\end{eqnarray}
The squared matrix elements are
\begin{eqnarray}
A_{\phi_{A^{\prime} \eta} \phi_{A^{\prime} \eta} \rightarrow \chi_i \chi_i, t_1 t_2 } &=& \frac{1}{2} \sum_{r,s} \mathcal{M}_{A^{\prime} \phi_{A^{\prime} \eta}, \chi_i r s} \mathcal{M}_{A^{\prime} \phi_{A^{\prime} \eta} \chi_i r s}^{*}  \nonumber \\
& & f_B(\frac{p_{A^{\prime}}^0}{T}) f_B(\frac{p_{\phi_{A^{\prime}}}^0}{T}) Z_{\text{GS}}(p_{\phi_{A^{\prime} \eta} 1}) Z_{\text{GS}}(p_{\phi_{A^{\prime} \eta} 2}),
\end{eqnarray}
so the production rates are expressed as
\begin{eqnarray}
\gamma_{\phi_{A^{\prime} \eta} \phi_{A^{\prime} \eta}, i } = & & \int \frac{d^3 \vec{p}_{\phi_{A^{\prime} \eta } 1} d^3 \vec{p}_{\phi_{A^{\prime} \eta} 2} d^3 \vec{p}_{\chi_1} d^3 \vec{p}_{\chi_2}}{(2 \pi)^{12} (2 p_{\phi_{A^{\prime} \eta} 1}^0) (2 p_{\phi_{A^{\prime} \eta} 2}^0)  (2 p_{\chi_i 1}^0)  (2 p_{\chi_i 2}^0)} A_{\phi_{A^{\prime} \eta} \phi_{A^{\prime} \eta} \rightarrow \chi_i \chi_i}  \nonumber \\
& \times & (2 \pi)^4 \delta^4(p_{\phi_{A^{\prime} \eta } 1} + p_{\phi_{A^{\prime} \eta } 2} - p_{\chi_i 1}-p_{\chi_i 2}).  \label{gammaPhiAPrimePhiAPrime}
\end{eqnarray}

Fig.~\ref{GADiagrams} and Fig.~\ref{GGDiagrams} also stand for the $A^{\prime} \phi_{G\eta} \leftrightarrow \chi_i \chi_i$, $\phi_{A^{\prime} \eta} \phi_{G\eta} \leftrightarrow \chi_i \chi_i$ and $\phi_{G\eta} \phi_{G\eta} \leftrightarrow \chi_i \chi_i$ channels. The basic steps are exactly the same with those in (\ref{MAPrimePhiAPrime})-(\ref{gammaPhiAPrimePhiAPrime}) to calculate the $\gamma_{A^{\prime} \phi_{G \eta}, it}$, $\gamma_{\phi_{A^{\prime} \eta} \phi_{G \eta}, i}$ and $\gamma_{\phi_{G \eta} \phi_{G \eta}, i}$ except that one needs to modify the momentum symbols, the coupling constants, and abolish the $Z_{\text{GS}}$ when a $\phi_{G\eta}$ is replacing the $\phi_{A^{\prime}\eta}$.

The processes $\phi_{1,2} A^{\prime} \leftrightarrow \chi_1 \chi_2$,  $\phi_{A^{\prime}\eta, G\eta} \phi_{1,2} \leftrightarrow \chi_1 \chi_2$ and $\phi_{1,2} \phi_{1,2} \leftrightarrow \chi_i \chi_i$, as indicated in Fig.~\ref{HADiagrams}, \ref{GHDiagrams} and \ref{HHDiagrams} also contribute to the production rates $\gamma_{\phi_i A^{\prime}, t}$, $\gamma_{G\eta A^{\prime}, t}$ and $\gamma_{\phi_{i} \phi_{j}, k}$, and the evaluation processes are very similar to the previous channels we discussed, except that one needs to select the appropriate couplings from Tab.~\ref{CouplingsTable}, and to manipulate the ``renormalization factors'' properly. In the $X \leftrightarrow \chi_1 \chi_2$ processes, the identical particle factor $\frac{1}{2}$ is also discarded. We are not going to enumerate all of the detailed processes in this paper, however, we would like to point out that the second diagrams in both Fig.~\ref{HADiagrams} and \ref{GHDiagrams} are special, since the intermediate s-channel particle is the mixed propagators (\ref{FullPropagatorV}) among $\phi_{A^{\prime} \eta}$ and $A^{\prime}$. As an example, we calculate the corresponding matrix elements of $\phi_{1,2} A^{\prime} \leftrightarrow \chi_1 \chi_2$ to show the detailed evaluation processes.

The matrix elements of this diagram is denoted as $i \mathcal{M}_{A^{\prime} \phi_{i}, sV, rs}^M$, where subscript ``$sV$'' means ``the vector boson mediated s-channel''. The vector part of it is given by
\begin{eqnarray}
i \mathcal{M}_{A^{\prime} \phi_{i}, sV, rs}^{\mu} &=& i g_{\chi} G_{A^{\prime} A^{\prime} i} \overline{u}_r(p_{\chi_1}) \gamma^{\nu} v_s(p_{\chi_2}) D_{0, \nu \lambda}^{\text{full}}(p_{\chi_1}+p_{\chi_2}) g^{\lambda \mu} \nonumber \\
&+& g_{\chi} g_{i A^{\prime}} \overline{u}_r(p_{\chi_1}) \gamma^{\nu} v_s(p_{\chi_2}) D_{0, \nu 4}^{\text{full}}(p_{\chi_1}+p_{\chi_2}) (p_{\chi_1} + p_{\chi_2}+p_{\phi_i})^{\mu} \label{MAPrimePhiMu} \\
&-& y_{A^{\prime} \eta} G_{A^{\prime} A^{\prime} i} \overline{u}_r(p_{\chi_1}) v_s(p_{\chi_2}) D_{0, 4 \lambda}^{\text{full}}(p_{\chi_1}+p_{\chi_2}) g^{\lambda \mu}  \nonumber \\
&+& i y_{A^{\prime} \eta} g_{i A^{\prime}} \overline{u}_r(p_{\chi_1}) v_s(p_{\chi_2}) D_{0, 44}^{\text{full}}(p_{\chi_1}+p_{\chi_2}) (p_{\phi_i} + p_{\chi_1} + p_{\chi_2})^{\mu} \nonumber
\end{eqnarray}
The Goldstone part of them is given by
\begin{eqnarray}
i \mathcal{M}_{A^{\prime} \phi_{i}, sV, rs}^{4} &=& i g_{\chi}  \overline{u}_r(p_{\chi_1}) \gamma^{\nu} v_s(p_{\chi_2}) D_{0, \nu 4}^{\text{full}}(p_{\chi_1}+p_{\chi_2}) A_{i G A^{\prime}} \nonumber \\
&-& y_{A^{\prime} \eta} \overline{u}_r(p_{\chi_1}) v_s(p_{\chi_2}) D_{0, 44}^{\text{full}}(p_{\chi_1}+p_{\chi_2}) A_{i G A^{\prime}}.  \label{MAPrimePhi4}
\end{eqnarray}
(\ref{MAPrimePhiMu})-(\ref{MAPrimePhi4}) participate in composing $i \mathcal{M}_{A^{\prime} \phi_{i}, rs}^{4}$, and just like the processes from (\ref{AAPrimePhiAPrimeEta})-(\ref{gammaAPrimePhiAPrime}),  one finally arrives at $\gamma_{\phi_i A^{\prime}, t}$. 

\subsection{$2 \leftrightarrow 2$ processes with the SM external states}

The $2 \leftrightarrow 2$ Processes with the SM external states include $\overline{f} f \leftrightarrow \chi_1 \chi_2$ and $H^+ H^-  \leftrightarrow \chi_1 \chi_2$, where $f = e_L$,  $\mu_L$,  $\tau_L$, $e_R$,  $\mu_R$, $\tau_R$,  $\nu_{i L}$,  $u_L$,  $d_L$,  $s_L$, $c_L$,  $t_L$,  $b_L$, $u_R$,  $d_R$,  $s_R$, $c_R$,  $t_R$,  $b_R$. Notice that in this paper, we omit the freeze-in processes below the electroweak phase-transition temperature, so the left-handed and right-handed fermions decouple and receive different thermal corrections. The Higgs doublet are also degenerate and the hyper-charge symmetry guarantees hyper-charge conservation. This always produces the Higgs particles in pairs with the opposite hyper-charges respectively. The corresponding diagrams are listed in Fig.~\ref{ChiChi2SMSM}. Both these two channels involve the mixing between the $A^{\prime}$ and the hyper-charge gauge boson $B$.

Let us calculate the $\overline{f} f \leftrightarrow \chi_1 \chi_2$ processes at first. The matrix elements are
\begin{eqnarray}
i \mathcal{M}_{\overline{f} f, pq,rs}^{\pm, \pm} &=& - g_1 Y_f \epsilon \overline{v}_p(\tilde{p}_{\overline{f}}^{\pm}) \gamma^{\mu} u_q(\tilde{p}_f^{\pm}) \left[ D_{0, \mu \nu}^{B, \text{full}} (p_f+p_{\tilde{f}})_{\lambda} - D_{0, \mu \lambda}^{B, \text{full}} (p_f+p_{\tilde{f}})_{\nu} \right] \nonumber \\
&&\left\lbrace g_{\chi} \left[ D_{0}^{\text{full}, \nu \rho} (p_f+p_{\tilde{f}})^{\lambda} - D_{0}^{\text{full}, \lambda \rho} (p_f+p_{\tilde{f}})^{\nu} \right] \overline{u}_r(p_{\chi_1}) \gamma_{\rho} v_s(p_{\chi_2}) \right. \nonumber \\
&&+ \left. y_{A^{\prime} \eta} \left[ D_{0}^{\text{full}, \nu 4} (p_f+p_{\tilde{f}})^{\lambda} - D_{0}^{\text{full}, \lambda 4} (p_f+p_{\tilde{f}})^{\nu} \right] \overline{u}_r(p_{\chi_1}) v_s(p_{\chi_2}) \right\rbrace. \label{Mff}
\end{eqnarray}
Here, the definition $D_0^{\text{full}, MN}$ is given by (\ref{FullPropagatorV}), while $D_0^{B, \text{full}, \mu \nu}$ is the corresponding thermal propagator for the hyper-charge gauge boson $B$. Compared with the (\ref{FullPropagatorV}), in $D_0^{B, \text{full}, \mu \nu}$, the $m_A^{\prime}$ should be eliminated, and the $c_{A^{\prime}}(T)$ should be replaced with the $c_{B}(T)$ in (\ref{ThermalMass}). In (\ref{Mff}), we omit all the function parameter with the bracket, so $D_0^{(B), \text{full}, MN}$ are the abbreviations for the $D_0^{(B), \text{full}, MN}(p_f+p_{\overline{f}})$.  $Y_f$ is the hypercharge of the fermion $f$. $\epsilon$ is the mixing coupling of the $A^{\prime}$ and $B$ for the Lagrangian term
\begin{eqnarray}
\mathcal{L}_{\epsilon} = -\epsilon B_{\mu \nu} F^{\prime}_{\mu \nu}, \label{YExoticMixing}
\end{eqnarray}
where $B_{\mu \nu} = \partial_{\mu} B_{\nu} - \partial_{\nu} B_{\mu}$. The definitions of $\tilde{p}_{f/\overline{f}}^{\pm}$ are given by
\begin{eqnarray}
\tilde{p}_{f/\overline{f}}^{\pm} = p_{f/\overline{f}}^0 (1, \pm\frac{\vec{p}_{f/\overline{f}}}{|\vec{p}_{f/\overline{f}}}).
\end{eqnarray}
$p_{f/\overline{f}}$ should be a solution of (\ref{FDispersion}). A ``particle'' corresponds to $p_{f/\overline{f}}^0 > m_f^2$ and $p_{f/\overline{f}}^+$. A ``hole'' corresponds to $p_{f/\overline{f}}^0 < m_f^2$ and $p_{f/\overline{f}}^-$. The definition of $m_f$ for each SM fermions is expressed in Eq.~(\ref{LeptonThermalMasses}).

The summed squared amplitudes are given by
\begin{eqnarray}
A_{\overline{f} f} &=& \sum_{a,b=\pm} \sum_{p,q,  r,s} \mathcal{M}_{\overline{f} f, pq, rs}^{a,b} \mathcal{M}_{\overline{f} f, pq, rs}^{a,b*} f_F(\frac{p_{f}^0}{T}) f_F(\frac{p_{\overline{f}}^0}{T}) Z_{f}(p_{f}) Z_{f}(p_{f}),
\end{eqnarray}
and then again
\begin{eqnarray}
\gamma_{\overline{f} f} = N_f \int \frac{d^3 \vec{p}_{\overline{f}} d^3 \vec{p}_{f} d^3 \vec{p}_{\chi_1} d^3 \vec{p}_{\chi_2}}{(2 \pi)^{12} (2 p_{\overline{f}}^0) (2 p_{f}^0)  (2 p_{\chi_i 1}^0)  (2 p_{\chi_i 2}^0)} A_{\overline{f} f}  \times (2 \pi)^4 \delta^4(p_{\overline{f}} + p_f - p_{\chi_i 1}-p_{\chi_i 2}).  \label{gammaff}
\end{eqnarray}
where $N_f$ indicates the number of color of the fermion $f$.

Finally, for the $H^+ H^- \leftrightarrow \chi_1 \chi_2$, the matrix elements are expressed below,
\begin{eqnarray}
i \mathcal{M}_{H^+ H^- pq,rs} &=& - \frac{1}{2} g_1 \epsilon (p_{H^+} - p_{H^-})^{\mu} \left[ D_{0, \mu \nu}^{B, \text{full}} (p_f+p_{\tilde{f}})_{\lambda} - D_{0, \mu \lambda}^{B, \text{full}} (p_f+p_{\tilde{f}})_{\nu} \right] \nonumber \\
&&\left\lbrace g_{\chi} \left[ D_{0}^{\text{full}, \nu \rho} (p_f+p_{\tilde{f}})^{\lambda} - D_{0}^{\text{full}, \lambda \rho} (p_f+p_{\tilde{f}})^{\nu} \right] \overline{u}_r(p_{\chi_1}) \gamma_{\rho} v_s(p_{\chi_2}) \right. \nonumber \\
&&+ \left. y_{A^{\prime} \eta} \left[ D_{0}^{\text{full}, \nu 4} (p_f+p_{\tilde{f}})^{\lambda} - D_{0}^{\text{full}, \lambda 4} (p_f+p_{\tilde{f}})^{\nu} \right] \overline{u}_r(p_{\chi_1}) v_s(p_{\chi_2}) \right\rbrace. \label{MHH}
\end{eqnarray}
Here, $p_{H^{\pm}}^2 = m_H^2$, with the definition of $m_H^2$ given in (\ref{MH}). The factor $\frac{1}{2}$ stands for the hypercharge of the SM-Higgs doublet.

The corresponding summed squared amplitudes are
\begin{eqnarray}
A_{H^+ H^-} &=& \sum_{p,q,  r,s} \mathcal{M}_{H^+ H^-, pq, rs}\mathcal{M}_{H^+ H^-, pq, rs}^* f_B(\frac{p_{H^+}^0}{T}) f_B(\frac{p_{\overline{H^-}}^0}{T}),
\end{eqnarray}
so therefore
\begin{eqnarray}
\gamma_{H^+ H^-} &=& 2 \int \frac{d^3 \vec{p}_{H^+} d^3 \vec{p}_{H^-} d^3 \vec{p}_{\chi_1} d^3 \vec{p}_{\chi_2}}{(2 \pi)^{12} (2 p_{H^+}^0) (2 p_{H^-}^0)  (2 p_{\chi_i 1}^0)  (2 p_{\chi_i 2}^0)} A_{H^+ H^-}  \nonumber \\
& & \times (2 \pi)^4 \delta^4(p_{H^+} + p_{H^-} - p_{\chi_i 1}-p_{\chi_i 2}).  \label{gammaHpHm}
\end{eqnarray}
Here the additional factor of $2$ stands for summation of both the elements in the SM Higgs doublet.

\bibliography{darkmatter}

\end{document}